\documentclass[reqno,11pt]{article}
\pdfoutput=1
\usepackage{jheppub}
\graphicspath{ {./graphs1/} }
\usepackage{epsfig}
\usepackage{amssymb}
\usepackage{verbatim}
\usepackage{amsmath}
\usepackage{hyperref}
\usepackage{dsfont}
\usepackage{slashed}
\usepackage[dvipsnames]{xcolor}
\usepackage{epstopdf}
\usepackage{graphicx}
\usepackage{color}
\usepackage{shuffle}
\usepackage{bbold}

\def\sh{\shuffle}

\def\o{\omega}

\def\L{\Lambda}

\def\s{\sigma}
\def\half{{1 \over 2}}
\def\a{\alpha}
\def\b{\beta}
\def\d{\delta}
\def\g{\gamma}

\newcommand{\bc}{\begin{center}}
\newcommand{\ec}{\end{center}}
\newcommand{\PT}{{\rm PT}}
\newcommand{\PPT}{\widehat{{\rm PT}}}

\def\tr{{\rm tr}}
\def\non{\nonumber\\}
\def\T{T}
\def\Ti{{\tilde T}}

\newcommand{\ba}{\begin{eqnarray}}
\newcommand{\ea}{\end{eqnarray}}

\title{Non-planar one-loop Parke-Taylor factors  in the CHY approach for quadratic propagators}
\author{Naser Ahmadiniaz,${}^{a}$ Humberto Gomez,${}^{b}$ Cristhiam Lopez-Arcos${}^{b}$}

\affiliation{${}^{a}$ Center for Relativistic Laser Science, Institute for Basic Science (IBS), 61005 Gwangju, Korea\\
${}^{b}$ Universidad Santiago de Cali, Facultad de Ciencias Basicas,\\
Campus Pampalinda, Calle 5 No. 62-00, C\'{o}digo postal 76001, Santiago de Cali, Colombia}

\emailAdd{ahmadiniaz@ibs.re.kr, humgomzu@gmail.com, crismalo@gmail.com}

\abstract{In this work we have studied the Kleiss-Kuijf relations for the recently introduced Parke-Taylor factors at one-loop in the CHY approach, that reproduce quadratic Feynman propagators. By doing this, we were able to identify the non-planar one-loop Parke-Taylor factors. In order to check that, in fact, these new factors can describe non-planar amplitudes, we applied  them to  the bi-adjoint $\Phi^3$ theory. As a byproduct, we found a new type of graphs that we called {\it the non-planar CHY-graphs}. These graphs encode all the information for the subleading order at one-loop, and there is not an equivalent of these in the Feynman formalism.}

\begin{document}

\maketitle


\section{Introduction} \label{sect:intro}

Among the recent developments that apply on-shell methods to the calculation of amplitudes, following Witten's work in 2003 \cite{Witten:2003nn}, the proposal by Cachazo-He-Yuan (CHY) \cite{Cachazo:2013gna,Cachazo:2013hca} offers some advantages. The CHY formalism applies to several dimensions and also to a large array of theories \cite{Cachazo:2013iea,Cachazo:2014xea,Cachazo:2014nsa,Cachazo:2016njl}, that go even beyond field theory \cite{Gomez:2013wza,Mizera:2017sen,Mizera:2017cqs,Azevedo:2017lkz,Azevedo:2017yjy}. The formalism is based on the scattering equations (at tree-level) 
\ba  
E_a := \sum_{b \ne a}\frac{k_a\cdot k_b}{\s_{ab}}=0,  \quad  \s_{ab}:=\s_a - \s_b , \quad  a=1,2,\ldots, n, 
\ea
with the $\s_a$'s denoting the local coordinates on the moduli space of n-punctured Riemann spheres and $k_a^2=0$. To obtain the tree-level S-matrix we have to perform a contour integral localized over solutions of these equations, i.e. 
\ba
{\cal A}_n = \int_{\Gamma} \,d\mu_n^{\rm tree}\,\, \cal{I}_{\rm tree}^{\rm CHY}(\s), 
\ea
where the integration measure, $d\mu_n^{\rm tree}$, is given by
\begin{equation}
d\mu_n^{\rm tree} =\frac{\prod_{a=1}^n d\s_a}{{\rm Vol}\,({\rm PSL}(2,\mathbb{C}))} \times \frac{(\s_{ij} \s_{jk} \s_{ki})}{    \prod_{b\neq i,j,k}^{n} E_b  },
\end{equation}
and the contour $\Gamma$ is defined by the $n-3$ independent scattering equations
\begin{equation}
E_b=0, \quad b\neq i,j,k\,.
\end{equation}
A different integrand, $\cal{I}_{\rm tree}^{\rm CHY}$, describes a different theory. In the study of the scattering equations at tree-level and the development of techniques for integration, many approaches have been formulated \cite{Cachazo:2013iaa,Cachazo:2013iea,Kalousios:2013eca,Lam:2014tga,Dolan:2014ega,Cachazo:2015nwa,Kalousios:2015fya,Dolan:2015iln,Huang:2015yka,Cardona:2015ouc,Cardona:2015eba, Sogaard:2015dba,Baadsgaard:2015ifa,Baadsgaard:2015voa,Cachazo:2016sdc,He:2016vfi,Cachazo:2016ror,Bosma:2016ttj,Zlotnikov:2016wtk,Chen:2016fgi,Mafra:2016ltu,Huang:2016zzb,Cardona:2016gon,Chen:2017edo,Zhou:2017mfj}. In particular, the method of integration that we use in the calculations for the present paper was developed by one of the authors in \cite{Gomez:2016bmv}, which is called the $\L-$algorithm.

The following step for the CHY formalism is going to loop corrections. A prescription that allows to go to higher genus Riemann surfaces was developed in \cite{Mason:2013sva,Berkovits:2013xba,Geyer:2015bja,Geyer:2015jch,Adamo:2015hoa}, which is called the ambitwistor and pure spinor ambitwistor string theory. Another alternative approach employing an elliptic curve was proposed in \cite{Cardona:2016wcr, Cardona:2016bpi} by one of the authors. Additionally, in \cite{Baadsgaard:2015hia,He:2015yua,Cachazo:2015aol} they took an approach from tree-level, by introducing the forward limit with two additional massive particles  that played the role of the loop momenta.

The premise for the previously mentioned prescriptions at one-loop, is that from the CHY formalism a new representation for the Feynman propagators arise, the so called linear propagators, which look like $(2\ell\cdot K + K^2)^{-1}$ \cite{Geyer:2015jch, Casali:2014hfa}. Many interesting developments in one-loop integrands identities (like the Kleiss-Kuijf (KK) identities \cite{Kleiss:1988ne}) and dualities (e.g. the the Bern-Carrasco-Johansson (BCJ) color-kinematics duality \cite{Bern:2008qj}) have been found from the CHY approach, suported over the tree-level scattering equations with two extra massive particles \cite{He:2016mzd, He:2017spx, Geyer:2017ela}. 

The linear propagators approach still leaves several open questions. First, there is no direct way to relate some of the results with the ones from traditional field theory, like the BCJ numerators for example. Another question is about direct loop integration, and to see if it is more efficient to compute these new integrals that appear, in comparison with many well known and long time developed integration techniques for the traditional Feynman propagators.  

Currently, we are developing a program to obtain the traditional quadratic Feynman propagators, $(\ell + K)^{-2}$, directly from CHY. The first proposal came by one of the authors in \cite{Gomez:2017lhy} for the $\Phi^3$ scalar theory. Following, two of the authors in \cite{Gomez:2017cpe} presented a reformulation for the one-loop Parke-Taylor factors that gives quadratic propagators, but it was just made for leading (or planar) contributions. These new Parke-Taylor factors were successfully tested in the massless bi-adjoint $\Phi^3$ theory. 

\begin{figure}[!h]
	\centering
	\includegraphics[width=1.5in]{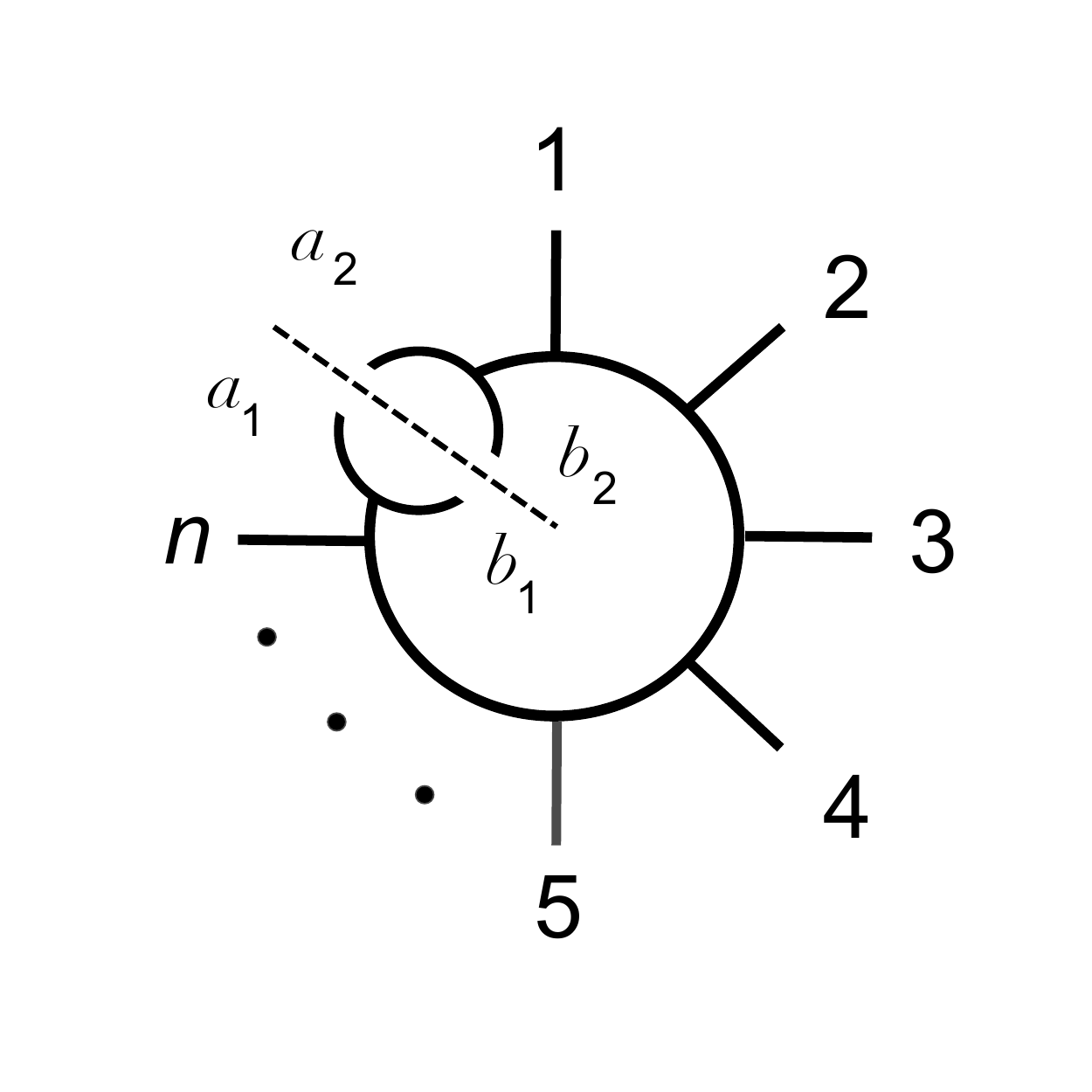}
	\vskip-0.7cm
	\caption{Unitary cut on a two-loop diagram}\label{UC_2-l}
\end{figure}

The proposal to obtain the traditional quadratic Feynman propagators from the CHY approach lies in the use of $n+4$ massless scattering equations instead of $n+2$. The extra particles come from splitting the extra two massive loop momenta $(\ell^+,\ell^-)$ into four massless ones $((a_1,b_1),(b_2,a_2))$. The splitting was motivated by taking an unitary cut on a $n$-point Feynman diagram at two-loop (see Figure \ref{UC_2-l}), that leave us with a $(n+4)$-point tree-level diagram . It was also proved that the four auxiliary loop momenta will always combine in order to give the loop momentum in the forward limit \cite{Gomez:2016cqb,Gomez:2017lhy,Gomez:2017cpe}. 
\vskip-0.45cm

This work was motivated by the tree-level KK relations, which were found for gluons at tree-level in 1988. Let us remind the algebraic relations that the partial amplitudes satisfy
\ba  
A^{{\rm tree}}(\beta_1,...,\beta_s,1,\alpha_1,...,\alpha_r,n)=(-1)^s\sum_{\sigma\in{\rm OP}\{\beta\}\{\alpha\}}A^{{\rm tree}}(1,\sigma,n),
\ea
where the order preserving product (OP) merges the ordering $\alpha^{{\rm T}}$ into the ordering $\beta$.

Going to quantum corrections, in the CHY approach, we performed a study of these relations using the recently defined Parke-Taylor factors at one-loop. From this analysis,  we have  been able to find the {\it non-planar one-loop Parke-Taylor factors}. In addition, we also found a more general KK relations that allow us to write a more general expression for such factors. 

Let us recall that the one-loop version of the KK relations was found by Bern-Dixon-Dunbar-Kosower in \cite{Bern:1994zx}. Their outline for the proof relies on the structure constants, so the result holds for any one-loop gauge theory amplitude where the external particles and the particles circulating around the loop are both in the adjoint representation of SU(N). This relation reads\footnote{The cyclic ordering preserving product (COP) merges the cyclic permutations of $\{\alpha\}=\{r-1,\ldots, 1 \}$ leaving the $\{\beta\}=\{ r,r+1,\ldots, n \}$ ordering fixed.}
\ba  
A_{n;r}^{{\rm 1-loop}}(1,...,{r-1}\,;\, r,..., n)=(-1)^r\sum_{\sigma\in{\rm COP}\{\alpha\}\{\beta\}}A_{n;1}^{{\rm 1-loop}}(\sigma).
\ea

The  new objects found in this paper, the non-planar Parke-Taylor factors at one-loop,  allow the construction of non-planar CHY graphs, which do not have an equivalent in the traditional formalism and encode all the information for the non-planar order.
The developed non-planar CHY-graphs are applied to the massless Bi-adjoint $\Phi^3$ theory, allowing us to explore the non-planar sector of the theory at one loop, where we obtained all the integrands that contribute to the amplitude at this quantum correction. 

{\bf Outline.} The present work is organised as follows. In section \ref{Parke-Taylor} we review the one-loop Parke-Taylor factors for planar corrections that were proposed in \cite{Gomez:2017cpe}, with a few changes in the notation and define the  {\it partial planar one-loop Parke-Taylor factor}, which will be studied in depth in the following section.

Section \ref{KK-relations} is divided in two parts. In the first one we take the previously presented partial Parke-Taylor factors and we find the one-loop KK relations for these. The development starts by analysing the tree-level relations for $n+4$ particles, we find how to obtain the simplest one-loop case from them, from there we will go to the most general relation possible at one-loop. The next part is where we find the KK relations for the full one-loop Parke-Taylor factors, which resemble the relations for gluon amplitudes.

In section \ref{BAsection} we apply the one-loop non-planar Parke-Taylor factors in the study of one loop amplitudes for the massless Bi-adjoint $\Phi^3$ theory. Since the CHY integrands for this theory are given by the product of two Parke-Taylor factors, the contributions for the one-loop amplitudes come from different sectors: the planar, the non-planar, and the mixed one (planar;non-planar).  

The non-planar CHY-graphs will be introduced in section \ref{ex-npgraphs} with some particular worked out examples for the four-point and five-point cases. In section \ref{nonplanarG} we present the general non-planar CHY-graphs for an arbitrary number of points. in section \ref{discussions} we conclude. In order to make the paper self contained, we provide the detailed calculations of the four-point amplitude from the CHY formalism in Appendix \ref{4-p-app-chy} and its exact equivalence with the results obtained from the standard methods based on the Feynman rules in Appendix \ref{4-p-app-fey}.

{\bf Notation.} For convenience, in this paper we use the following notation
\begin{align}\label{notations}
\s_{ij}  := \s_i - \s_j,  \qquad\qquad
\o_{i:j}^{a:b}  := \frac{\s_{ab}}{\s_{i a}\,\, \s_{j b}}.
\end{align}
Note that $\o_{i:j}^{a:b}$ is the generalization of the $(1,0)-$forms used in \cite{Gomez:2016cqb} to write the CHY integrands at two-loop.
In addition, we define the $\s_{ab}$'s and  $\o_{i:j}^{a:b}$'s  chains as
\begin{align}\label{notations-2}
&(i_1,i_2,\ldots, i_p)  := \s_{i_1i_2} \cdots \s_{i_{p-1}i_p}\s_{i_p i_1},    \\
&(i_1,i_2,\ldots, i_p)^{a:b}_\o  := \o^{a:b}_{i_1:i_2} \cdots \o^{a:b}_{i_{p-1} : i_p}\o_{i_p : i_1}^{a:b}  
=  \o^{a:b}_{i_1:i_1} \cdots \o^{a:b}_{i_{p-1} : i_{p-1}}\o_{i_p : i_p}^{a:b},\nonumber
\end{align}

\begin{figure}[!h]
	\centering
	\includegraphics[width=5.0in]{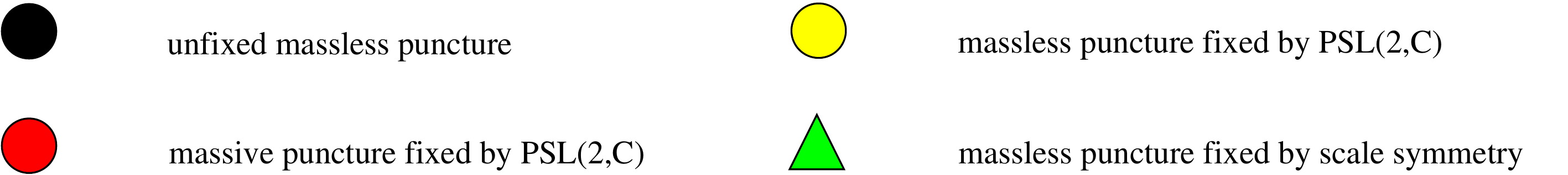}
	\caption{Vertex Color code in CHY-graphs for the $\L-$algorithm.}\label{color_codV}
\end{figure}

\begin{figure}[!h]
	\centering
	\includegraphics[width=4.5in]{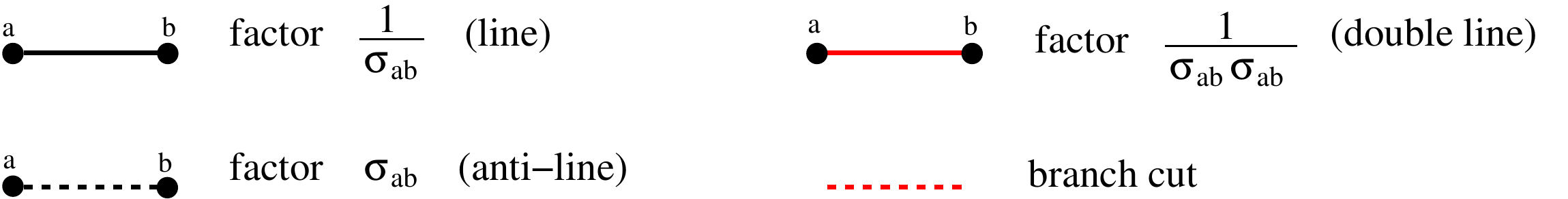}
	\caption{ Edges Color code in CHY-graphs for the $\L-$algorithm.}\label{color_codE}
\end{figure}

To have a graphical description for the CHY integrands on a Riemann sphere (CHY-graphs), it is useful to represent each $\s_a$ puncture as a vertex, the factor ${1 \over \s_{ab}}$ as a line and the factor $\s_{ab}$ as a dashed line that we call the {\it anti-line}. Additionally, since we often use the $\L-$algorithm\footnote{It is useful to recall that the $\L-$algorithm fixes four punctures, three of them by the ${\rm PSL}(2,\mathbb{C})$ symmetry and the last one by the scale invariance.} \cite{Gomez:2016bmv}, then  we introduce the color code  given in Fig. \ref{color_codV} and \ref{color_codE}
for a  mnemonic understanding.
\noindent

\noindent

Finally, we introduce the momenta notation
\begin{equation}
k_{ a_1,\ldots, a_m}:=\sum_{i=1}^m k_{a_i}= [a_1,\ldots, a_m], \quad
s_{a_1\ldots a_m}:=k_{ a_1,\ldots, a_m}^2, 
\quad \tilde s_{a_1\ldots a_m}:=\sum_{a_i<a_j}^m k_{a_i}\cdot k_{a_j}. \nonumber
\end{equation}

\section{Planar Parke-Taylor factors at one-Loop}\label{Parke-Taylor}

In a previous work, we found a reformulation for the Parke-taylor factors that leads to quadratic Feynman propagators by evaluating only massless scattering equations in the CHY prescription.

Let us remind the expression for the Parke-Taylor factor at tree level in the CHY approach, it is given by the following 
\begin{equation}
{\rm PT}[\pi]= \frac{1}{(\pi_1,\pi_2,\ldots , \pi_n)},
\end{equation}
where $``\pi"$ is a generic ordering and $n$ is the total number of particles.  In addition, it will be  useful to  define a reduced Parke-Taylor factor, which does not involve all $n$ particles, i.e.
\begin{equation}
\widehat{\rm PT}[i_1,\ldots, i_p]:=\frac{1}{(i_1,\ldots , i_p)},
\end{equation}
where $p<n$.

As it was shown in \cite{Gomez:2017cpe},  the planar one-loop Parke-Taylor factor (on the left-sector) for quadratic propagators  is given by\footnote{As it was explained in \cite{Gomez:2017lhy,Gomez:2017cpe}, in order to compute scattering amplitudes at one-loop we must perform the forward limit, $k_{a_1}=-k_{a_2}$ ($k_{b_1}=-k_{b_2}$). So, from the third line in \eqref{1lptdef}, it is clear this definition is  totally analog to the one given in \cite{He:2015yua,Baadsgaard:2015hia} by the expression 
\begin{align}
 {\rm PT}^{(1)} [\pi]:=
\sum_{\a\in {\rm cyc (\pi)}} {\rm PT}[\a_1,\ldots, \a_n,+,-]  ,\nonumber
\end{align}
which is just able to reproduce linear propagators. }  
\begin{align}
 {\rm PT}^{(1)}_{\, a_1:a_2} [\pi]:=&{\rm PT}[\pi_1,\pi_2,\ldots , \pi_n,a_1,b_1,b_2,a_2]+{\rm cyc}(\pi)\nonumber\\
 =& \frac{1}{(a_1,b_1,b_2,a_2)}\sum_{\a\in {\rm cyc (\pi)}}
\frac{1}{\s_{\a_1\a_2}\s_{\a_2\a_3}\cdots \s_{\a_{n-1}\a_n}} \o^{a_1:a_2}_{\a_n:\a_1} 
\label{1lptdef}\\
 =&
\frac{(a_1,a_2)}{(a_1,b_1,b_2,a_2)} \sum_{\a\in {\rm cyc (\pi)}} \widehat{\rm PT}[\a_1,\ldots, \a_n,a_1,a_2]\,. \nonumber
\end{align}

It is useful to remember that the Parke-Taylor factor on the right-sector (or $b-$sector) is defined to be
\begin{align}\label{b-sector}
 {\rm PT}^{(1)}_{\, b_1:b_2} [\rho]:
 = \frac{(b_1,b_2)}{(a_1,b_1,b_2,a_2)} \sum_{\b\in {\rm cyc (\rho)}} \widehat{\rm PT}[\b_1,\ldots, \b_n,b_1,b_2].
\end{align}
Such as we explained in appendix A in \cite{Gomez:2017cpe}, the CHY-integral of the product, $ {\rm PT}^{(1)}_{\, a_1:a_2} [\pi]\times  {\rm PT}^{(1)}_{\, b_1:b_2} [\rho]$, is always a function of the couples, $(k_{a_1}+k_{b_1})$ and $(k_{a_2}+k_{b_2})$. This means, these type of integrals always produce quadratic propagators at one-loop by the identification, $(k_{a_1}+k_{b_1})= -(k_{a_2}+k_{b_2})=\ell$.

Each one of the terms of the sum in \eqref{1lptdef} and \eqref{b-sector} is called {\it partial planar one-loop Parke-Taylor factor}. 
We are going to denoted them by ${\bf pt}^{(1)}_{\, a_1:a_2} [i_1,\ldots, i_p]$ (${\bf pt}^{(1)}_{\, b_1:b_2} [i_1,\ldots, i_p]$),  
namely
\ba
{\bf pt}^{(1)}_{\, a_1:a_2}[i_1,\ldots, i_p]:=
\frac{1}{\s_{i_1i_2}\s_{i_2 i_3}\cdots \s_{i_{p-1} i_p}} \o^{a_1:a_2}_{i_p: i_1},
\ea
in particular we define
\ba
{\bf pt}^{(1)}_{\, a_1:a_2}[i_1]:=
 \o^{a_1:a_2}_{i_1: i_1}.
\ea
Thus,
\begin{equation}
 {\rm PT}^{(1)}_{\, a_1:a_2} [\pi] \, = \frac{1}{(a_1,b_1,b_2,a_2)} \sum_{\a\in {\rm cyc (\pi)}} {\bf pt}^{(1)}_{\, a_1:a_2} [\a_1,\ldots, \a_n],
\end{equation}
and for the $b-$sector we must just perform the replacement, $(a_1,a_2)\rightarrow (b_1,b_2)$. Therefore, for the rest of this work it is enough to work on the $a-$sector.

In \cite{Gomez:2017cpe}, we found several algebraic manipulations on these Parke-Taylor factors that allowed us to obtain the contributing planar diagrams for the bi-adjoint scalar $\Phi^3$ theory. In order to go beyond the planar case, let us analyse a known relation between planar and non-planar integrands.

\section{General KK relation and non-planar Parke-Taylor factors at one-loop}\label{KK-relations}

Let us recall a particular case of the Kleiss-Kuijf (KK) relations for the Parke-Taylor factors \cite{Kleiss:1988ne,He:2017spx}. At tree-level we have the identity
\begin{align}
& {\rm PT}[ \{\{1\}\sh \{2,3,4,...,n-1\},n\}] :=
{\rm PT}[1,2,3,4,...,n] + {\rm PT}[2,1,3,4,...,n] \\
&+ {\rm PT}[2,3,1,4,...,n] + \cdots 
+{\rm PT}[2,3,4,...,n-1,1,n] =0,  \nonumber
\end{align}
where $``\sh"$ is the {\it shuffle} product\footnote{It is useful to remember the shuffle product, $\{\a_1,\ldots \a_p \} \sh \{\b_1,\ldots \b_q \}$,  has a total of $\frac{(p+q)!}{p!\, q!} $ terms.}. Note that we are making the following abuse in the notation
\begin{align}
{\rm PT} [A_1+A_2+\cdots + A_p] \equiv  \sum_{i=1}^p{\rm PT} [A_i] \, ,
\end{align}
where $A_i$ is an ordered list of $n$ different elements.

This relation will mark the starting point for our analysis. In this section, we are going to study and  introduce a more general KK relations for the partial and total one-loop Parke-Taylor factors.

\subsection{KK relations for the partial one-loop Parke-Taylor factors}\label{KK-relations-partial}

Translating the previous relation for the partial one-loop Parke-Taylor factors, it can be written as
\begin{align}
&\hspace{-0.5cm}\PT[1,2,...,n,a_1,b_1,b_2,a_2] + 
\sum_{i=2}^{n-1}\PT[2,...,i,1,i+1,...,n,a_1,b_1,b_2,a_2] +  \PT[2,...,n,1,a_1,b_1,b_2,a_2]\nonumber\\
& \hspace{-0.5cm}
+ \PT[2,...,n,a_1,1,b_1,b_2,a_2]+  \PT[2,...,n,a_1,b_1,1,b_2,a_2]=- \PT[2,...,n,a_1,b_1,b_2,1,a_2].\non
\label{kkplus}
\end{align}

Note that in the terms on the second line in \eqref{kkplus}  there is a splitting in the points assigned to the loop momenta. Those terms are perfectly normal at tree-level, but at loop level they would not lead to any quadratic Feynman propagator in the forward limit, in other words, the identification, $(k_{a_1}+k_{b_1})=- (k_{a_2}+k_{b_2}) = \ell$, is not well defined on them. In any case, we can analyse the terms where the splitting is not present.

Taking  the sector of \eqref{kkplus} where the loop momenta points are not split, we can find a different relation for the partial one-loop Parke-Taylor factors. The new relation is given by
\ba
&&\PT[1,2,3,...,n,a_1,b_1,b_2,a_2] + 
\sum_{i=2}^{n-1}\PT[2,3,...,i,1,i+1,...,n,a_1,b_1,b_2,a_2]\nonumber\\
&& + \PT[2,3,...,n,1,a_1,b_1,b_2,a_2] = \o_{1:1}^{a_1:a_2}\, \PPT[2,3,...,n,a_1,b_1,b_2,a_2].\non
\ea
This relation can be written in a compact and more legible way, as follows
\begin{align}
\hspace{-0.5cm}
{\bf pt}^{(1)}_{\, a_1:a_2}[\{1\} \sh \{2,\ldots, n\}]:&= {\bf pt}^{(1)}_{\, a_1:a_2}[1,2,\ldots, n]+{\bf pt}^{(1)}_{\, a_1:a_2}[2,1,\ldots, n]+\cdots  +{\bf pt}^{(1)}_{\, a_1:a_2}[2,\ldots, n,1] 
\nonumber\\
& = {\bf pt}^{(1)}_{\, a_1:a_2}[1] \times {\bf pt}^{(1)}_{\, a_1:a_2}[2,\ldots, n].\non
\label{newkk}
\end{align}

The result from above can be generalized. In fact, we have been able to find a more broad relation for the 
{\it partial planar one-loop Parke-Taylor factor}, which is given by the identity
\begin{center}
\begin{tabular}{| l |}
 \cline{1-1} 
\\
{\bf \quad General KK relations for the partial one-loop Parke-Taylor factors \quad}  
 \\
 \\
$
\hspace{1cm}
{\bf pt}^{(1)}_{\, a_1:a_2}[\{i_1,i_2,\ldots i_p\} \sh  \{i_{p+1},\ldots i_q\} \sh \cdots \sh   \{i_m,\ldots, i_n\}]
$
\\
$
\hspace{1cm}
 = {\bf pt}^{(1)}_{\, a_1:a_2}[i_1,i_2,\ldots i_p] \times {\bf pt}^{(1)}_{\, a_1:a_2}[i_{p+1},\ldots i_q] \times \cdots \times {\bf pt}^{(1)}_{\, a_1:a_2} [i_m,\ldots, i_n]\qquad
$
\\
\\
\cline{1-1}
\end{tabular}
\end{center}

In particular, it is straightforward to check
\begin{align}\label{n-omegas}
{\bf pt}^{(1)}_{\, a_1:a_2}[\{1\} \sh  \{2\} \sh \{3\} \sh \cdots \sh   \{ n\}]&= {\bf pt}^{(1)}_{\, a_1:a_2}[1] \times {\bf pt}^{(1)}_{\, a_1:a_2}[2] \times \cdots \times {\bf pt}^{(1)}_{\, a_1:a_2} [ n] \nonumber \\
& = \o_{1:1}^{a_1:a_2}\times \o_{2:2}^{a_1:a_2}\times \cdots \times \o_{n:n}^{a_1:a_2} \\
&= \sum_{\a\in {\rm S_{n-1}}}
\frac{1}{\s_{\a_1\a_2}\s_{\a_2\a_3}\cdots \s_{\a_{n-1}\a_n}} \o^{a_1:a_2}_{\a_n:\a_1} \nonumber ,
\end{align}
where $\a_1:=1$ and ${\rm S_{n-1}}$  is the set of all permutations of $\{2,3,\ldots ,n\}$. Let us recall that the expression in \eqref{n-omegas} reproduces the symmetrized {\it n-gon} at one-loop.

Finally,  by considering only two sets, for example 
\begin{align}\label{KKplanar-nonp}
{\bf pt}^{(1)}_{\, a_1:a_2}[\{1,2,\ldots p\} \sh  \{p+1,\ldots n\}]
=
{\bf pt}^{(1)}_{\, a_1:a_2}[1,2,\ldots p] \times {\bf pt}^{(1)}_{\, a_1:a_2}[p+1,\ldots n], 
\end{align}
we obtain a similar structure as the KK relation between the planar and non-planar amplitudes at one-loop \cite{Bern:1994zx,DelDuca:1999rs}, therefore, we define the {\it partial non-planar one-loop Parke-Taylor factors} as
\begin{align}\label{partial-nonpPT}
{\bf pt}^{(1)}_{\, a_1:a_2}[i_1,i_2,\ldots ,i_p \,| \, i_{p+1},\ldots , n ] := & {\bf pt}^{(1)}_{\, a_1:a_2}[\{i_1,i_2,\ldots i_p\} \sh  \{i_{p+1},\ldots i_n\}] \\
= & {\bf pt}^{(1)}_{\, a_1:a_2}[i_1,i_2,\ldots , i_p] \times {\bf pt}^{(1)}_{\, a_1:a_2}[i_{p+1},\ldots ,i_n]. \nonumber
\end{align}
and\footnote{Note that the partial non-planar one-loop Parke-Taylor factors for quadratic propagators can be written as
\begin{equation}
\hspace{-0.5cm}
\frac{1}{(a_1,b_1,b_2,a_2)}\,{\bf pt}^{(1)}_{\, a_1:a_2}[1,2,\ldots ,p \,| \,n,n-1,\ldots , p+1 ] = (-1)^{n-p}\, \frac{(a_1,a_2)}{(a_1,b_1,b_2,a_2)}  \times \widehat{\rm PT} [1,2,\ldots, p,a_1,  p+1,\ldots, n-1,n,a_2] . \nonumber
\end{equation}
This expression is totally similar to the one  given for non-planar  linear propagators in \cite{Cachazo:2015aol,He:2017spx}, where $a_1\rightarrow \, +$ and $a_2\rightarrow  \, -$. \label{footnote3}} 
the full {\it  non-planar one-loop Parke-Taylor factors} as
\begin{align}\label{nonpPT}
&{\rm PT}^{(1)}_{\, a_1:a_2}[\pi_1,\ldots  , \pi_{p} \,  | \,  \rho_{p+1},\ldots , \rho_n] := \frac{1}{(a_1,b_1,b_2,a_2)} \sum_{\a\in {\rm cyc (\pi)} \atop \b\in {\rm cyc (\rho)} } {\bf pt}^{(1)}_{\, a_1:a_2}[\a_1,\ldots , \a_{p} \,| \, \b_{p+1},\ldots , \b_n ] \nonumber \\ 
&
\hspace{3cm}
=
\frac{1}{(a_1,b_1,b_2,a_2)} \sum_{\a\in {\rm cyc (\pi)} \atop \b\in {\rm cyc (\rho)} } {\bf pt}^{(1)}_{\, a_1:a_2}[\a_1,\ldots , \a_{p}] \times  {\bf pt}^{(1)}_{\, a_1:a_2}[\b_{p+1},\ldots , \b_n] , \non
\end{align}
where $\{\pi_1,\ldots , \pi_{p} \}$ and $\{\rho_{p+1},\ldots , \rho_{n} \}$ are two different generic orderings such that
$\{\pi_1,\ldots , \pi_{p} \} \cap \{\rho_{p+1},\ldots , \rho_{n} \} = \emptyset$ and
 $\{\pi_1,\ldots , \pi_{p} \} \cup \{\rho_{p+1},\ldots , \rho_{n} \} = \{ 1,2,\ldots, n \}$.

The partial and full non-planar one-loop Parke-Taylor factors defined in \eqref{partial-nonpPT} and \eqref{nonpPT} can be generalized in the following way
\begin{align}
&\hspace{-0.3cm}
{\bf pt}^{(1)}_{\, a_1:a_2}[ i_1,\ldots , i_p \,|\, i_{p+1},\ldots ,  i_q \,|\,   \cdots \,|\,  i_m,\ldots, i_n   ]
:={\bf pt}^{(1)}_{\, a_1:a_2}[\{i_1,\ldots , i_p\} \sh  \{i_{p+1}, \ldots , i_q\} \sh \cdots \sh   \{i_m,\ldots, i_n\}] \nonumber \\
&
 = {\bf pt}^{(1)}_{\, a_1:a_2}[i_1,i_2,\ldots i_p] \times {\bf pt}^{(1)}_{\, a_1:a_2}[i_{p+1},\ldots i_q] \times \cdots \times {\bf pt}^{(1)}_{\, a_1:a_2} [i_m,\ldots, i_n],\qquad\non
\end{align}
and we can write the full version in terms of these
\begin{align}\label{genPT}
&
{\rm PT}^{(1)}_{\, a_1:a_2}[ \pi_1,\ldots ,\pi_{p} \,|\,\rho_{p+1}, \ldots , \rho_{q} \,|\,   \cdots \,|\,  \g_{m},\ldots, \g_n ]
 \non
&
 := \frac{1}{(a_1,b_1,b_2,a_2)} \sum_{\a\in {\rm cyc (\pi)}  } 
\sum_{ \b\in {\rm cyc (\rho)} } \cdots  \sum_{ \d\in {\rm cyc (\g)} }
 {\bf pt}^{(1)}_{\, a_1:a_2}[ \a_1,\ldots ,\a_{p} \,|\, \b_{p+1},\ldots ,  \b_{q} \,|\,   \cdots \,|\, \d_{m},\ldots, \d_n   ] ,
 \non
\end{align}
where $\{\pi_1,\ldots , \pi_{p} \} , \{\rho_{p+1},\ldots , \rho_{n} \}, \ldots \{\g_{m},\ldots , \g_{n} \}$,  are different generic orderings such that
$\{\pi_1,\ldots , \pi_{p} \} \cup \{\rho_{p+1},\ldots , \rho_{n} \}\cup \cdots \cup \{\g_{p+1},\ldots , \g_{n} \} = \{ 1,2,\ldots, n \}$ and they are disjoint, namely, they have no element in common. Although these generalizations are well defined in the CHY side, we have yet the task to understand what would be their physical meaning.

\subsection{KK relation for the full one-loop Parke-Taylor factors}\label{KK-relations-total}

In the previous section we found general relations for the partial one-loop Parke-Taylor factors. In the following section our interest lies in finding KK relations that involve the full one-loop Parke-Taylor factors defined in \eqref{1lptdef} and \eqref{nonpPT}.

To obtain full one loop Parke-Taylor factors on the right hand side of \eqref{nonpPT} we have to expand the sum and collect the terms related by a cyclic permutation. The result can be arranged again in a sum as follows
\ba\label{fullNP-P} 
\PT^{(1)}_{\, a_1:a_2}[\pi_1,\ldots  , \pi_{p} \,  | \,  \rho_{p+1},\ldots , \rho_n] = \sum_{\a\in {\rm cyc (\pi)}\sh \rho/\rho_n }\PT^{(1)}_{\, a_1:a_2}[\a_1,...,\a_n],
\ea
where ${\rm cyc (\pi)}\sh \rho/\rho_n \equiv \{{\rm cyc (\pi)}\sh\{\rho_{p+1},...,\rho_{n-1}\},\rho_n\}$. As an example, for the non-planar ordering $[1,2|3,4,5]$ we have the relation
\ba\label{ex-expansion}
\PT^{(1)}_{\, a_1:a_2}[1,2|3,4,5] &=& \PT^{(1)}_{\, a_1:a_2}[1, 2, 3, 4, 5] + \PT^{(1)}_{\, a_1:a_2}[1, 3, 2, 4, 5] + \PT^{(1)}_{\, a_1:a_2}[1, 3, 4, 2, 5]\nonumber\\
&& + \PT^{(1)}_{\, a_1:a_2}[2, 1, 3, 4, 5] + \PT^{(1)}_{\, a_1:a_2}[3, 1, 2, 4, 5] + \PT^{(1)}_{\, a_1:a_2}[3, 1, 4, 2, 5]\nonumber\\
&& + \PT^{(1)}_{\, a_1:a_2}[2, 3, 1, 4, 5] + \PT^{(1)}_{\, a_1:a_2}[3, 4, 1, 2, 5] + \PT^{(1)}_{\, a_1:a_2}[2, 3, 4, 1, 5]\nonumber\\
&& + \PT^{(1)}_{\, a_1:a_2}[3, 2, 4, 1, 5] + \PT^{(1)}_{\, a_1:a_2}[3, 4, 2, 1, 5] + \PT^{(1)}_{\, a_1:a_2}[3, 2, 1, 4, 5].\nonumber\\
\ea

This product for two orderings has been usually denoted in the literature as ${\rm COP}\{\pi^{{\rm T}}\}\{\rho\}$, where $\pi^{{\rm T}}$ is the reversed ordering of $\pi$. The proof of this result for gluons subamplitudes can be found in \citep{Bern:1994zx}, where they approach it from the string theory and field theory sides, here we have obtained it as a consequence of the shuffle product that appears at the tree-level KK relations. 

Finding the relation between the full one-loop Parke-Taylor allows us to see more clearly the bridge between CHY and the subamplitudes from the traditional field theory approach. Another important point is that a result in \cite{Gomez:2017cpe} shows an expansion of the planar Parke-Taylor factors at one loop in terms of $\omega$'s, which displays an advantage from a computational point of view.

Going to the more general case in \eqref{genPT}, the same analysis can be performed to find a relation with the full one-loop Parke-Taylor factors. After expanding and collecting the terms we are left with
\vskip-2.5cm
\begin{center}
\begin{tabular}{| l |}
 \cline{1-1} 
\\
{\bf \qquad\qquad General KK relations for the full one-loop Parke-Taylor factors \quad}  
 \\
 \\
$
{\rm PT}^{(1)}_{\, a_1:a_2}[ \pi_1,\ldots ,\pi_{p} \,|\,\rho_{p+1}, \ldots , \rho_{q} \,|\,   \cdots \,|\,  \g_{m},\ldots, \g_n ] = \sum\limits_{\a\in {\rm cyc (\pi)}\sh{\rm cyc (\rho)}\sh...\sh \gamma/\gamma_n }\PT^{(1)}_{\, a_1:a_2}[\a_1,...,\a_n],
$
\\
\\
\cline{1-1}
\end{tabular}
\end{center}
where the product of orderings follows the same definition given after \eqref{fullNP-P}. Here even the simplest non-trivial example (that could be $[1,2|3,4|5,6]$) gives a considerable number of terms to fit on a single page.



\section{Bi-adjoint $\Phi^3$ scalar theory}\label{BAsection}

In this section we propose a non-planar CHY prescription  for the  S-matrix at one-loop for the bi-adjoint $\Phi^3$ scalar theory. In this proposal we will be able reproduce directly the quadratic propagators, in the same way as the traditional Feynman approach.


\subsection{Full bi-adjoint $\Phi^3$ amplitude at one-loop}\label{fullAmp}

Along the line of reasoning of \cite{Cachazo:2015aol} and with the partial Parke-Taylor factors defined in section \ref{Parke-Taylor}, we define the full  bi-adjoint $\Phi^3$ amplitude at one-loop with flavor group $U(N)\times U(\tilde N)$ as 
\begin{align}\label{full-m}
&{\bf m}^{\rm 1-loop}_n=\int d^D\ell  \sum_{\pi\in S_{n+2}/\mathbb{Z}_{n+2} \atop  \rho\in S_{n+2}/\mathbb{Z}_{n+2}  }{\rm Tr}(T^{i_{\pi_1}} T^{i_{\pi_2}} \cdots T^{i_{\pi_n}} T^{i_{\pi_{a_1}}} T^{i_{\pi_{a_2}}} ) \times \left\{
\frac{1}{2^{n+1}}\int d\Omega \, s_{a_1b_1} \,   \right. \nonumber \\
& \left.
\times
\int_\Gamma d\mu^{t}_{n+4} \,\,
\frac{(a_1,a_2)}{(a_1,b_1,b_2,a_2)}\, \widehat{\rm PT}[\pi_1,\ldots, \pi_n,\pi_{a_1},\pi_{a_2}] 
\times \frac{(b_1,b_2)}{(a_1,b_1,b_2,a_2)} \,
\widehat{\rm PT}[\rho_1,\ldots, \rho_n,\rho_{b_1},\rho_{b_2}] \right\}\nonumber\\
&\times
{\rm Tr}(\tilde T^{i_{\rho_1}} \tilde T^{i_{\rho_2}} \cdots  \tilde T^{i_{\rho_n}} \tilde T^{i_{\rho_{b_1}}} \tilde T^{i_{\rho_{b_2}}}  )  ,\non
\end{align}
where
the measures, $d\Omega$ and $d\mu^{\rm t}_{n+4}$,  are given by the expressions\footnote{In this paper we are considering that the $D-$dimensional momentum space is real, i.e. $k_i\in \mathbb{R}^{D-1,1}$. Therefore, the Dirac delta functions in \eqref{dOmega} are well defined.} \cite{Cachazo:2013hca,Gomez:2017lhy} 
\begin{align}\label{dOmega}
d\Omega :=& \, d^D(k_{a_1}+k_{b_1})\, \delta^{(D)}(k_{a_1}+k_{b_1} - \ell)\, d^{D}k_ {a_2}\, d^{D}k_{ b_2}\,\delta^{(D)} (k_{a_2}+k_{a_1})\,\,\delta^{(D)} (k_{b_2}+k_{b_1}),
\end{align}
and 
\begin{align}
d\mu_{n+4}^{\rm t}:=& \,
 \frac{   \prod_{A=1}^{n+4}\,d\s_A          }{{\rm Vol}\,\,({\rm PSL}(2,\mathbb{C}))} \times \frac{(\s_{1b_1}\,\s_{b_1b_2}\, \s_{b_21})}{ \prod_{ A\neq 1, b_1, b_2}^{n+4}  E_A}\nonumber\\
 &   ^{\underrightarrow{\,\quad{\rm fixing\,\, PSL}(2,\mathbb{C})\,\quad }}\,\,\,
\frac{d\s_{a_1}}{E_{a_1}} \times
\frac{d\s_{a_2}}{E_{a_2}} \times
 \prod_{i=2}^{n} \frac{d\s_{i}}{E_{i}}\times (\s_{1b_1}\,\s_{b_1b_2}\, \s_{b_21}  )^2. \non
 \label{dmu}
\end{align} 
The  $\Gamma$  contour is  defined by the massless scattering equations
\begin{align} 
E_A := \sum_{B=1\atop B\ne A }^{n+4}\frac{k_A\cdot k_B}{\s_{AB}}=0,  \qquad A=1,2,\ldots, n+4, \quad {\rm with}\,\,\, \sum_{A=1}^{n+4}k_A=0, 
 \label{Sequations}
\end{align}
where we are making the following identifications
\begin{align}\label{Lidentification}
& n+1\rightarrow a_1, \, \, \,  n+2 \rightarrow a_2, \,\,\,  n+3 \rightarrow b_1 , \,\,\, n+4 \rightarrow b_2\, .
\end{align}
Finally, without loss of generality, note that in \eqref{dmu}  we have fixed $\{\s_{1},\s_{b_1},\s_{b_2} \}$
and  $\{ E_{1},E_{b_1},E_{b_2} \}$. 

It is very important to remark that the proposal given in \eqref{full-m} is well defined. After performing the contour integral, $\int_\Gamma d\mu_{n+4}^{\rm t}\, $, the result obtained has a functional dependence of the loop momenta couples\footnote{It is guaranteed by the KK  relations in \eqref{KKplanar-nonp} and the footnote \ref{footnote3}. Since the partial non-planar one-loop Parke-Taylor factors can always be written as a linear combination of the partial planar one-loop Parke-Taylor factors then, as it was explained in appendix A in \cite{Gomez:2017cpe}, this implies the CHY-integral is always a function of the momenta $(k_{a_1}+k_{b_1})$ and $(k_{a_2}+k_{b_2})$.} like $(k_{a_1}+k_{b_1})$ and $(k_{a_2}+k_{b_2})$,  therefore, the Dirac delta functions in  $d\Omega$ can be carried out without problems.

Such as it has been argued in \cite{Cachazo:2015aol}, since we are looking for the forward limit with the measure $\delta^{(D)}(k_{a_1}+k_{a_2}) \times \delta^{(D)}(k_{b_1}+k_{b_2})$ in \eqref{dOmega}, then it requires that when summing over the $U(N)$ (and $U(\tilde N)$) degrees of freedom of the two internal particles they must be identified. Being more precise, we must introduce the sum
\begin{align}
\sum_{i_{a_1},i_{a_2}=1}^{N^2} \delta_{i_{a_1}i_{a_2}} \times \sum_{i_{b_1},i_{b_2}=1}^{\tilde N^2} \delta_{i_{b_1}i_{b_2}},
\end{align}
where $N^2$ ($\tilde N^2$) is the dimension of $U(N)$ $(U(\tilde N ))$. Now, by using the  identities\footnote{The second line in \eqref{sum-identities} is known as the reflection identity, which is simple to check from the adjoint representation.} 
\begin{align}
& \sum_{i_{a_1}=1}^{N^2} {\rm Tr} (X\,T^{i_{a_1}} \, Y\, T^{i_{a_1}})  = {\rm Tr} (X) \,{\rm Tr} (Y)  , \quad
\sum_{i_{a_1}=1}^{N^2} {\rm Tr} (X\,Y\,\,T^{i_{a_1}} \, T^{i_{a_1}})  = N\,{\rm Tr} (XY) \non
& {\rm Tr} (T^{m_1}\, T^{m_2}\, \cdots \, T^{m_{p-1}} \, T^{m_{p}})  =  ( -1)^p\, {\rm Tr} ( T^{m_{p}} \, T^{m_{p-1}}  \, \cdots \,  T^{m_2} \, T^{m_1}), \non
\label{sum-identities}
\end{align}
the full amplitude, ${\bf m}_n^{\rm 1-loop}$, becomes
\begin{eqnarray}\label{full-amp}
&& 
\hspace{-0.5cm}
{\bf m}^{\rm 1-loop}_n  = 4\times\,\bigg\{\left.\,( N\, \tilde N) \sum_{\pi\in S_{n}/\mathbb{Z}_{n} \atop  \rho\in S_{n}/\mathbb{Z}_{n}  }{\rm Tr}(T^{i_{\pi_1}} \cdots T^{i_{\pi_n}}  ) \times 
m_{n}^{(\rm 1-P;P)}[\pi \, ; \, \rho]
\times
{\rm Tr}(\tilde T^{i_{\rho_1}} \cdots  \tilde T^{i_{\rho_n}}  ) \right. \nonumber \\
&& 
\hspace{-0.7cm}
+ ( N)  \sum_{p=1 }^{n-1}
 \sum_{\pi\in S_{n}/\mathbb{Z}_{n}   }
  \sum_{\g\in S_{p}/\mathbb{Z}_{p} \atop  \d\in S_{n-p}/\mathbb{Z}_{n-p}  }
{\rm Tr}(T^{i_{\pi_1}} \cdots T^{i_{\pi_n}}  ) \times 
m_{n}^{(\rm 1-P;NP)}[\pi \, ; \, \g | \d ]
\times
{\rm Tr}(\tilde T^{i_{\g_1}} \cdots  \tilde T^{i_{\g_p}}  ) \times {\rm Tr}(\tilde T^{i_{\d_{p+1}}} \cdots  \tilde T^{i_{\d_n}}  ) \nonumber \\
&& 
\hspace{-0.7cm}
+ ( \tilde N)  \sum_{p=1 }^{n-1}
  \sum_{\pi\in S_{p}/\mathbb{Z}_{p} \atop  \rho\in S_{n-p}/\mathbb{Z}_{n-p}  }
   \sum_{\g\in S_{n}/\mathbb{Z}_{n}   }
{\rm Tr}(T^{i_{\pi_1}} \cdots T^{i_{\pi_p}}  ) \times 
{\rm Tr}(T^{i_{\rho_{p+1}}} \cdots T^{i_{\rho_n}}  ) \times
m_{n}^{(\rm 1-NP;P)}[\pi | \rho \, ; \, \g  ]
\times
{\rm Tr}(\tilde T^{i_{\g_1}} \cdots  \tilde T^{i_{\g_n}}  )  \nonumber \\
&& 
\hspace{-0.7cm}
+ \sum_{p=1 \atop q=1 }^{n-1}
  \sum_{\pi\in S_{p}/\mathbb{Z}_{p}, \, \g\in S_{q}/\mathbb{Z}_{q} \atop  \rho\in S_{n-p}/\mathbb{Z}_{n-p}  , \, \d\in S_{n-q}/\mathbb{Z}_{n-q}}
{\rm Tr}(T^{i_{\pi_1}} \cdots T^{i_{\pi_p}}  ) \times 
{\rm Tr}(T^{i_{\rho_{p+1}}} \cdots T^{i_{\rho_n}}  ) \times
m_{n}^{(\rm 1-NP;NP)}[\pi | \rho \, ; \, \g | \d ] \nonumber\\
&& \left.
\hspace{4.3cm}
\times
{\rm Tr}(\tilde T^{i_{\g_1}} \cdots  \tilde T^{i_{\g_q}}  ) \times {\rm Tr}(\tilde T^{i_{\d_{q+1}}} \cdots  \tilde T^{i_{\d_n}}  ) \right.\bigg\} ,\non
\end{eqnarray}
where ${\rm P} \,  ({\rm NP})$ means planar (non-planar), $m_{n}^{(\rm 1-P;P)}[\pi \, ; \, \rho]$ is the same amplitude defined in \cite{Gomez:2017cpe} as $\mathfrak{M}_n^{\rm 1-loop}[\pi | \rho]$, and the non-planar contributions, $m_{n}^{(\rm 1-P;NP)}[\pi  \, ; \, \g | \d ]$ ($m_{n}^{(\rm 1-NP;P)}[\pi | \rho \, ; \, \g ]$) and $m_{n}^{(\rm 1-NP;NP)}[\pi | \rho \, ; \, \g | \d ]$ are given by 
\begin{align}
& m_{n}^{(\rm 1-P;NP)}[\pi_1,\ldots, \pi_n  \, ; \, \g_1,\dots,\g_p | \d_{p+1},\ldots, \d_n ] :=
{1 \over 2^{n+1}} \int d^D\ell  \int d\Omega \times s_{a_1b_1} \nonumber  \\
&
\hspace{1cm}
\times \int d\mu_{n+4}^{\rm t} \times
\mathbf{I}_{\rm (1-P;NP)}^{\rm CHY}[\pi_1,\ldots, \pi_n  \, ; \, \g_1,\dots,\g_p | \d_{p+1},\ldots, \d_n ] ,  \label{1-p-np}\\
& m_{n}^{(\rm 1-NP;NP)}[\pi_1,\ldots, \pi_p | \rho_{p+1},\ldots, \rho_n  \, ; \, \g_1,\dots,\g_q | \d_{q+1},\ldots, \d_n ] :=
{1 \over 2^{n+1}} \int d^D\ell  \int d\Omega \times s_{a_1b_1}  \nonumber \\
&
\hspace{1cm}
\times \int d\mu_{n+4}^{\rm t} \times
\mathbf{I}_{\rm (1-NP;NP)}^{\rm CHY}[\pi_1,\ldots, \pi_p | \rho_{p+1},\ldots, \rho_n  \, ; \, \g_1,\dots,\g_q | \d_{q+1},\ldots, \d_n ] \label{1-np-np},
\end{align}
with 
\begin{align}\label{}
&\mathbf{I}_{\rm (1-P;NP)}^{\rm CHY}[\pi_1,\ldots, \pi_n  \, ; \, \g_1,\dots,\g_p | \d_{p+1},\ldots, \d_n ] 
:=     \frac{(a_1,a_2)\, (b_1,b_2)}{(a_1,b_1,b_2,a_2)^2} 
\\
&\times 
\sum_{\a \in {\rm cyc}(\pi)     }
\widehat{\rm PT}[\a_1,\ldots, \a_n ,a_1, a_2] \times
\sum_{\xi \in {\rm cyc}(\g)    \atop  \zeta\in {\rm cyc}(\d)  }
\widehat{\rm PT}[\xi_1,\ldots, \xi_q,b_1, \zeta_{q+1}, \ldots, \zeta_n  , b_2] , \nonumber\\
&\mathbf{I}_{\rm (1-NP;NP)}^{\rm CHY}[\pi_1,\ldots, \pi_p | \rho_{p+1},\ldots, \rho_n  \, ; \, \g_1,\dots,\g_q | \d_{q+1},\ldots, \d_n ] 
:=      
\frac{(a_1,a_2)\, (b_1,b_2)}{(a_1,b_1,b_2,a_2)^2} 
\\
&\times 
\sum_{\a \in {\rm cyc}(\pi)    \atop  \b\in {\rm cyc}(\rho)  }
\widehat{\rm PT}[\a_1,\ldots, \a_p,a_1, \b_{p+1}, \ldots, \b_n  , a_2] \times
\sum_{\xi \in {\rm cyc}(\g)    \atop  \zeta\in {\rm cyc}(\d)  }
\widehat{\rm PT}[\xi_1,\ldots, \xi_q,b_1, \zeta_{q+1}, \ldots, \zeta_n  , b_2] . \nonumber
\end{align}
Finally, the amplitude, $m_{n}^{(\rm 1-NP;P)}[\pi | \rho \, ; \, \g ]$, is defined in a similar way to the one given in \eqref{1-p-np}. 

In the following sections we are going to focus on these non-planar amplitudes, we will give a few examples and define the {\it non-planar CHY graphs}.

\subsection{Non-planar bi-adjoint $\Phi^3$ amplitudes at one-loop}\label{section-def}

In the previous section we have defined the non-planar $\Phi^3$ amplitudes at one-loop. In order to use the technology developed in \cite{Gomez:2017cpe} and  section \ref{KK-relations},  it is useful to manipulate the integrands $\mathbf{I}_{\rm (1-P;NP)}^{\rm CHY}[\pi  \, ; \, \g | \d ] $ and $\mathbf{I}_{\rm (1-NP;NP)}^{\rm CHY}[\pi | \rho \, ; \, \g | \d ] $.

From the identity in footnote \ref{footnote3}, it is straightforward to see
\begin{align}\label{}
&\mathbf{I}_{\rm (1-P;NP)}^{\rm CHY}[\pi_1,\ldots, \pi_n  \, ; \, \g_1,\dots,\g_p | \d_{p+1},\ldots, \d_n ] 
=    \nonumber\\ 
&{\rm PT}^{(1)}_{a_1:a_2}[\pi_1,\ldots , \pi_n] 
\times
{\rm PT}^{(1)}_{b_1:b_2}[\g_1,\ldots , \g_p \, | \, \d_{n}, \d_{n-1}, \ldots, \d_{p+1}],
\\
&\mathbf{I}_{\rm (1-NP;NP)}^{\rm CHY}[\pi_1,\ldots, \pi_p | \rho_{p+1},\ldots, \rho_n  \, ; \, \g_1,\dots,\g_q | \d_{q+1},\ldots, \d_n ] 
=  \nonumber \\
&    
{\rm PT}^{(1)}_{a_1:a_2}[\pi_1,\ldots , \pi_p \, | \, \rho_{n}, \rho_{n-1}, \ldots, \rho_{p+1}]
\times 
{\rm PT}^{(1)}_{b_1:b_2}[\g_1,\ldots , \g_q \, | \, \d_{n}, \d_{n-1}, \ldots, \d_{q+1}].
\end{align}
So, for convenience we define:
\\

\noindent {\bf Definition:} {\sl The non-planar partial amplitudes, $\mathfrak{M}_n^{\rm (1-P;NP)} [\pi  \, ; \, \g | \d ]$ and $\mathfrak{M}_n^{\rm (1-NP;NP)} [\pi | \rho  \, ; \, \g | \d ]$, for bi-adjoint $\Phi^3$ scalar theory are defined as}\,
\begin{align}\label{}
& \mathfrak{M}_n^{\rm (1-P;NP)}[\pi_1, \ldots, \pi_n  \, ; \, \g_1\ldots, \g_p\, | \, \d_{p+1},\ldots, \d_n ]:=\nonumber\\
& m_n^{\rm (1-P;NP)}[\pi_1, \ldots, \pi_n \, \, ; \, \g_1\ldots, \g_p\, | \, \d_{n},\ldots, \d_{p+1} ] ,  \\
& \mathfrak{M}_n^{\rm (1-NP;NP)}[\pi_1, \ldots, \pi_p | \rho_{p+1}, \ldots, \rho_n   \, ; \, \g_1\ldots, \g_q\, | \, \d_{q+1},\ldots, \d_n ]:=  \nonumber \\
& m_n^{\rm (1-NP;NP)}[\pi_1, \ldots, \pi_p | \rho_n, \ldots , \rho_{p+1} \, \, ; \, \g_1\ldots, \g_q\, | \, \d_{n},\ldots, \d_{q+1} ] .
\end{align}
In a similar way we define $\mathfrak{M}_n^{\rm (1-NP;P)} [\pi | \rho  \, ; \, \g  ]$. 

In the next section we will present some examples, which can be directly compared with the field theory results.

\section{Examples and non-planar CHY-graphs}\label{ex-npgraphs}

In this section we compute some particular examples for lower number of points, $n = 4,5$ by  using the new proposal given in section \ref{section-def}. In addition, we define and understand the more general structure of the {\it non-planar  CHY-graphs at one-loop} in a simple way, which will be presented in the next section.

Before giving the examples, it is useful to introduce the {\it line} notation\footnote{This line is the same {\it square} defined in \cite{Gomez:2017cpe}.}
\vskip-1.5cm
\begin{eqnarray}\label{}
 \parbox[c]{13em}{\includegraphics[scale=0.3]{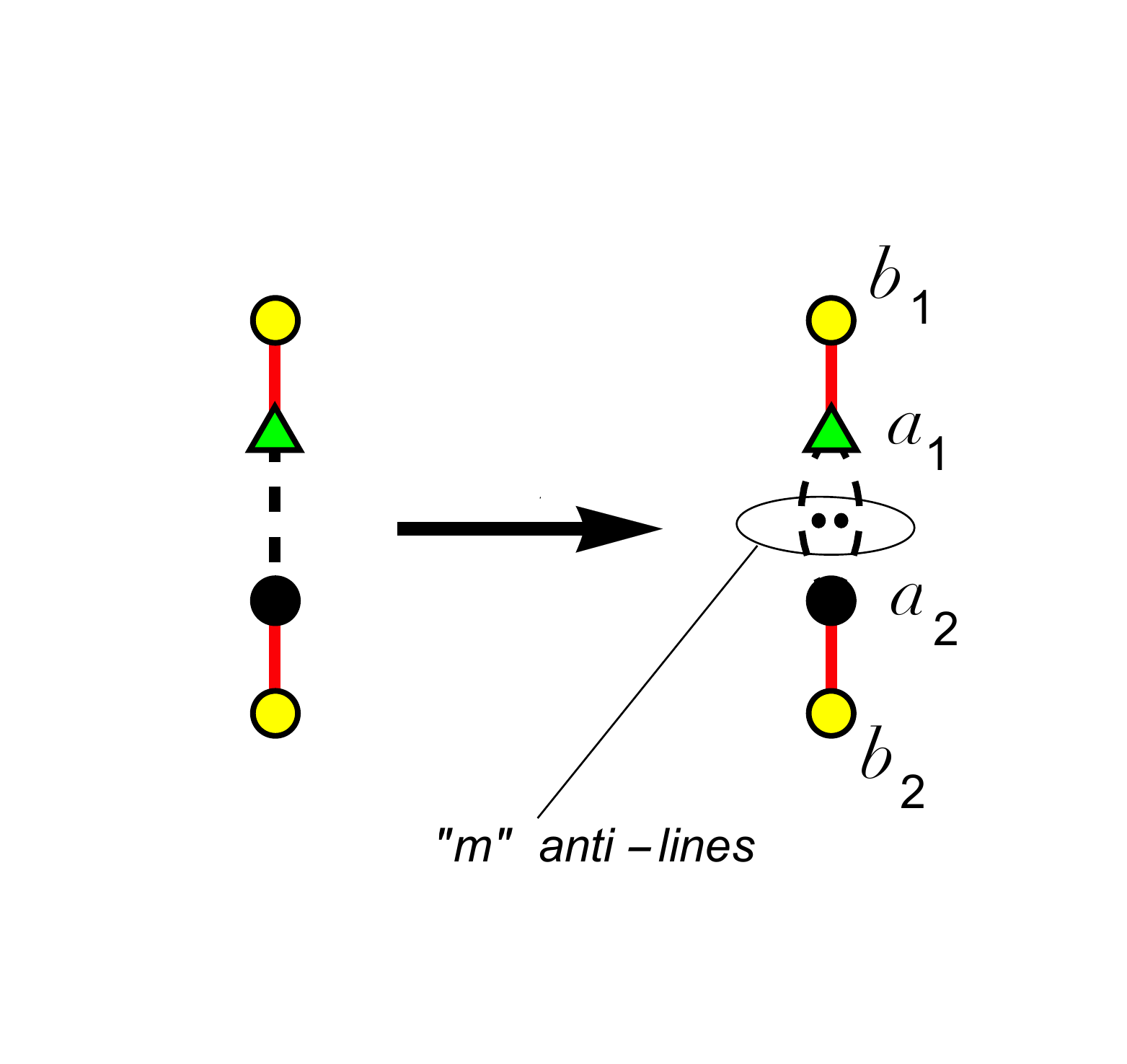}} ,
\end{eqnarray}
\vskip-0.7cm\noindent
where the number of {\it anti-lines} ``$m$"  is equal to, $ m=\#\,{\it lines} - 4 $. 
Finally, in order to obtain a more compact notation, we bring in the following definitions
\vskip-0.9cm
\begin{eqnarray}
\overrightarrow{ \,\,\left[1,2,\ldots ,n \right] \,\,}:=
\hspace{-0.62cm}
\parbox[c]{7.5em}{\includegraphics[scale=0.20]{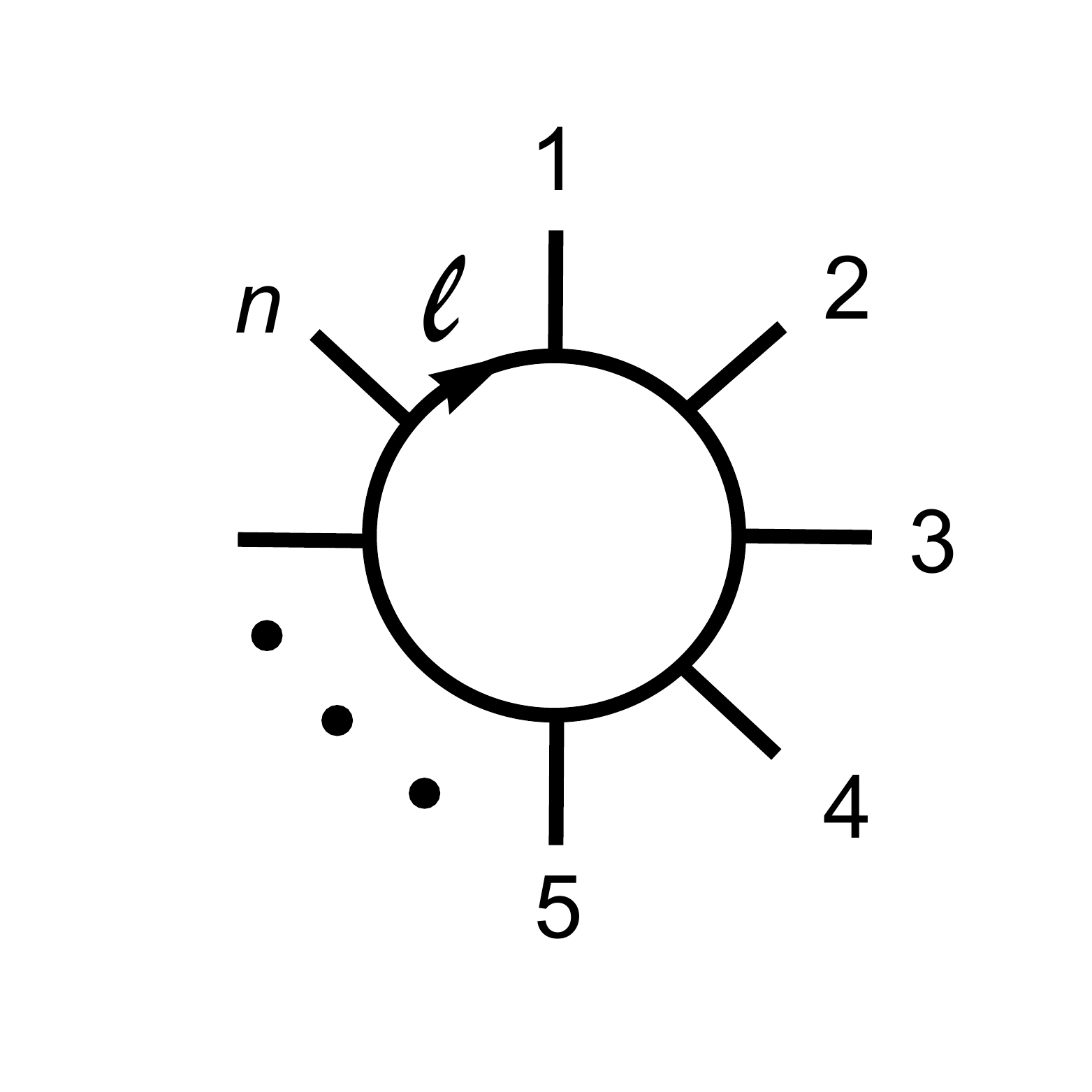}} = \,\, \frac{1}{\ell^2 (\ell+k_1)^2 (\ell+k_1 + k_2)^2 \cdots (\ell+k_1 + k_2+ \cdots + k_{n-1})^2 }\,,\quad\quad
\label{}
\end{eqnarray}
\vskip-0.6cm\noindent
and
\vskip-0.8cm
\begin{eqnarray}\label{fey-1loop-T} 
&&
\parbox[c]{8.8em}{\includegraphics[scale=0.24]{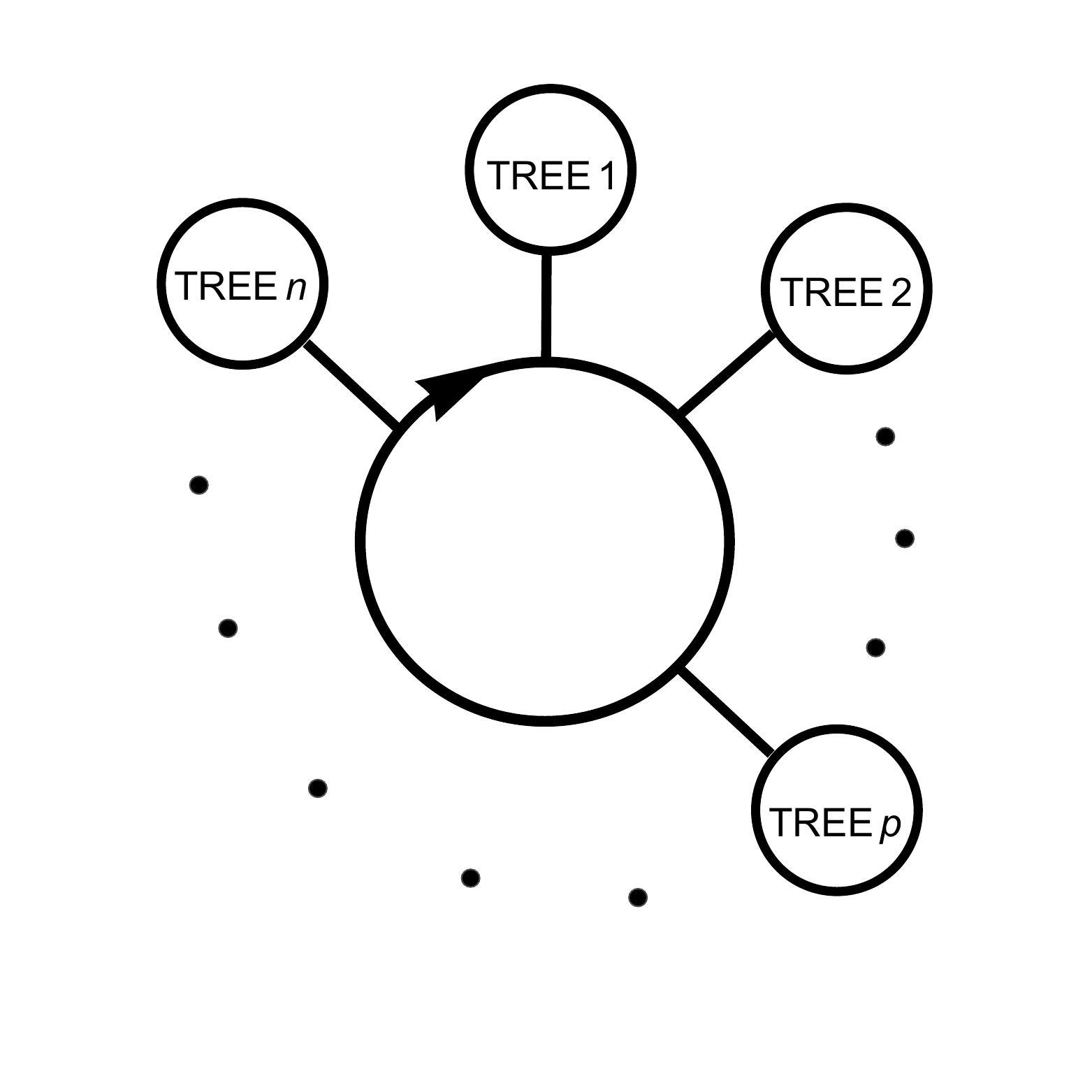}} \equiv  
\hspace{-0.4cm}
\parbox[c]{8.7em}{\includegraphics[scale=0.24]{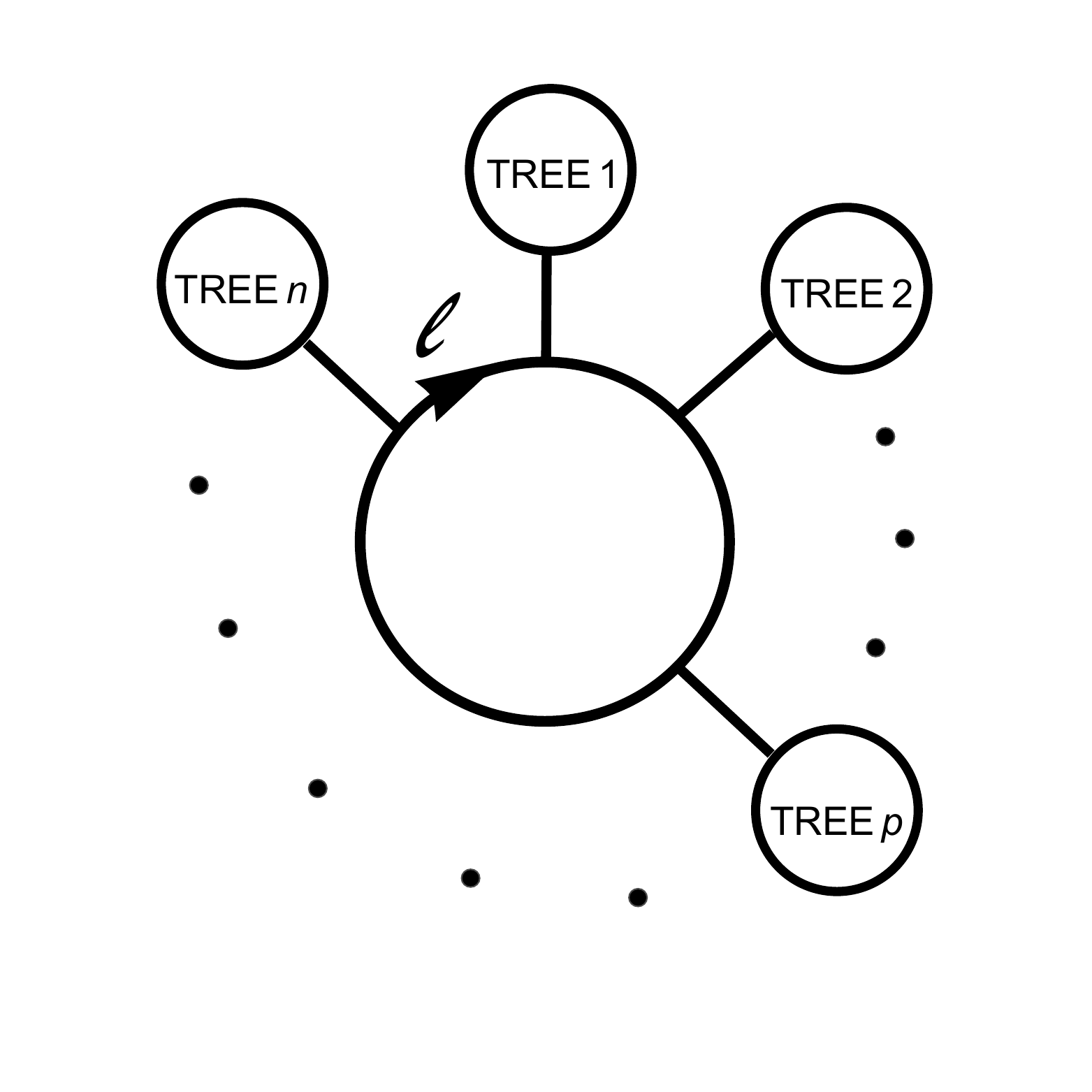}} +
\hspace{-0.4cm}
\parbox[c]{8.7em}{\includegraphics[scale=0.24]{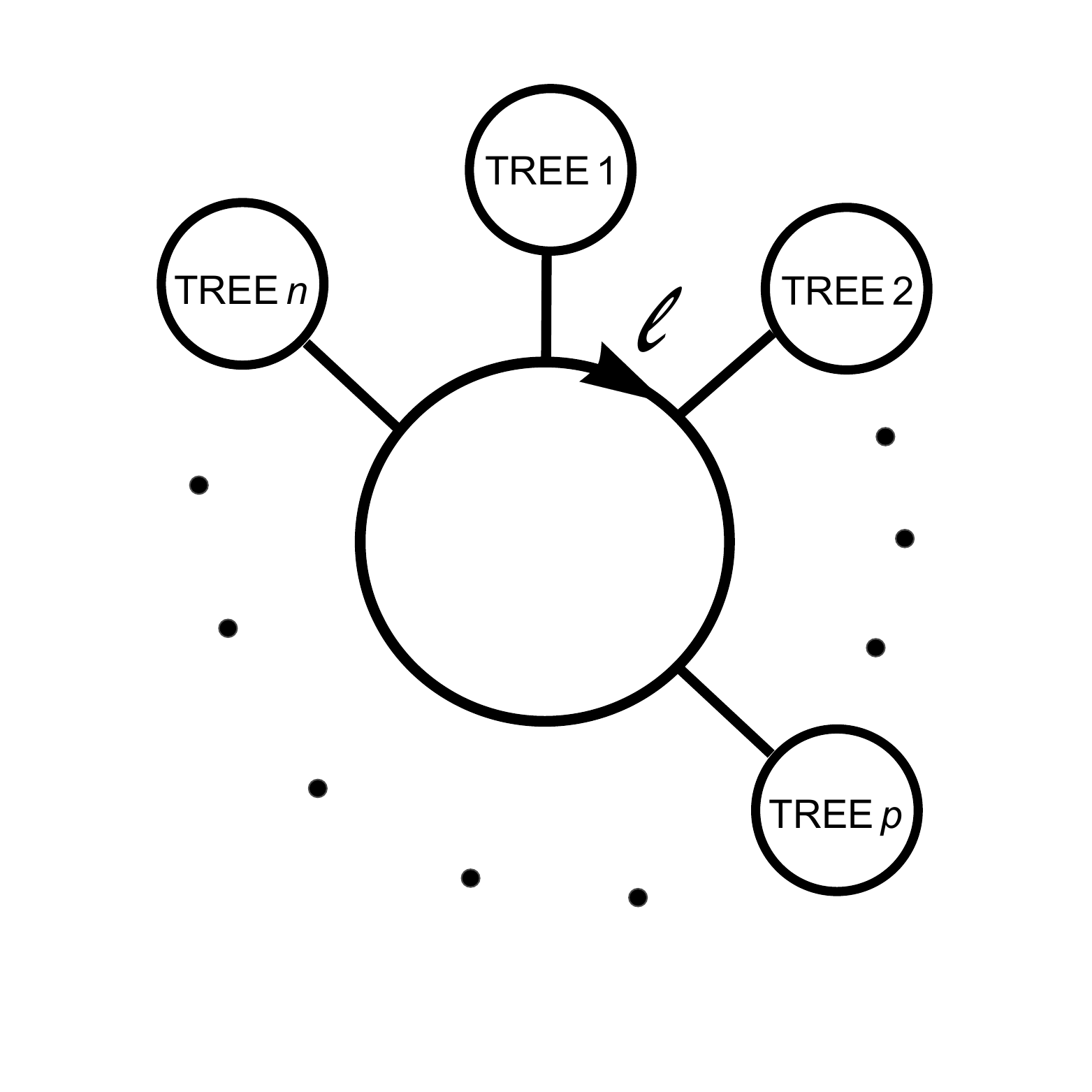}} +
\cdots +
\hspace{-0.5cm}
\parbox[c]{10em}{\includegraphics[scale=0.24]{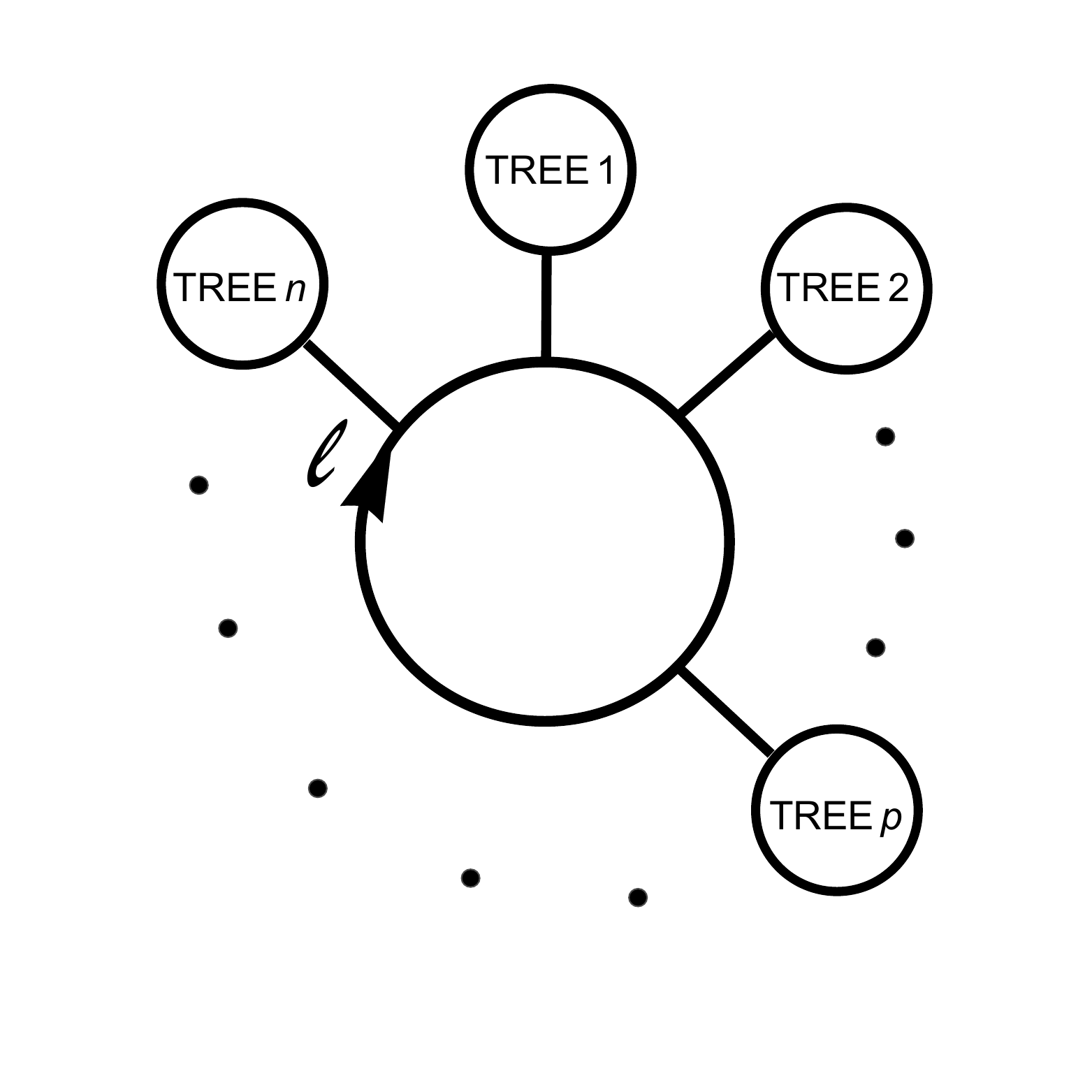}}
\quad\non
&&= \overrightarrow{ \,\left[ {\rm tree\,1,tree\,2,\ldots ,tree\, n }\right] \,} +  \overrightarrow{ \,\left[ {\rm tree\,2,\ldots, tree\,n, tree\, 1 }\right] \,} +
\cdots +   \overrightarrow{ \,\left[ {\rm tree\,n, tree\,1, \ldots,  tree\, (n-1) }\right] \,} ,\non
\end{eqnarray}
\vskip-0.1cm\noindent
where the arrow over the bracket,  $\overrightarrow{ \,\left[ {\rm \ldots }\right] \,}$,  means the transit of the loop momentum ``$\ell$" and  
``${\rm tree\,} i$" is a generic Feynman diagram at tree level. For simplicity in the notation we omit the integral, $\int d^{D}\ell$.  

Finally, in order to obtain a correspondence among the CHY-graphs for non-planar bi-adjoint scalar theory and Feynman diagrams at one-loop, we follow the same procedure performed in \cite{Gomez:2017cpe}, i.e. we will carry out a power expansion\footnote{Note that the lowest power of that expansion for the non-planar Parke-Taylor factors is four, it is a consequence of the {\it Theorem 1} in \cite{Gomez:2017cpe}. } of ${\rm PT}^{(1)}_{\, a_1:a_2}[\cdots ]$ in terms of $ \o^{a_1:a_2}_{i:j} $, while for ${\rm PT}^{(1)}_{\, b_1:b_2}[\cdots ]$ we will use its original definition.

\subsection{Four-point}\label{four-point-chy}

First of all, we analyze the simplest example, the four-point computation\footnote{In appendix \ref{4-p-app-fey} we carried out the four-point computation using the Feynman rules.}.  

Let us consider the ${\rm NP;P}$ contribution, which is given by the expression
\begin{align}\label{4p-NP-P}
 \mathfrak{M}_4^{\rm (1-NP;P)}[1,2| 3,4 \, ; \, 1,2 ,3,4 ] &= {1 \over 2^{4+1}}\int d\Omega \times s_{a_1b_1} \non
&\times 
\int d\mu_{4+4}^{\rm t} \,\,\,
{\rm PT}^{(1)}_{\, a_1:a_2}[ 1,2|3,4   ]\times {\rm PT}^{(1)}_{\, b_1:b_2}[ 1,2,3,4 ].\non
\end{align}
Since we have the identity, ${\rm PT}^{(1)}_{\, a_1:a_2}[ 1,2|3,4   ] = \frac{1}{(a_1,b_1,b_2,a_2)}\times \o^{a_1:a_2}_{1:1}  \o^{a_1:a_2}_{2:2}  \o^{a_1:a_2}_{3:3}   \o^{a_1:a_2}_{4:4}$, which has been shown in \cite{Gomez:2017cpe}, it is simple to draw the CHY-graphs
\vskip-0.3cm
\begin{eqnarray}\label{4pts-1ex}
\mathfrak{M}_4^{\rm (1-NP;P)}[1,2 | 3,4 \, ;  \, 1,2,3,4 ]= 
\frac{1}{2^{5}} \int d\Omega \, s_{a_1 b_1} \,
 \int d\mu^{\rm t}_{4+4}  
\left\{
\hspace{-0.3cm}
 \parbox[c]{8.5em}{\includegraphics[scale=0.22]{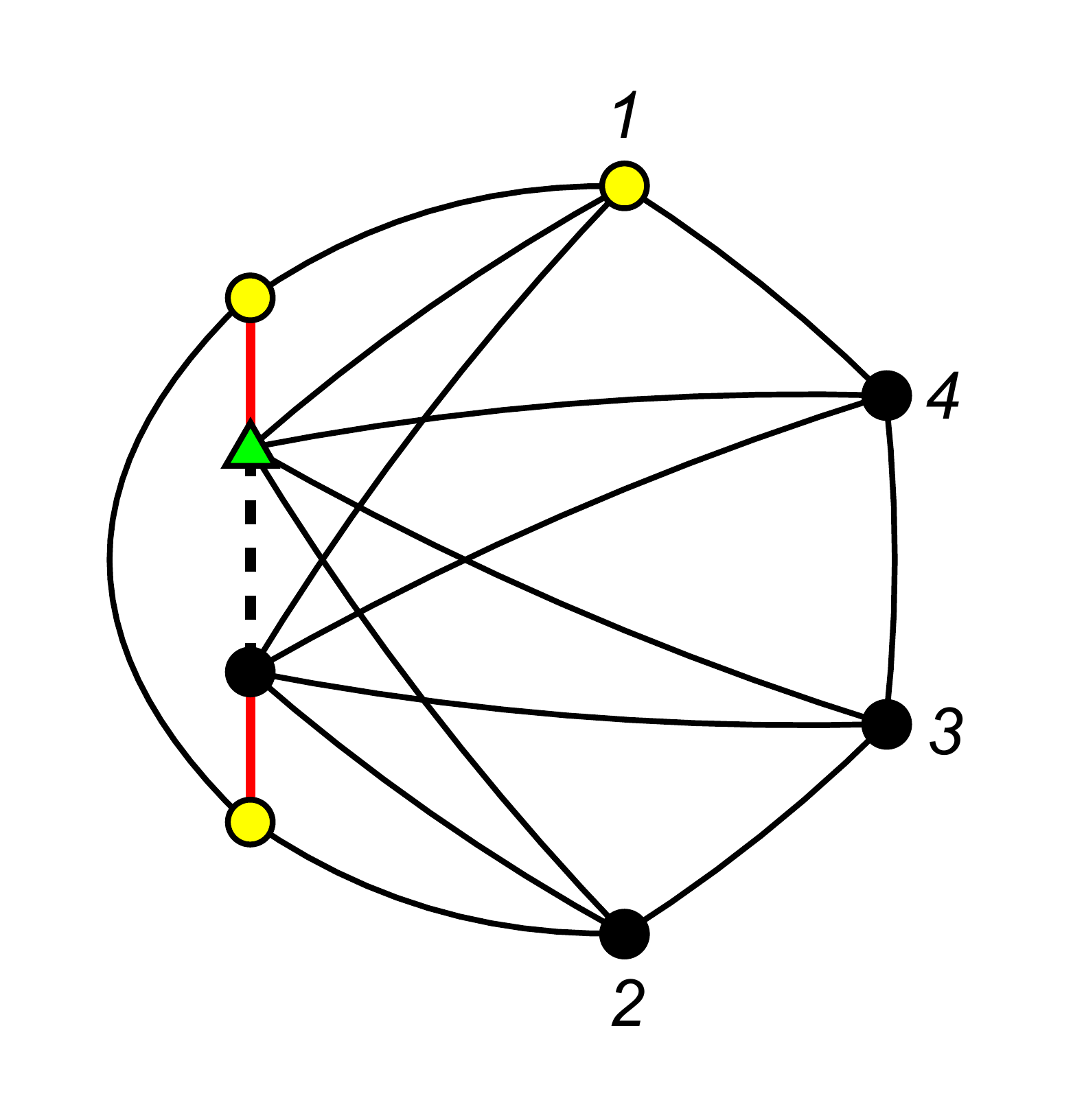}}
 + \,\,
 {\rm cyc(1,2,3,4)} 
 \right\} . \qquad\non 
\end{eqnarray}
\vskip-0.5cm\noindent
These kind of graphs were already computed in \cite{Gomez:2017cpe} and the answer is
\vskip-1.6cm
\begin{eqnarray}
\mathfrak{M}_4^{\rm (1-NP;P)}[1,2 | 3,4 \, ;  \, 1,2,3,4 ] \,\,= 
\hspace{-1.6cm}
 \parbox[c]{11.5em}{\includegraphics[scale=0.37]{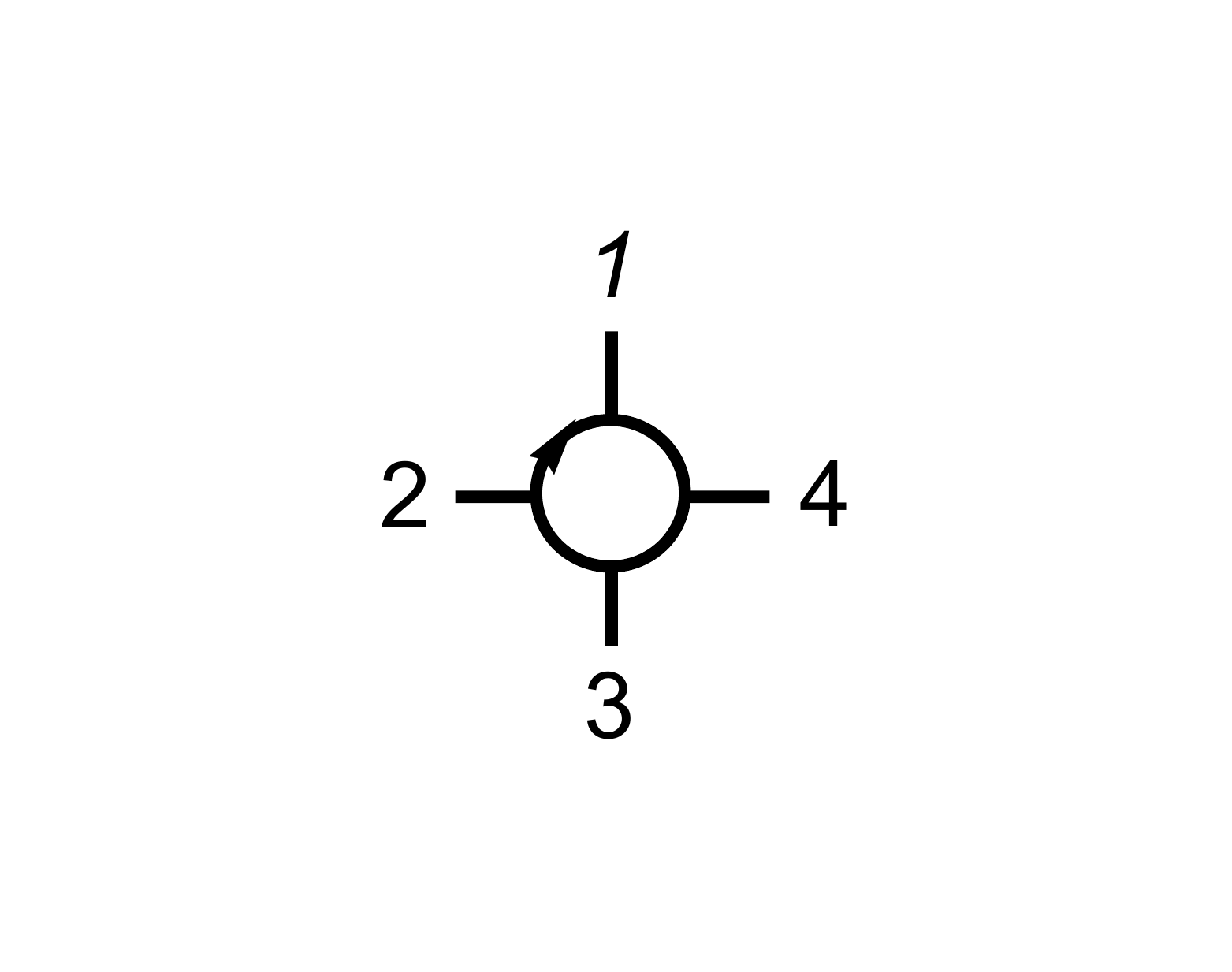}}.
 \label{4p-NP-P2}
 \qquad 
\end{eqnarray}
\vskip-1.0cm\noindent

Now, we compute the ${\rm NP;NP}$ part of the four-point case. Let us consider the amplitude 
\begin{align}\label{4p-NP-NP}
 \mathfrak{M}_4^{\rm (1-NP;NP)}[1,2| 3,4 \, ; \, 1,2 |3,4 ] &= {1 \over 2^{4+1}}\int d\Omega \times s_{a_1b_1} \non 
&\times 
\int d\mu_{4+4}^{\rm t} \,\,\,
{\rm PT}^{(1)}_{\, a_1:a_2}[ 1,2|3,4   ]\times {\rm PT}^{(1)}_{\, b_1:b_2}[ 1,2| 3,4 ] \nonumber\\
& 
\hspace{-4.8cm}=
{1 \over 2^{5}}\int d\Omega \, s_{a_1b_1}
\int d\mu_{4+4}^{\rm t} \,\frac{1}{(a_1,b_1,b_2,a_2)^2}\,
(\o^{a_1:a_2}_{1:1}  \o^{a_1:a_2}_{2:2}   \o^{a_1:a_2}_{3:3}  \o^{a_1:a_2}_{4:4}) \times
(\o^{b_1:b_2}_{1:1}   \o^{b_1:b_2}_{2:2}   \o^{b_1:b_2}_{3:3}  \o^{b_1:b_2}_{4:4})\, \non,
\end{align}
where we have used the same identity as in the above example. This CHY-integral was computed by one of the authors in \cite{Gomez:2017lhy} and the result is  
\vskip-1.6cm
\begin{eqnarray}\label{4pt-NP-NP}
\mathfrak{M}_4^{\rm (1-NP;NP)}[1,2 | 3,4 \, ;  \, 1,2|3,4 ] \,\,= 
\hspace{-1.6cm}
 \parbox[c]{11.0em}{\includegraphics[scale=0.37]{fey-4p-NP-P.pdf}} + \,\,{\rm per} (2,3,4) \, \, .
 \qquad 
\end{eqnarray}
\vskip-0.9cm\noindent
Next, in order to compare this example with the previous one and to obtain more information from it,  we use the same procedure as in \eqref{4pts-1ex}. Thus,  on the $a$-sector we apply the identity, ${\rm PT}^{(1)}_{\, a_1:a_2}[ 1,2|3,4   ] = (a_1,b_1,b_2,a_2)^{-1}\times  \o^{a_1:a_2}_{1:1}  \o^{a_1:a_2}_{2:2}  \o^{a_1:a_2}_{3:3}   \o^{a_1:a_2}_{4:4}$, but, on the $b$-sector we use the definition given in \eqref{nonpPT},  therefore we have the graph expansion 
\begin{eqnarray}\label{4pts-2ex}
\hspace{-0.3cm}
\mathfrak{M}_4^{\rm (1-NP;NP)}[ 1,2\, | \,3,4 \, ; \, 1,2\, | \,3,4 ]= \frac{1}{2^{5}} \int d\Omega \, s_{a_1 b_1} 
 \int d\mu^{\rm t}_{4+4}  
\left\{
\hspace{-0.5cm}
 \parbox[c]{7.7em}{\includegraphics[scale=0.23]{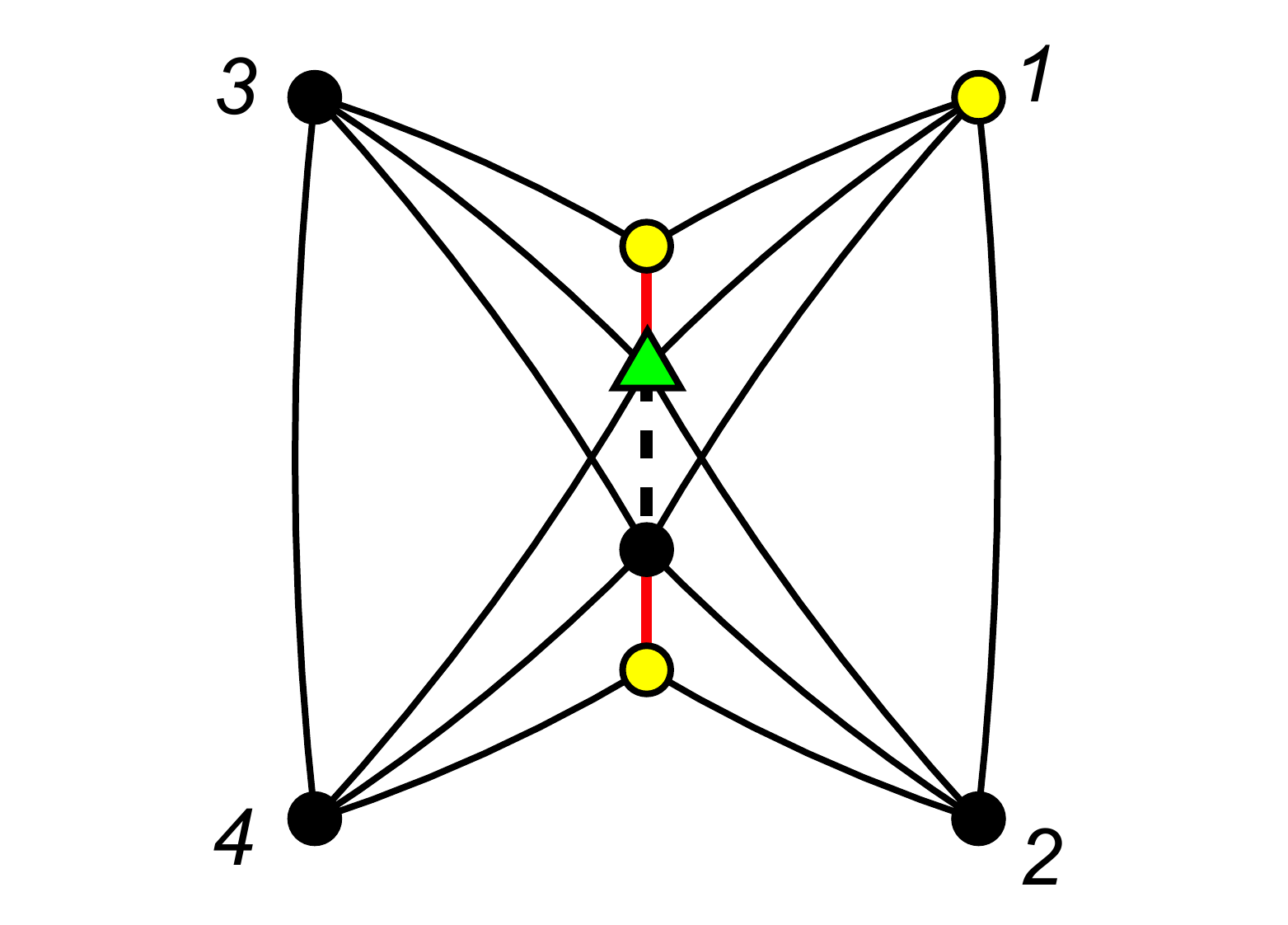}}
\begin{matrix}
 +&  \hspace{-1.8cm} (1\leftrightarrow 2) \\
+&  \hspace{-1.8cm}   (3\leftrightarrow 4)  \\
+&  (1\leftrightarrow  2) \times   (3\leftrightarrow 4)
\end{matrix}
 \right\}  . \qquad\non
\end{eqnarray}
The CHY-graphs obtained above are a new type of graphs, which will be called\footnote{Note that the graph in \eqref{4pts-2ex} is like a double copy of the graph in \eqref{4pts-1ex} without the connection between $\s_{b_1}$ and $\s_{b_2}$.} {\it the non-planar CHY-graphs (or butterfly graphs)}.  Using the $\L-$algorithm developed by one of the authors in \cite{Gomez:2016bmv}, we can compute this butterfly graph in a simple way, and the result is  (see appendix \ref{4-p-app-chy})
\vskip-1.5cm
\begin{eqnarray}\label{}
 \frac{1}{2^{5}} \int d\Omega \times s_{a_1 b_1}\int d\mu^{\rm t}_{4+4}   
 \hspace{-0.5cm}
 \parbox[c]{7.7em}{\includegraphics[scale=0.23]{4pts_12-34.pdf}}
 &=& 
 \hspace{-1.6cm}
\parbox[c]{11em}{\includegraphics[scale=0.37]{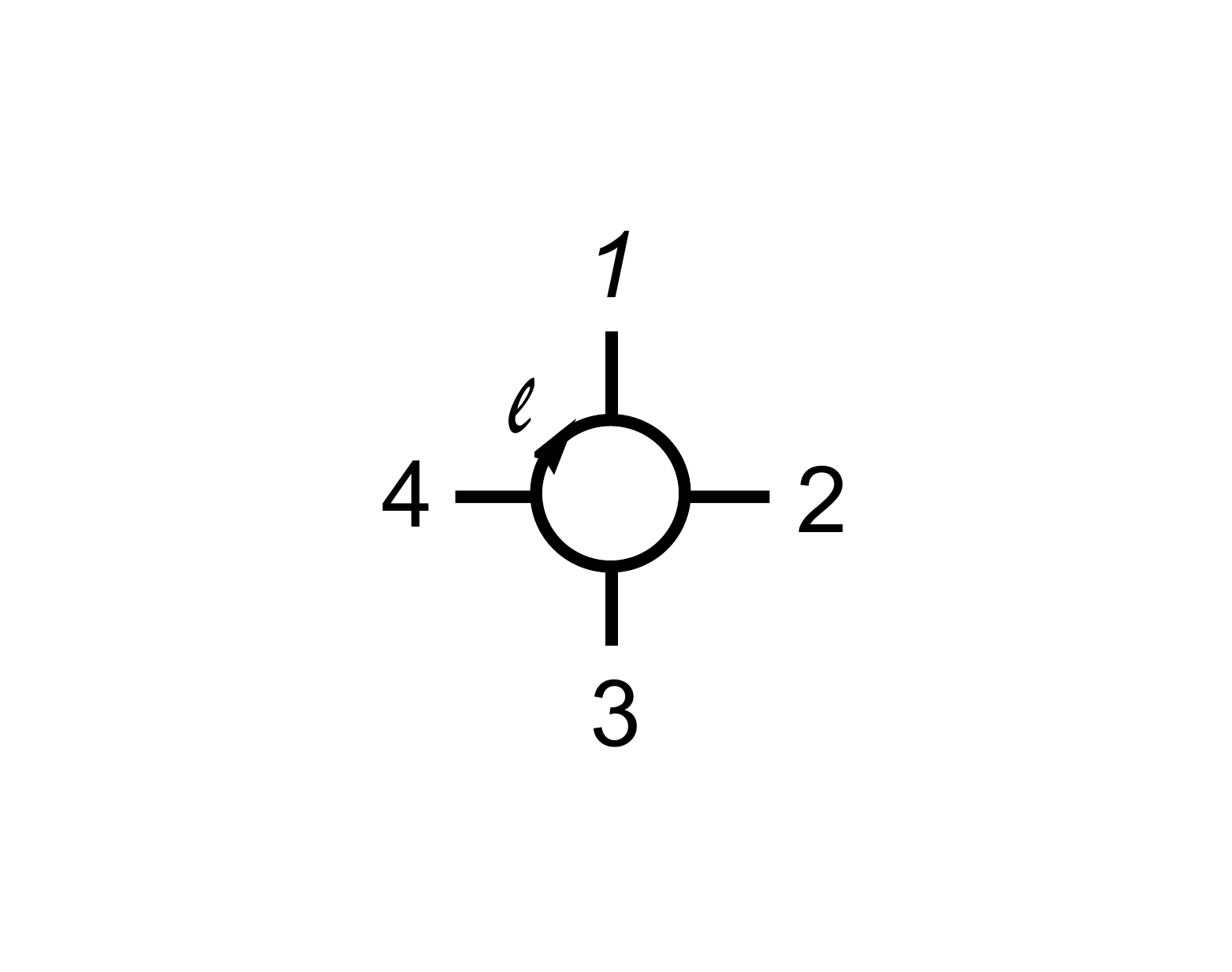}}
 +
  \hspace{-1.6cm}
\parbox[c]{11em}{\includegraphics[scale=0.37]{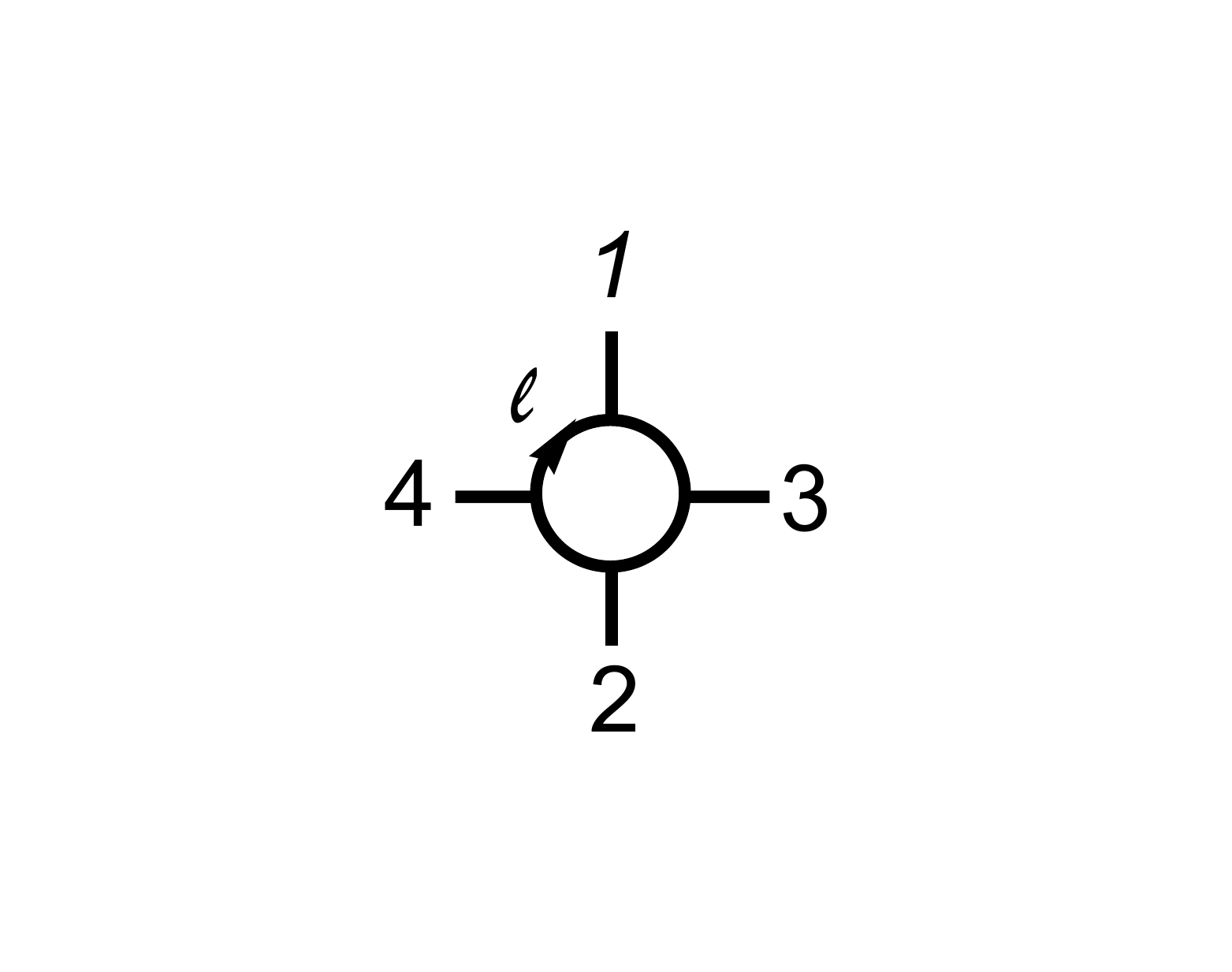}}
+
 \hspace{-1.6cm}
\parbox[c]{11em}{\includegraphics[scale=0.37]{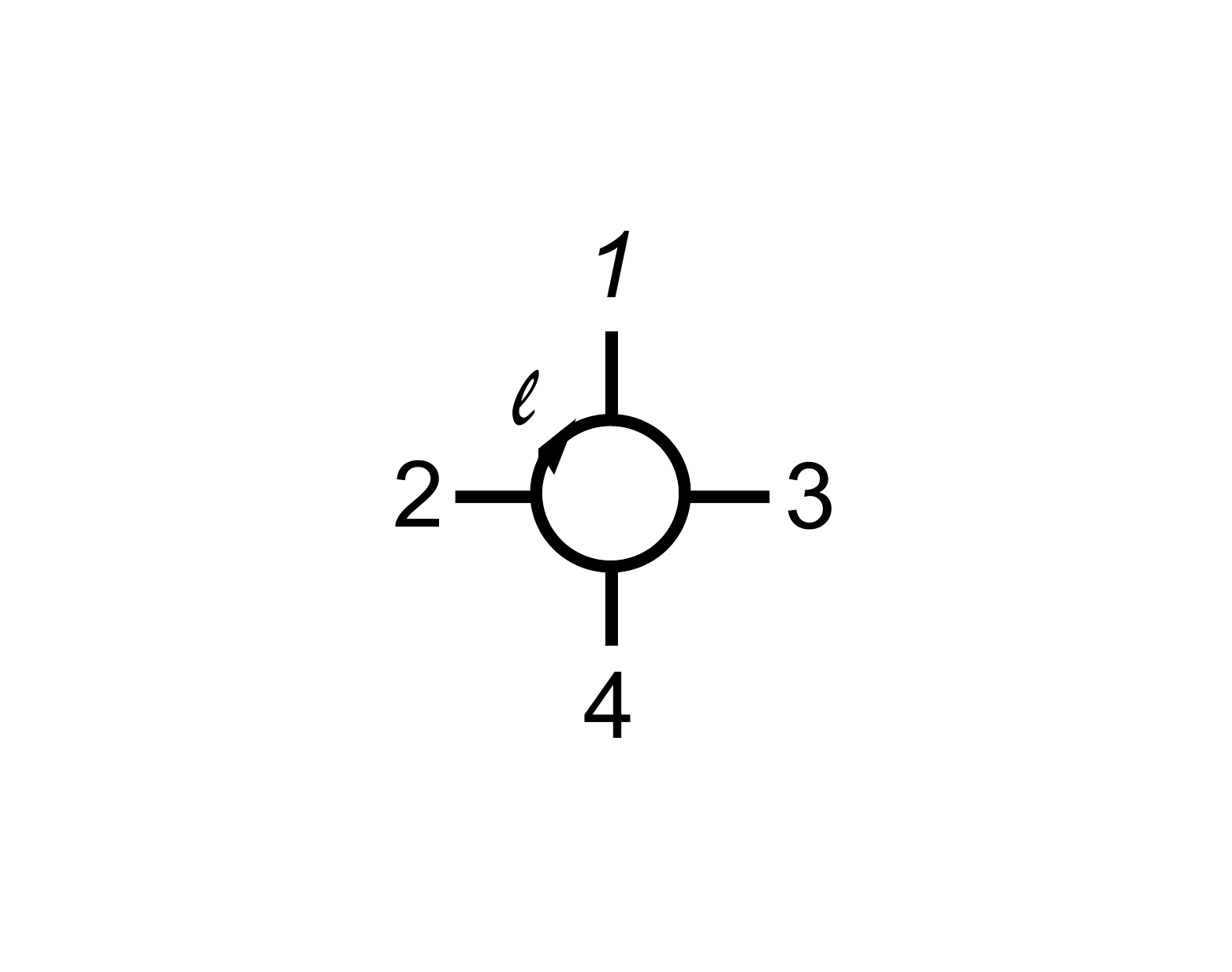}} 
\nonumber
\end{eqnarray}
\vskip-2.4cm
\begin{eqnarray}
\hspace{6.4cm}
 && +
 \hspace{-1.6cm}
\parbox[c]{11em}{\includegraphics[scale=0.37]{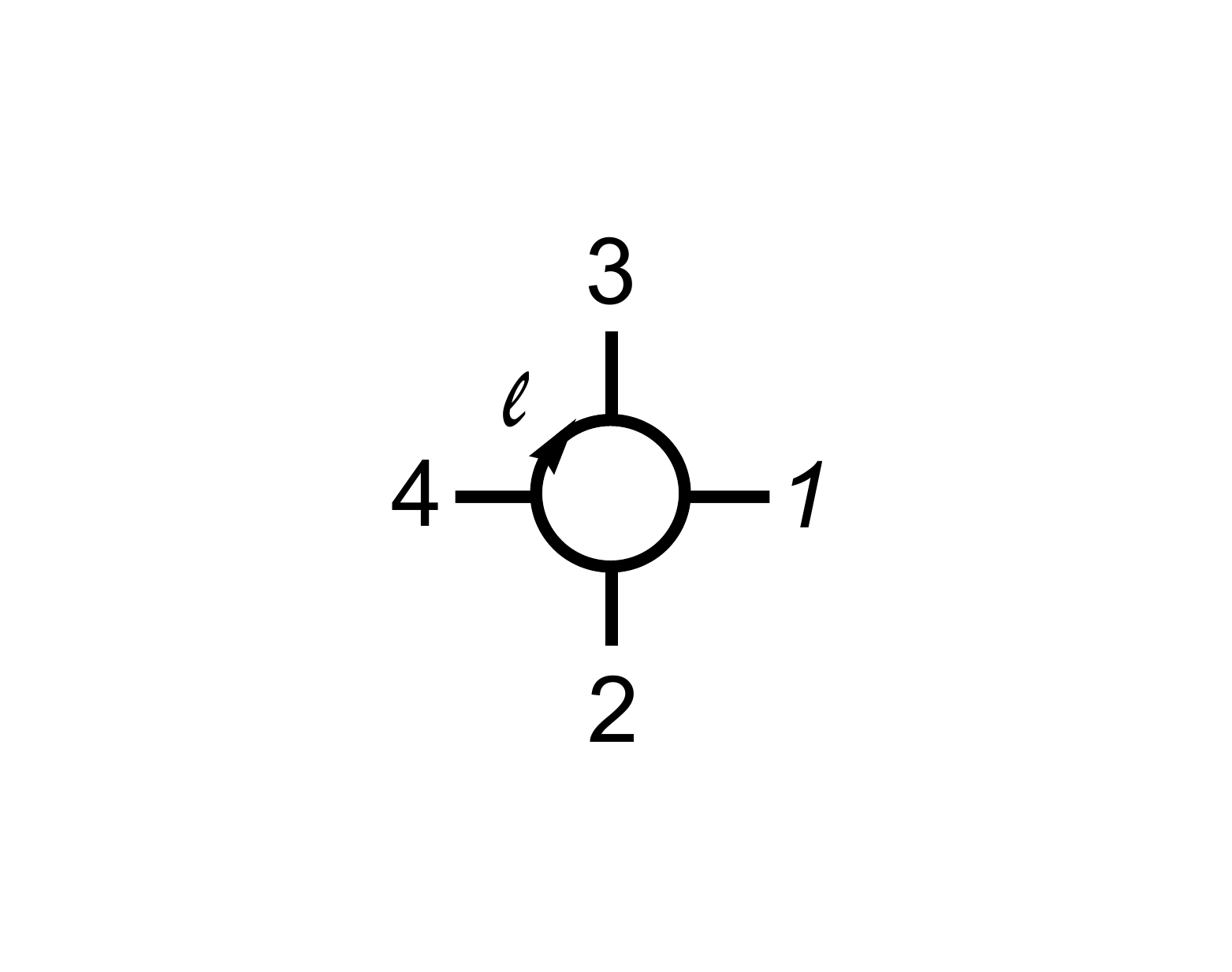}}
+
 \hspace{-1.6cm}
\parbox[c]{11em}{\includegraphics[scale=0.37]{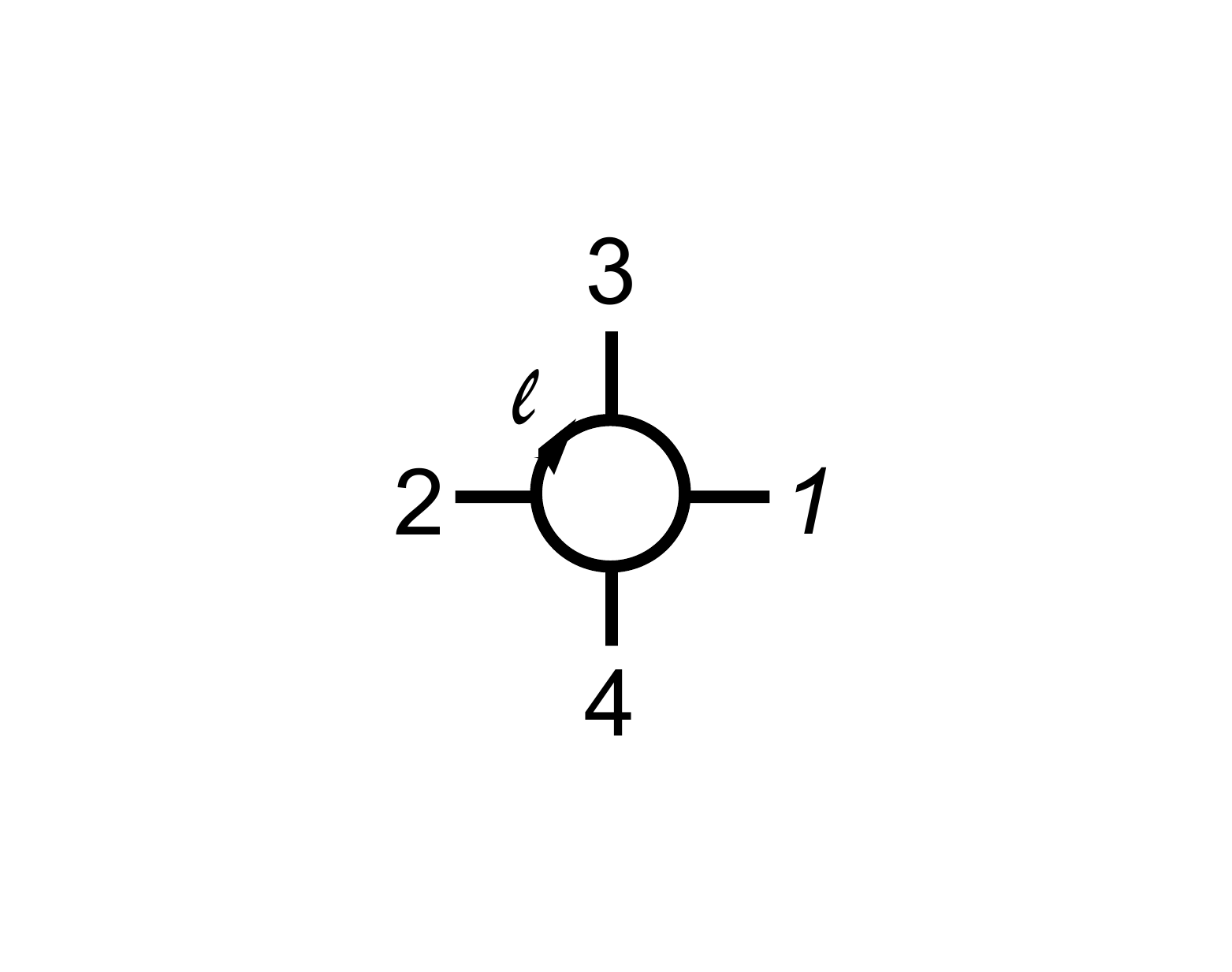}}
+
 \hspace{-1.6cm}
\parbox[c]{11em}{\includegraphics[scale=0.37]{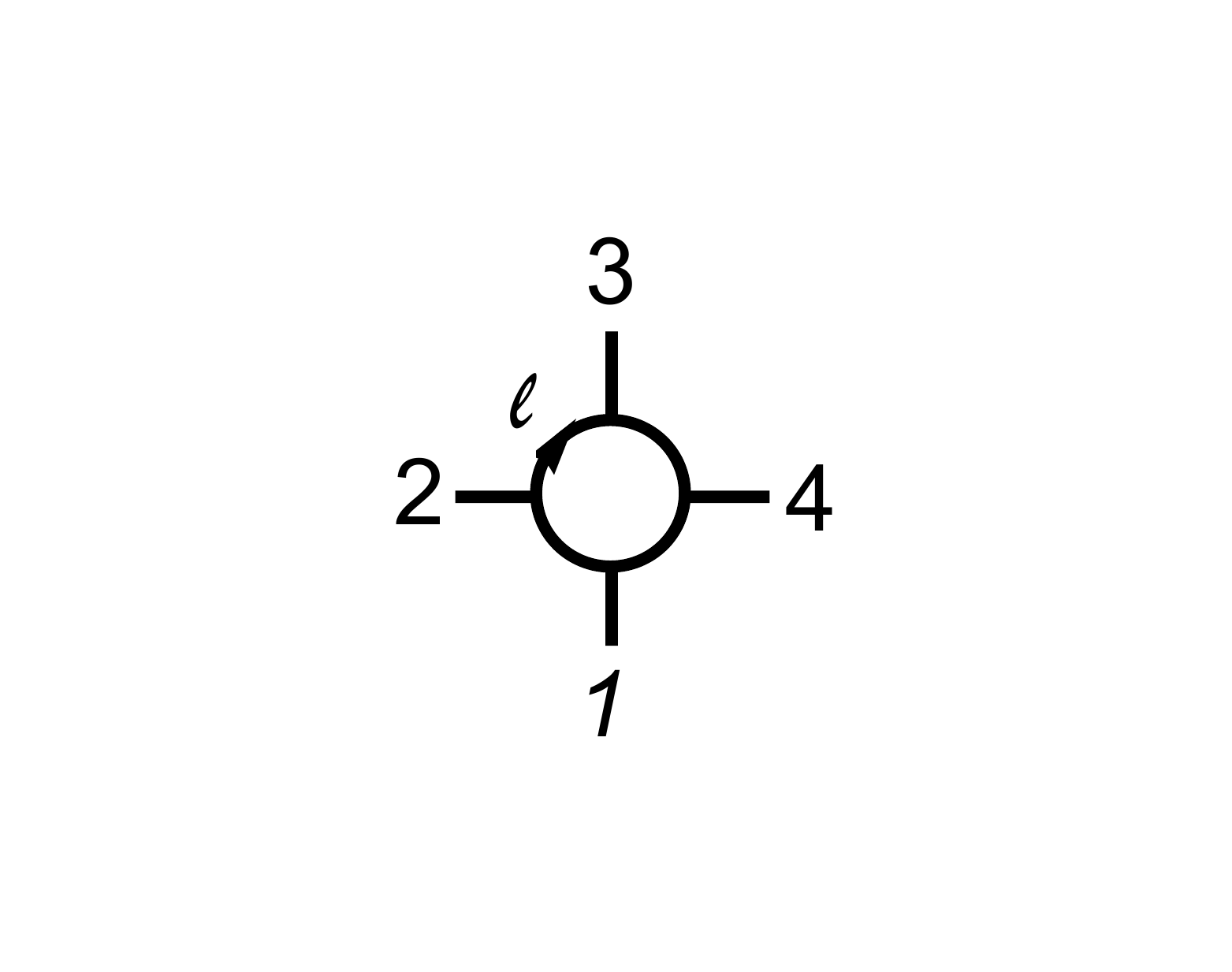}} 
\qquad\quad
\nonumber
\end{eqnarray}
\vskip-1.4cm
\begin{eqnarray}\label{4ptB-r}
&&
\hspace{2cm}
 =
\overrightarrow{ \,\left[ 1,2, 3,4 \right] \,}+ \overrightarrow{ \,\left[ 1, 3,2,4 \right] \,} + \overrightarrow{ \,\left[ 1, 3,4,2 \right] \,} + \overrightarrow{ \,\left[ 3,1,2,4 \right] \,} + \overrightarrow{ \,\left[ 3,1,4,2 \right] \,} + \overrightarrow{ \,\left[ 3,4,1,2 \right] \,} \nonumber\\
&&
\hspace{2cm}
=  \overrightarrow{ \, \, \left[ 1,2\right] \sh \left[3,4\right] \, \,} \,  \qquad\qquad .
\end{eqnarray}
This interesting result is generalized in section \ref{nonplanarG}. 
Certainly, by summing the four butterfly graphs from \eqref{4pts-2ex}, one obtains the same answer as in \eqref{4pt-NP-NP}.  

Notice that we have not used the Parke-Taylor identities found in section \ref{KK-relations-total} in the above examples, for instance
\begin{align}
{\rm PT}^{(1)}_{\, a_1:a_2}[ 1,2|3,4   ] = & {\rm PT}^{(1)}_{\, a_1:a_2}[ 1,2, 3,4   ]+ {\rm PT}^{(1)}_{\, a_1:a_2}[ 1,3,2,4   ]+ {\rm PT}^{(1)}_{\, a_1:a_2}[ 3,1,2,4   ] \\
 & + {\rm PT}^{(1)}_{\, a_1:a_2}[ 2,1, 3,4   ]+ {\rm PT}^{(1)}_{\, a_1:a_2}[ 2,3,1,4   ]+ {\rm PT}^{(1)}_{\, a_1:a_2}[ 3,2,1,4   ] \,\,\, .\nonumber
\end{align}
We could do that and then to apply the technology developed in \cite{Gomez:2017cpe}. However, this procedure is  longer and tedious, we would also lose the correspondence between the non-planar CHY-graphs and Feynman diagrams at one-loop.

\subsection{Five-point}

In this section we consider less trivial examples, the five-point case computations. The first one  is given by
\begin{align}\label{5p-NP-P}
 \mathfrak{M}_5^{\rm (1-NP;P)}[1,2| 3,4,5 \, ; \, 1,2 ,3,4,5 ] &= {1 \over 2^{5+1}}\int d\Omega \times s_{a_1b_1} \non 
&\times 
\int d\mu_{5+4}^{\rm t} \,\,\,
{\rm PT}^{(1)}_{\, a_1:a_2}[ 1,2|3,4,5   ]\times {\rm PT}^{(1)}_{\, b_1:b_2}[ 1,2,3,4,5 ].\nonumber\\
\end{align}
From the results found in \cite{Gomez:2017cpe}, it is straightforward to check  the identity  
\begin{align}\label{pt1-5-np}
& {\rm PT}^{(1)}_{\, a_1:a_2}[ 1,2|3,4,5   ] =  \frac{1}{(a_1,b_1,b_2,a_2)}\times \left(\o^{a_1:a_2}_{1:1}  \o^{a_1:a_2}_{2:2}\right) \times \left.\Big(  
2\,\o^{a_1:a_2}_{3:3}  \o^{a_1:a_2}_{4:4}  \o^{a_1:a_2}_{5:5}   \right. \nonumber\\
& \left.
+ \frac{\s_{34} \, \s_{45} \,  \o^{a_1:a_2}_{5:3} }{(3,4,5)}     \,\,  \o^{a_1:a_2}_{4:4}
+\frac{\s_{45} \, \s_{53} \,  \o^{a_1:a_2}_{3:4} }{(3,4,5)}     \,\,  \o^{a_1:a_2}_{5:5}
+\frac{\s_{53} \, \s_{34} \,  \o^{a_1:a_2}_{4:5} }{(3,4,5)}     \,\,  \o^{a_1:a_2}_{3:3}
\right),\non
\end{align}
so, using the definition, ${\rm PT}^{(1)}_{\, b_1:b_2}[ 1,2,3,4,5 ]=\frac{1}{(a_1,b_1,b_2,a_2)} \left[  \o^{b_1:b_2}_{1:2}\frac{1}{\s_{23}\s_{34}\s_{45}\s_{51}}+{\rm cyc}(1,2,3,4,5) \right] $,
 we obtain the graph expansion
\vskip-0.5cm
\begin{eqnarray}\label{5pts-1ex}
&&\mathfrak{M}_5^{\rm (1-NP;P)}[1,2 | 3,4,5 \, ;  \, 1,2,3,4,5 ]= 
\frac{1}{2^{6}} \int d\Omega \times  s_{a_1 b_1} \times
 \int d\mu^{\rm t}_{5+4}  \non
&&
\left\{ 2\times
\hspace{-0.3cm}
 \parbox[c]{8.5em}{\includegraphics[scale=0.22]{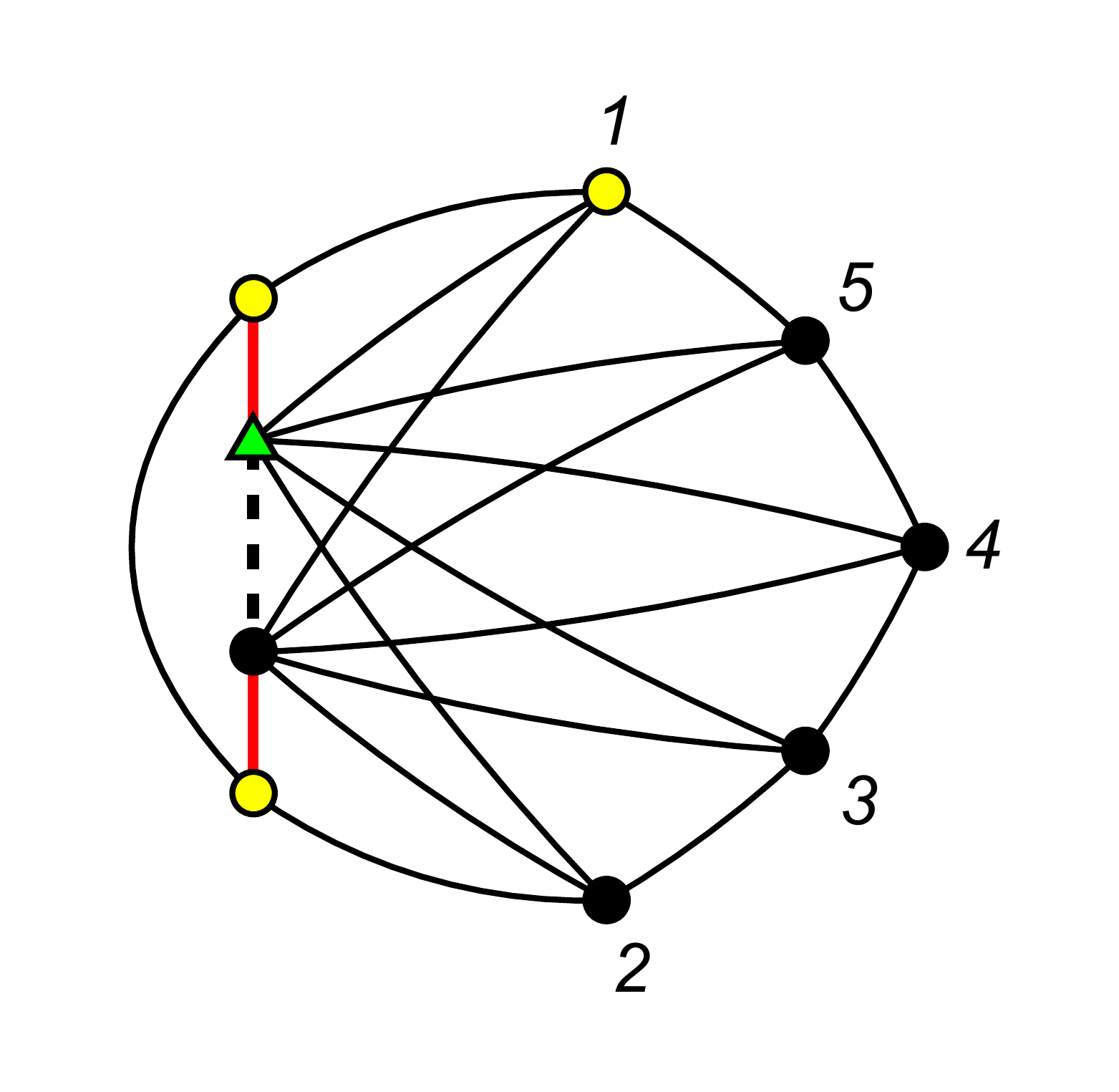}}
 + 
 \hspace{-0.3cm}
    \parbox[c]{8.5em}{\includegraphics[scale=0.22]{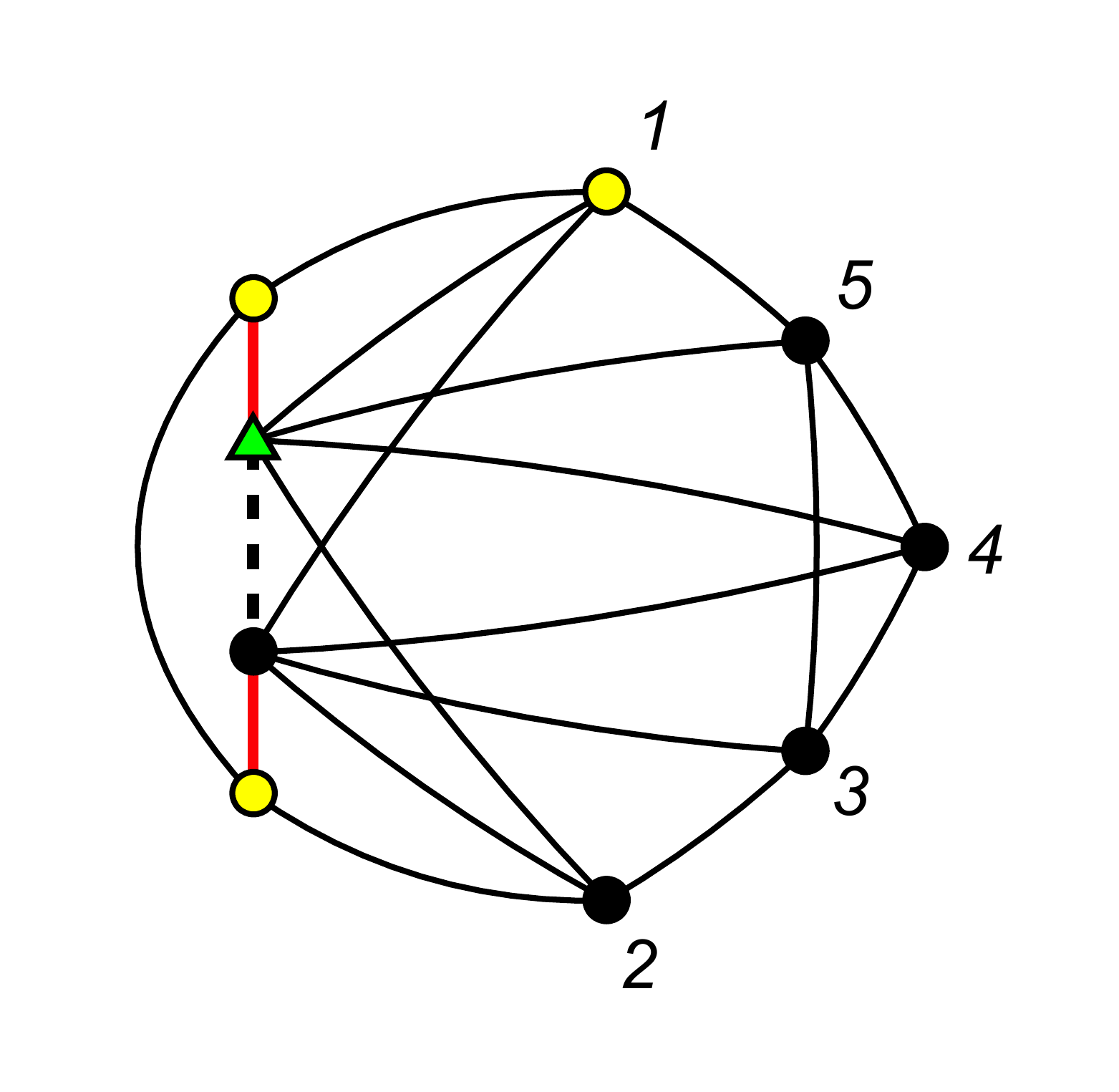}} 
    +  
\hspace{-0.3cm}
 \parbox[c]{8.1em}{\includegraphics[scale=0.22]{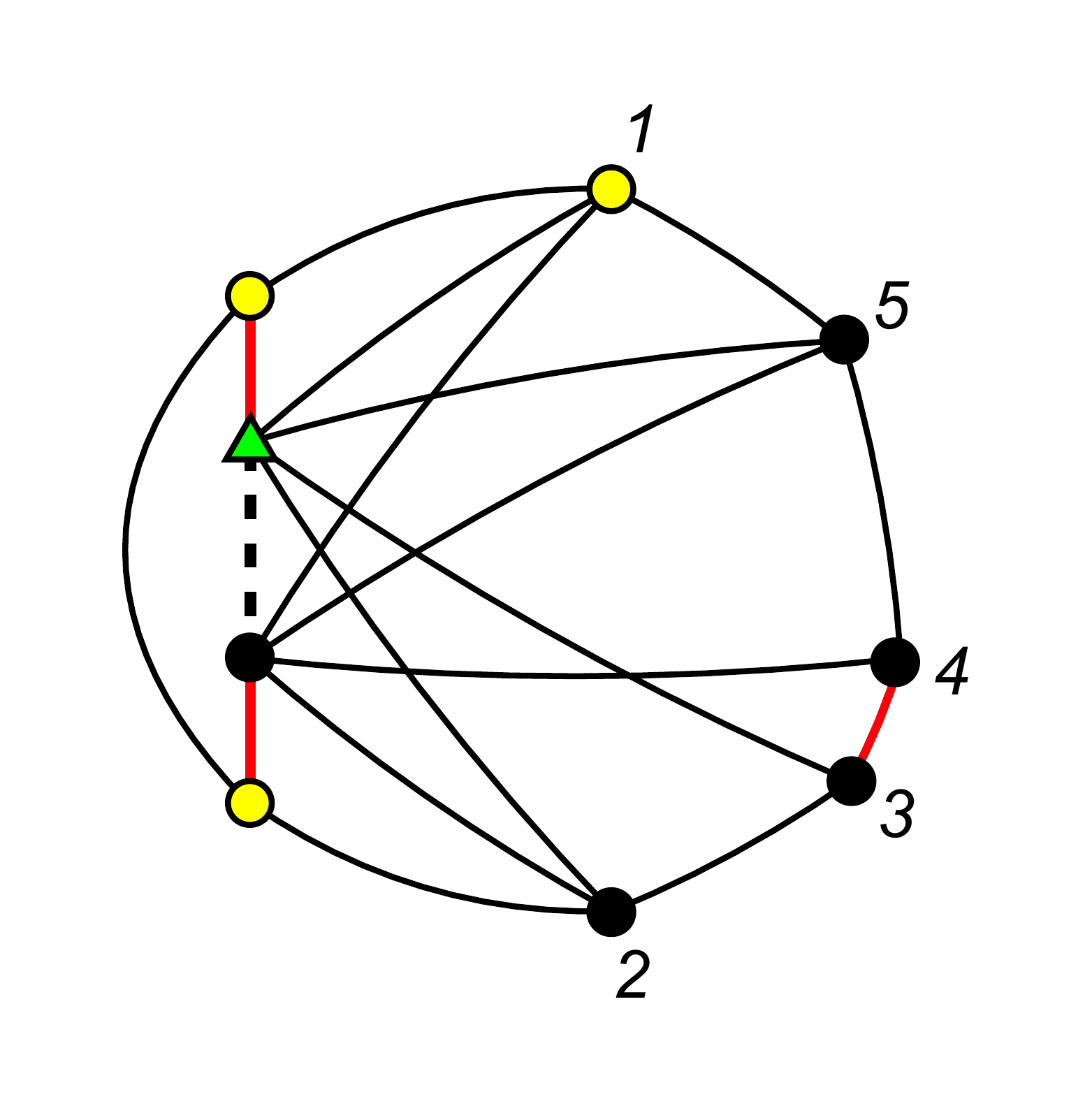}} +
\hspace{-0.3cm}
  \parbox[c]{8.1em}{\includegraphics[scale=0.22]{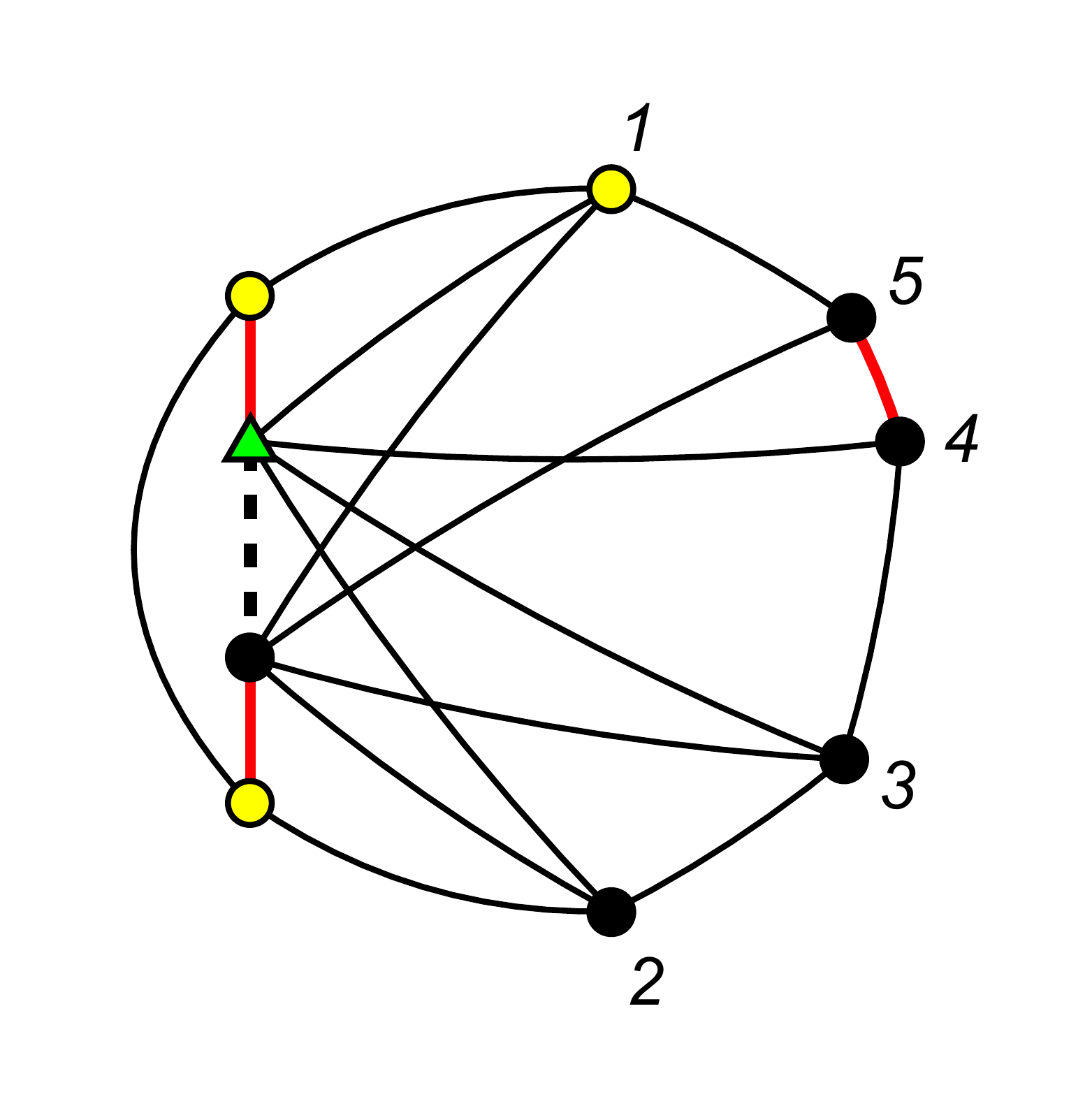}} + {\rm cyc}\,\,
 \right\}\,,
 \non 
\end{eqnarray}
\vskip-0.1cm\noindent
here ``${\rm cyc}$" means cyclic permutations of the labels, $(1,2,3,4,5)$, but by keeping the connection among them. Note that the graphs in \eqref{5pts-1ex} are totally similar to the ones obtained in \cite{Gomez:2017cpe}, equation $(5.3)$. In fact, the only new graph is the second one, which is a generalization  of the graph in {\it proposition 3} of  \cite{Gomez:2017cpe}.  Following the same procedure that was applied there, we multiply this graph  by the cross-ratio identity, $\mathbb{1} = -\s_{53}\, \o^{a_1:a_2}_{3:5} + \frac{\s_{5a_1} \s_{3a_2}}{\s_{3a_1} \s_{5a_2}}$, therefore  the second term in \eqref{pt1-5-np} becomes      
\begin{equation}
\frac{\s_{34} \, \s_{45} \,  \o^{a_1:a_2}_{5:3} }{(3,4,5)}     \,\,  \o^{a_1:a_2}_{4:4} \times \mathbb{1} = -
 \o^{a_1:a_2}_{3:3} \,  \o^{a_1:a_2}_{4:4}  \, \o^{a_1:a_2}_{5:5} 
 + \frac{\s_{34} \, \s_{45} \,  \o^{a_1:a_2}_{3:5} }{(3,4,5)}   \,\,  \o^{a_1:a_2}_{4:4},  
\end{equation}
and the graph turns into
\vskip-1.0cm
\begin{eqnarray}\label{5pts-secG}
    \parbox[c]{8.5em}{\includegraphics[scale=0.22]{5pts-NP-P-4.pdf}} 
    =  \,-
\hspace{-0.3cm}
 \parbox[c]{8.5em}{\includegraphics[scale=0.22]{5pts-NP-P-1.pdf}} +
\hspace{-0.3cm}
  \parbox[c]{8.1em}{\includegraphics[scale=0.22]{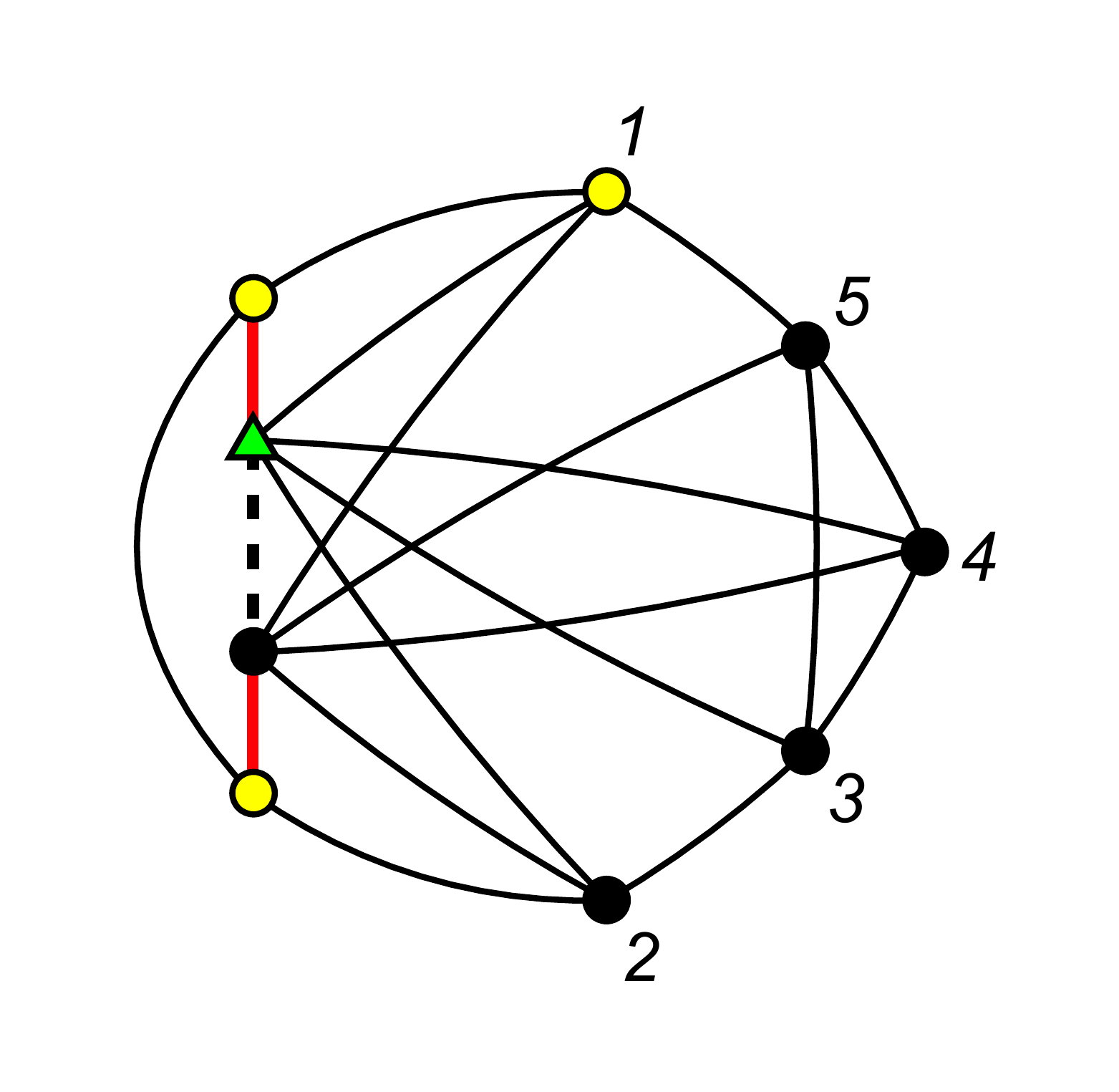}} 
\,\, \,\, .
\end{eqnarray}
\vskip-0.1cm\noindent
The second resulting graph is a generalization of the one given  in {\it proposition 2} of  \cite{Gomez:2017cpe}. This graph is simple to compute using the $\L-$algorithm and its result is zero.  So,  the  $\mathfrak{M}_5^{\rm (1-NP;P)}$ amplitude can now be written as 
\vskip-0.5cm
\begin{eqnarray}\label{5pts-1ex-2}
&&\mathfrak{M}_5^{\rm (1-NP;P)}[1,2 | 3,4,5 \, ;  \, 1,2,3,4,5 ]= 
\frac{1}{2^{6}} \int d\Omega \times s_{a_1 b_1} \times
\non
&&
 \int d\mu^{\rm t}_{5+4}  
\left\{
\hspace{-0.1cm}
 \parbox[c]{8.5em}{\includegraphics[scale=0.22]{5pts-NP-P-1.pdf}}
    +  
\hspace{-0.3cm}
 \parbox[c]{8.1em}{\includegraphics[scale=0.22]{5pts-NP-P-2.pdf}} +
\hspace{-0.3cm}
  \parbox[c]{8.1em}{\includegraphics[scale=0.22]{5pts-NP-P-3.pdf}} + {\rm cyc}\,\,
 \right\} \,,\non
\end{eqnarray}
\vskip-0.1cm\noindent
where all these graphs were already mapped to Feynman diagrams in  \cite{Gomez:2017cpe}.  Thus, the final answer is
\vskip-1.6cm
\begin{eqnarray}\label{5pts-1ex-fey}
\hspace{-0.5cm}
\mathfrak{M}_5^{\rm (1-NP;P)}[1,2 | 3,4,5 \, ;  \, 1,2,3,4,5 ]= 
\hspace{-0.3cm}
 \parbox[c]{8.8em}{\includegraphics[scale=0.24]{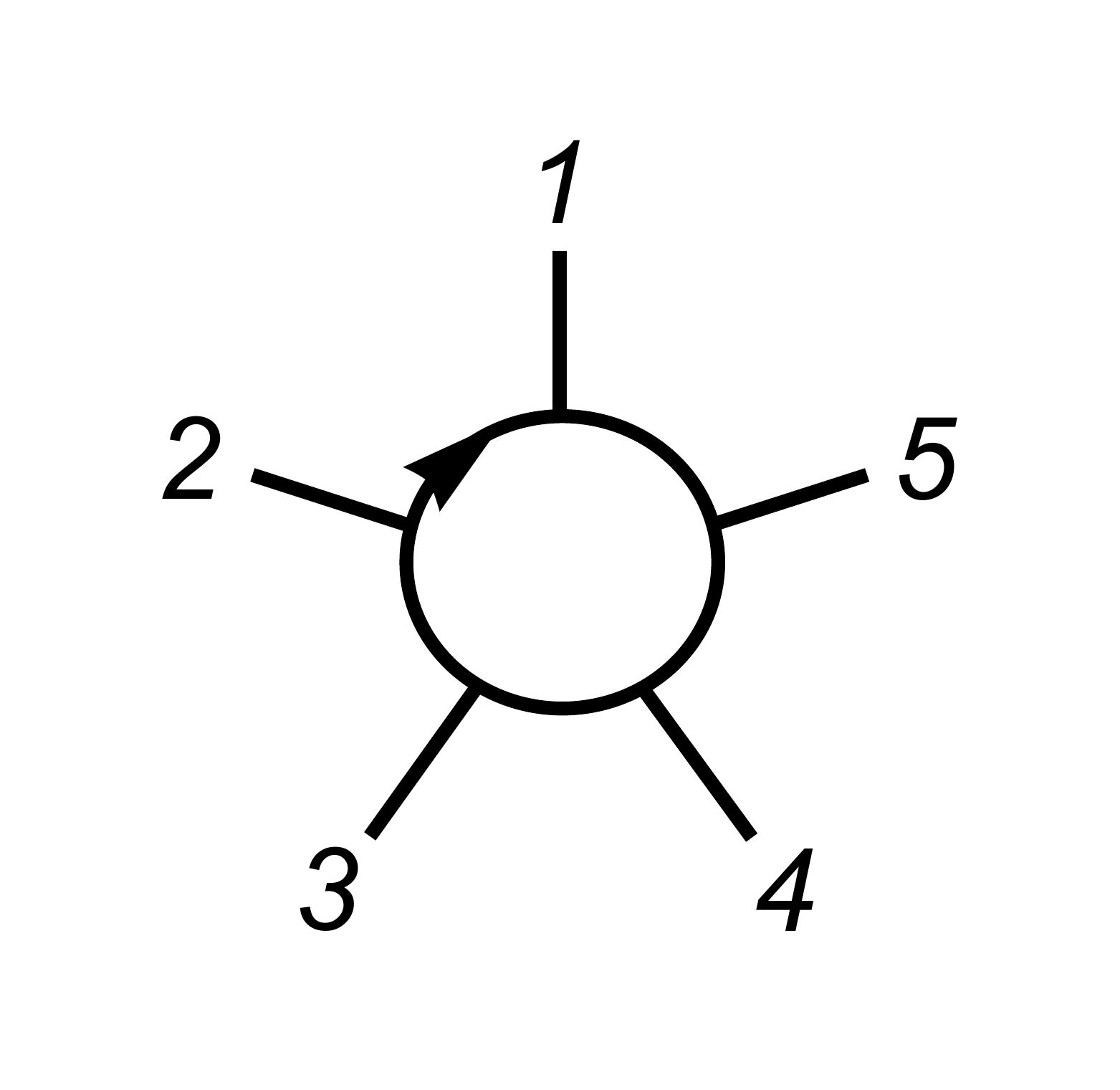}}
    +  
\hspace{-0.3cm}
 \parbox[c]{8.8em}{\includegraphics[scale=0.24]{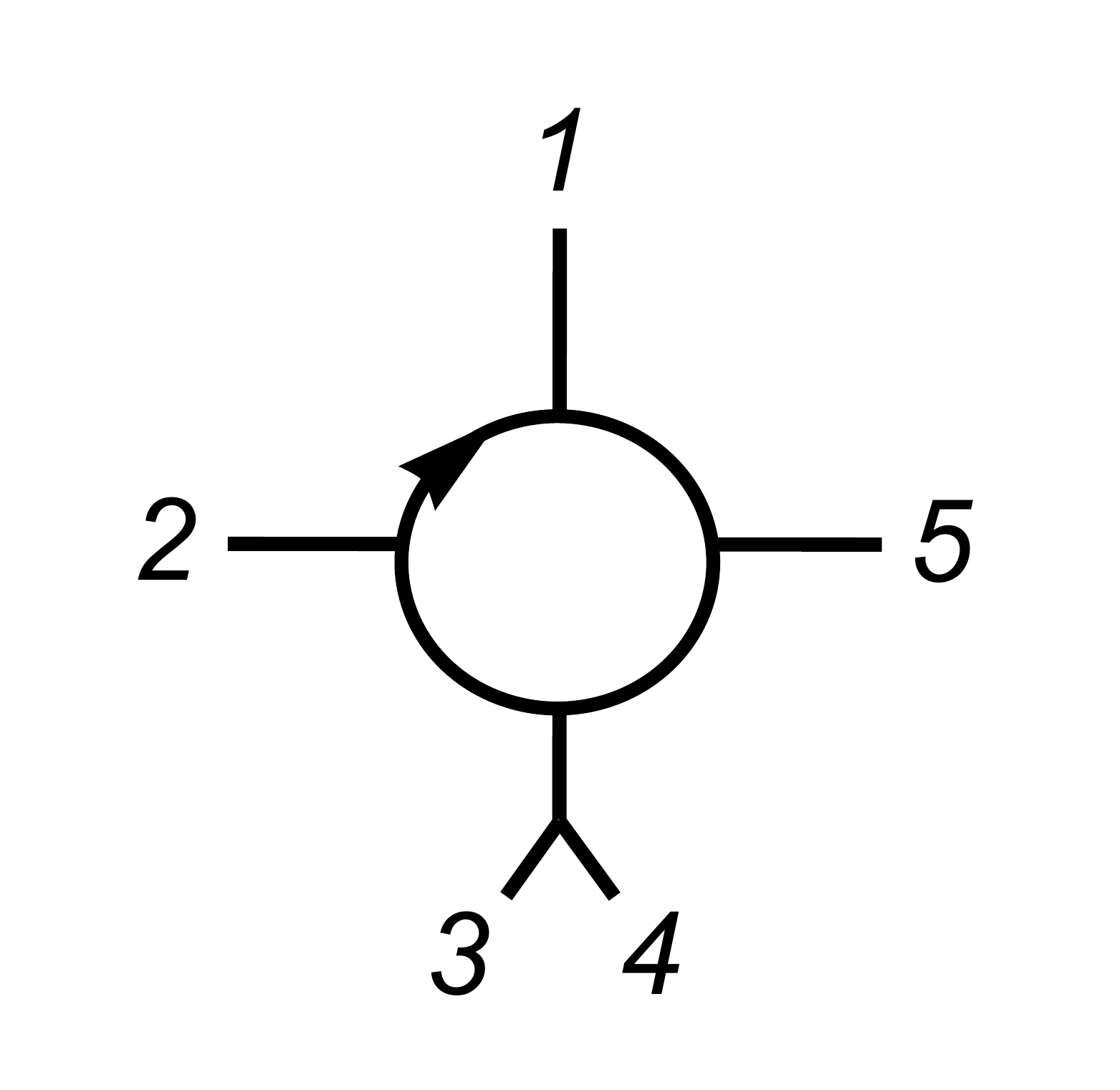}} +
\hspace{-0.3cm}
  \parbox[c]{9.4em}{\includegraphics[scale=0.24]{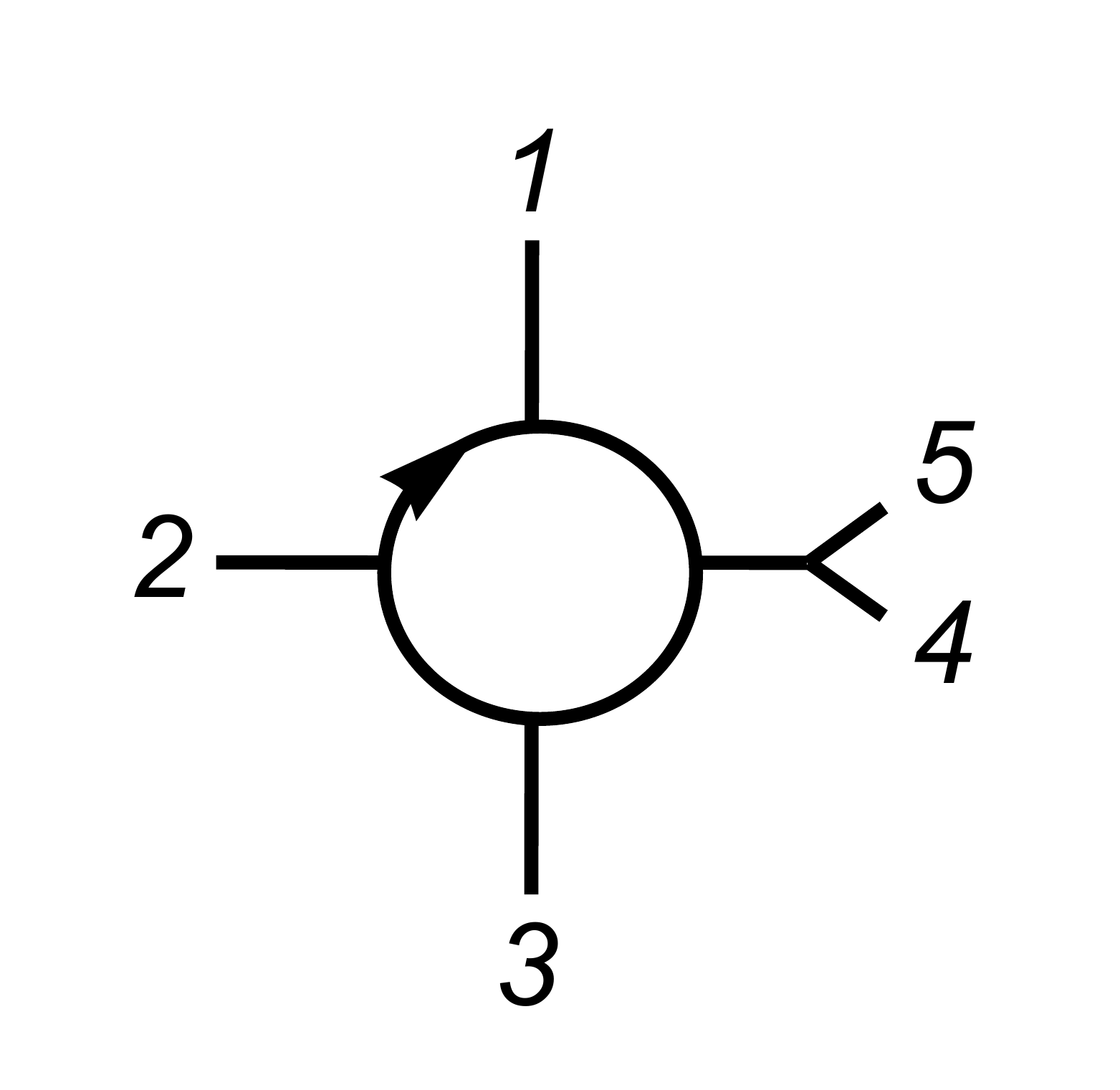}} 
. \qquad\,\,\non
\end{eqnarray}
\vskip-0.2cm\noindent

Next, we compute the ${\rm NP;NP}$ contribution to the five-point case. Let us consider the amplitude 
\begin{align}\label{5p-NP-NP}
 \mathfrak{M}_5^{\rm (1-NP;NP)}[1,2| 3,4,5 \, ; \, 1,2 |3,4,5 ] &= {1 \over 2^{5+1}}\int d\Omega \times s_{a_1b_1} \non  
&\times 
\int d\mu_{5+4}^{\rm t} \,\,\,
{\rm PT}^{(1)}_{\, a_1:a_2}[ 1,2|3,4 ,5  ]\times {\rm PT}^{(1)}_{\, b_1:b_2}[ 1,2| 3,4 ,5]\,.  \non
\end{align}
Such as it was done previously, we use the expansion given in \eqref{pt1-5-np} for ${\rm PT}^{(1)}_{\, a_1:a_2}[ 1,2|3,4 ,5  ]$ and the original definition  for ${\rm PT}^{(1)}_{\, b_1:b_2}[ 1,2|3,4 ,5]$ written in \eqref{nonpPT}, i.e.
\begin{align}
& {\rm PT}^{(1)}_{\, a_1:a_2}[ 1,2|3,4,5   ] =  \frac{1}{(a_1,b_1,b_2,a_2)}\times \left(\o^{a_1:a_2}_{1:1}  \o^{a_1:a_2}_{2:2}\right) \label{pt1a}\\
& 
\times\left(
2\,\o^{a_1:a_2}_{3:3}  \o^{a_1:a_2}_{4:4}  \o^{a_1:a_2}_{5:5} 
+\frac{\s_{45} \, \s_{53} \,  \o^{a_1:a_2}_{3:4} }{(3,4,5)}     \,\,  \o^{a_1:a_2}_{5:5}
+ \frac{\s_{34} \, \s_{45} \,  \o^{a_1:a_2}_{5:3} }{(3,4,5)}     \,\,  \o^{a_1:a_2}_{4:4}
+\frac{\s_{53} \, \s_{34} \,  \o^{a_1:a_2}_{4:5} }{(3,4,5)}     \,\,  \o^{a_1:a_2}_{3:3}
\right), \nonumber\\
& {\rm PT}^{(1)}_{\, b_1:b_2}[ 1,2|3,4 ,5]  =  \frac{1}{(a_1,b_1,b_2,a_2)}\times \left( \frac{1}{\s_{21}}\, \o^{b_1:b_2}_{1:2} + \frac{1}{\s_{12}}\, \o^{b_1:b_2}_{2:1} \right) \nonumber\\
 &
\hspace{3cm} 
 \times \left( 
 \frac{1}{\s_{45}\s_{53}}\, \o^{b_1:b_2}_{3:4}   +  \frac{1}{\s_{53}\s_{34}}\, \o^{b_1:b_2}_{4:5} + \frac{1}{\s_{34}\s_{45}}\, \o^{b_1:b_2}_{5:3}
 \right)\,\, . 
 \end{align}
Thus, one obtains the graph expansion
\vskip-0.5cm
\begin{eqnarray}\label{5pts-2ex}
&&
\hspace{-1.3cm}
\mathfrak{M}_5^{\rm (1-NP;NP)}[1,2 | 3,4,5 \, ;  \, 1,2|3,4,5 ]= 
\frac{1}{2^{6}} \int d\Omega \times  s_{a_1 b_1} \times
 \int d\mu^{\rm t}_{5+4}  \non
&&
\hspace{-1.3cm}
\left\{ 2\times
\hspace{-0.3cm}
 \parbox[c]{8.0em}{\includegraphics[scale=0.22]{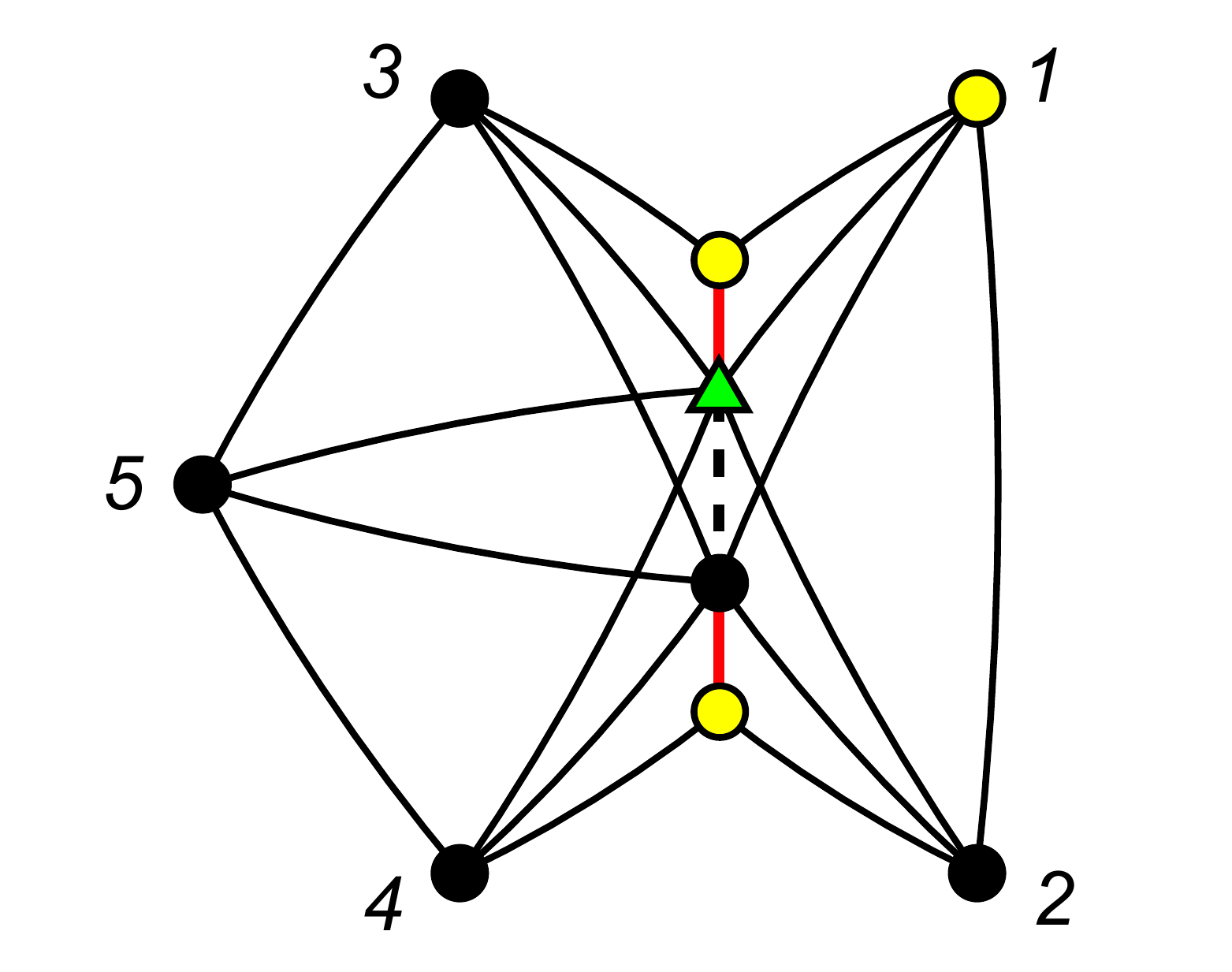}}
 + 
    \parbox[c]{7.6em}{\includegraphics[scale=0.22]{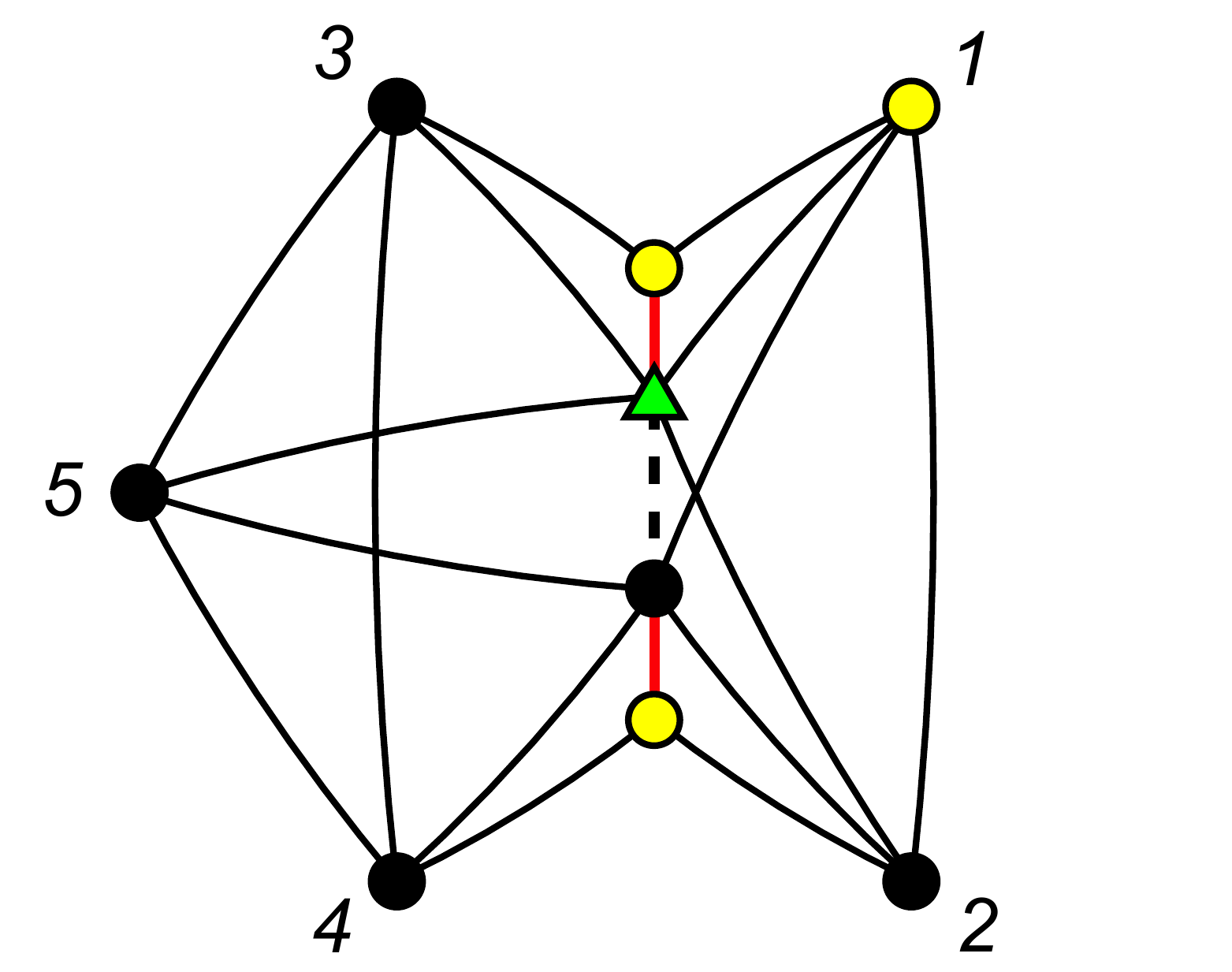}} 
    +  
\hspace{-0.3cm}
 \parbox[c]{7.5em}{\includegraphics[scale=0.21]{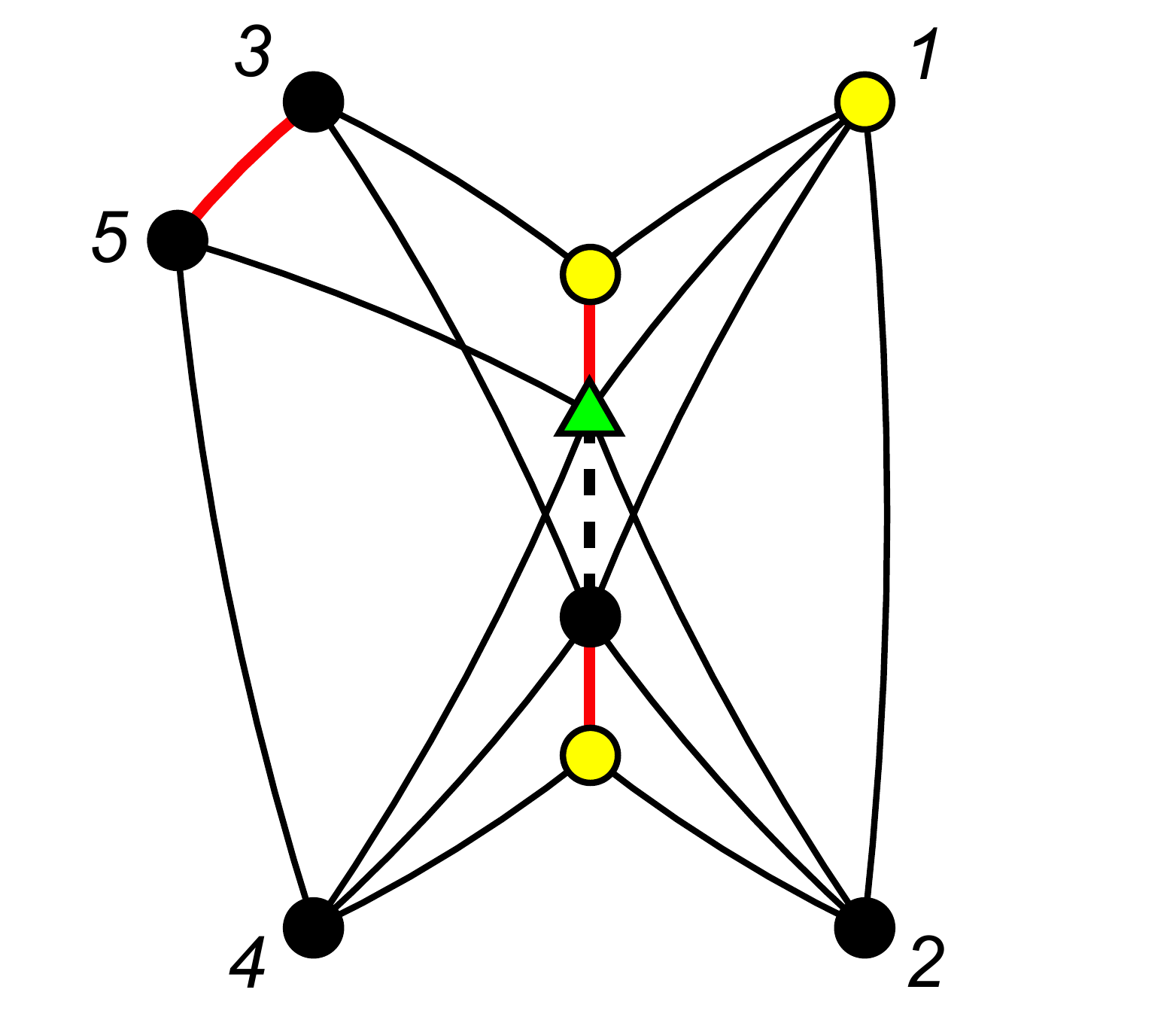}} +
\hspace{-0.3cm}
  \parbox[c]{7.5em}{\includegraphics[scale=0.21]{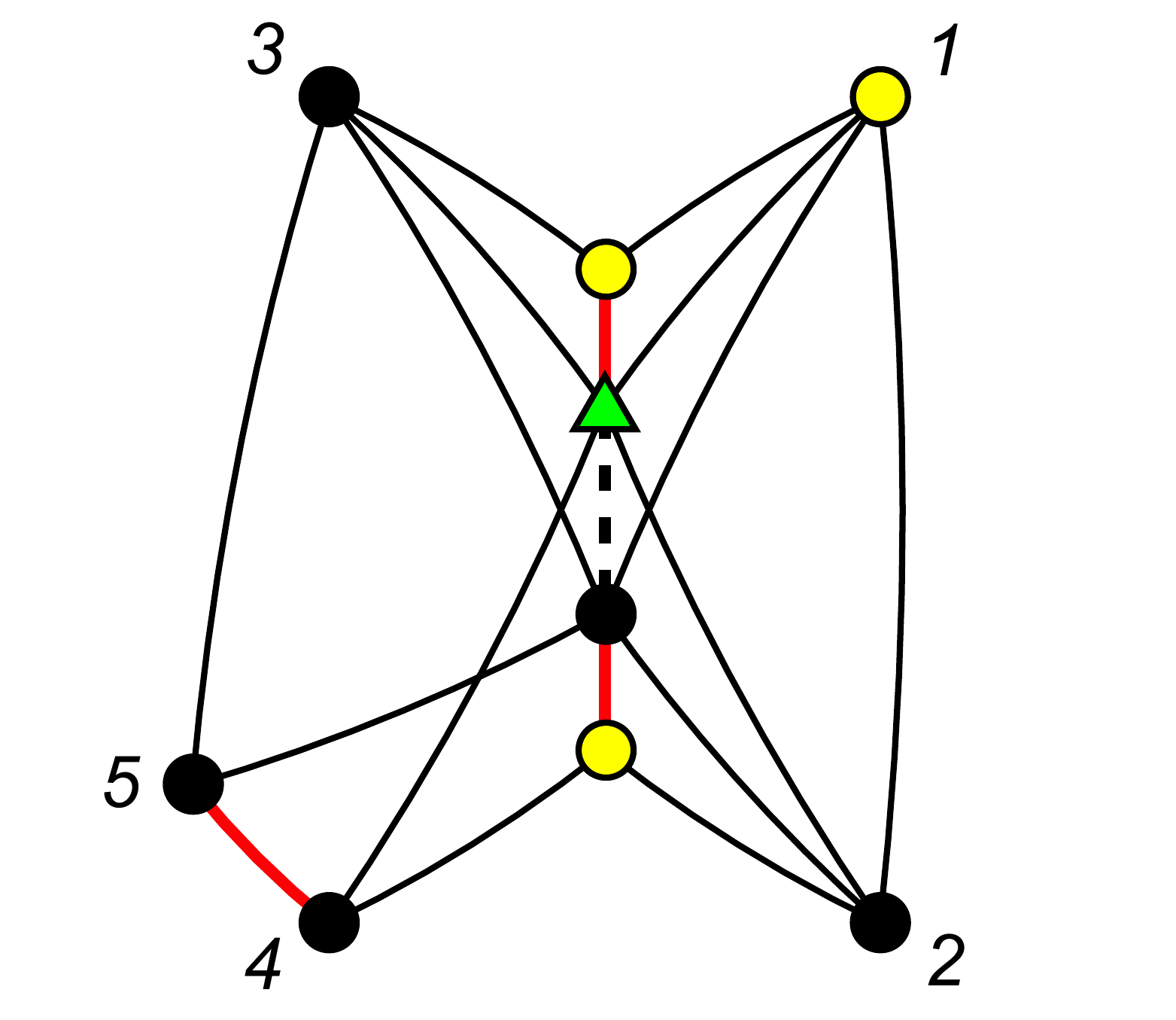}} \,\,
\begin{matrix}
 +&  \hspace{-2.2cm} (1\leftrightarrow 2) \\
+&  \hspace{-1.8cm}  {\rm cyc}(3,4,5)  \\
+&  (1\leftrightarrow  2) \times  {\rm cyc}(3,4,5)
\end{matrix}
\right\}\,.
 \non
\end{eqnarray}
\vskip-0.05cm\noindent
The first, third and fourth CHY-graphs  are a generalization of the butterfly graph that was found in the above section, equations  \eqref{4pts-2ex}. On the other hand, the second CHY-graph is a  combination between the butterfly graph and the graph in {\it proposition 3} of  \cite{Gomez:2017cpe}. 
So, by multiplying this graph by the cross-ratio identity, $\mathbb{1} = -\s_{34}\, \o^{a_1:a_2}_{4:3} + \frac{\s_{3a_1} \s_{4a_2}}{\s_{4a_1} \s_{3a_2}}$, 
 the second term in \eqref{pt1a} becomes      
\begin{equation}
\frac{\s_{45} \, \s_{53} \,  \o^{a_1:a_2}_{3:4} }{(3,4,5)}     \,\,  \o^{a_1:a_2}_{5:5} \times \mathbb{1} = -
 \o^{a_1:a_2}_{3:3} \,  \o^{a_1:a_2}_{4:4}  \, \o^{a_1:a_2}_{5:5} 
 + \frac{\s_{45} \, \s_{53} \,  \o^{a_1:a_2}_{4:3} }{(3,4,5)}     \,\,  \o^{a_1:a_2}_{5:5},  
\end{equation}
and the graph turns into
\vskip-0.7cm
\begin{eqnarray}\label{5ptsNP-secG}
    \parbox[c]{7.6em}{\includegraphics[scale=0.22]{5pts-NP-NP-4.pdf}} 
    =  \,-
\hspace{-0.2cm}
 \parbox[c]{8.1em}{\includegraphics[scale=0.22]{5pts-NP-NP-1.pdf}} +
\hspace{-0.1cm}
  \parbox[c]{8.1em}{\includegraphics[scale=0.22]{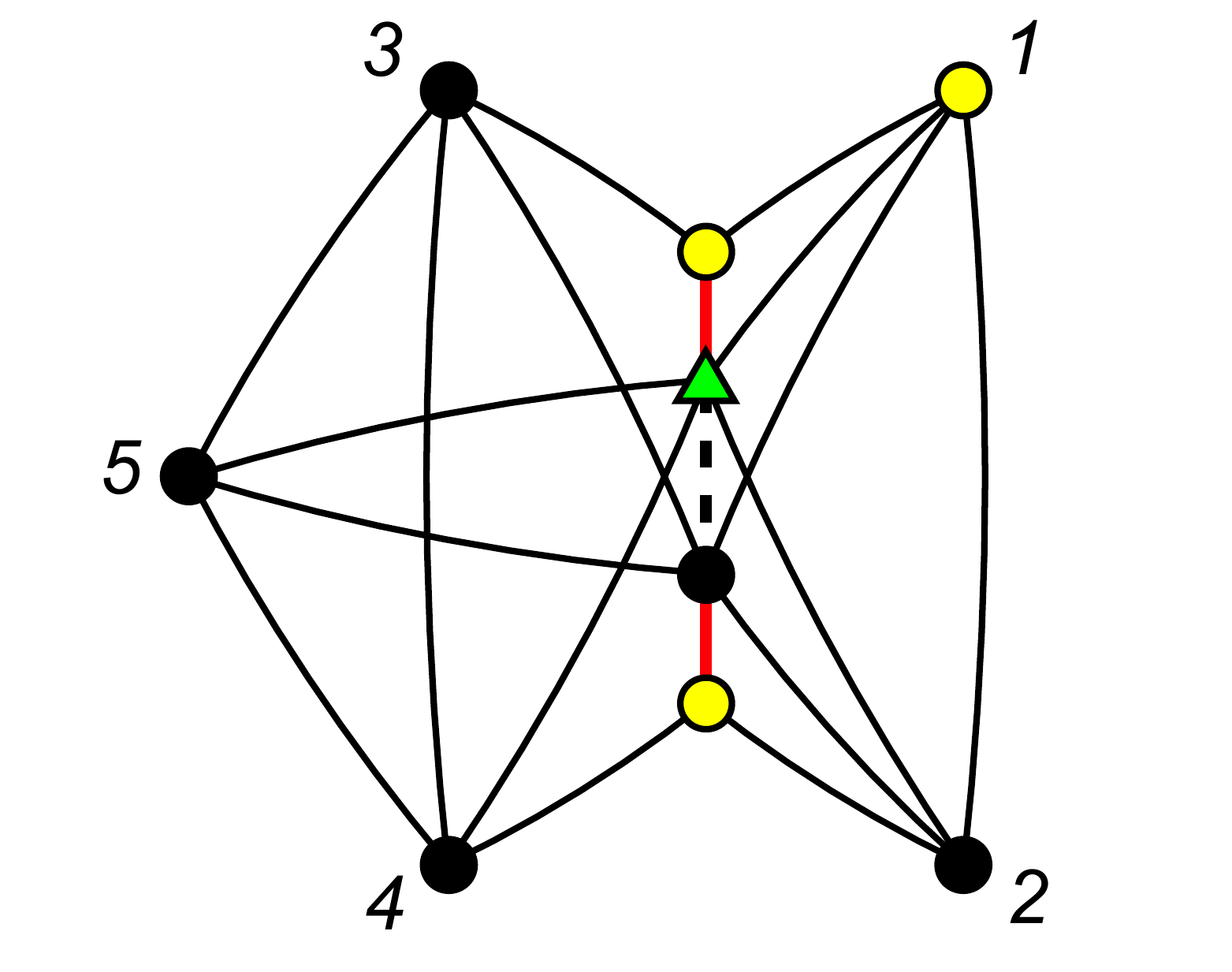}} 
\,\, \,\, .
\end{eqnarray}
\vskip-0.1cm\noindent
Like it happened in the previous example, the second resulting graph is a mixing between the butterfly graph and the one given in {\it proposition 2} of  \cite{Gomez:2017cpe}. The result for this graph is again zero.  Therefore, we can now write  the  $\mathfrak{M}_5^{\rm (1-NP;NP)}$ amplitude as 
\vskip-0.5cm
\begin{eqnarray}\label{5ptsNP-2ex}
&&
\mathfrak{M}_5^{\rm (1-NP;NP)}[1,2 | 3,4,5 \, ;  \, 1,2|3,4,5 ]= 
\frac{1}{2^{6}} \int d\Omega \times  s_{a_1 b_1} \times
 \int d\mu^{\rm t}_{5+4}  \non
&&
\left\{ 
\hspace{-0.3cm}
 \parbox[c]{8.0em}{\includegraphics[scale=0.22]{5pts-NP-NP-1.pdf}}
    +  
\hspace{-0.3cm}
 \parbox[c]{7.3em}{\includegraphics[scale=0.21]{5pts-NP-NP-2.pdf}} +
\hspace{-0.3cm}
  \parbox[c]{7.8em}{\includegraphics[scale=0.21]{5pts-NP-NP-3.pdf}}
\begin{matrix}
 +&  \hspace{-2.2cm} (1\leftrightarrow 2) \\
+&  \hspace{-1.8cm}  {\rm cyc}(3,4,5)  \\
+&  (1\leftrightarrow  2) \times  {\rm cyc}(3,4,5)
\end{matrix}
\right\}\,.
 \non
\end{eqnarray}
\vskip-0.05cm\noindent
Certainly, we have been able to rewrite  the $\mathfrak{M}_5^{\rm (1-NP;NP)}$ amplitude just in terms of butterfly graphs (non-planar CHY-graphs),  in the same way as it was made for $\mathfrak{M}_4^{\rm (1-NP;NP)}$. The computation of these graphs is completely similar to the one performed in \eqref{4ptB-r}, and we obtain the following results  
\vskip-0.9cm
\begin{eqnarray}\label{}
\hspace{-1.5cm}
 \frac{1}{2^{6}} \int d\Omega \, s_{a_1 b_1}\int d\mu^{\rm t}_{5+4}   
 \hspace{-0.2cm}
 \parbox[c]{7.8em}{\includegraphics[scale=0.23]{5pts-NP-NP-1.pdf}}
 &=& 
 \hspace{-0.5cm}
\parbox[c]{9.0em}{\includegraphics[scale=0.25]{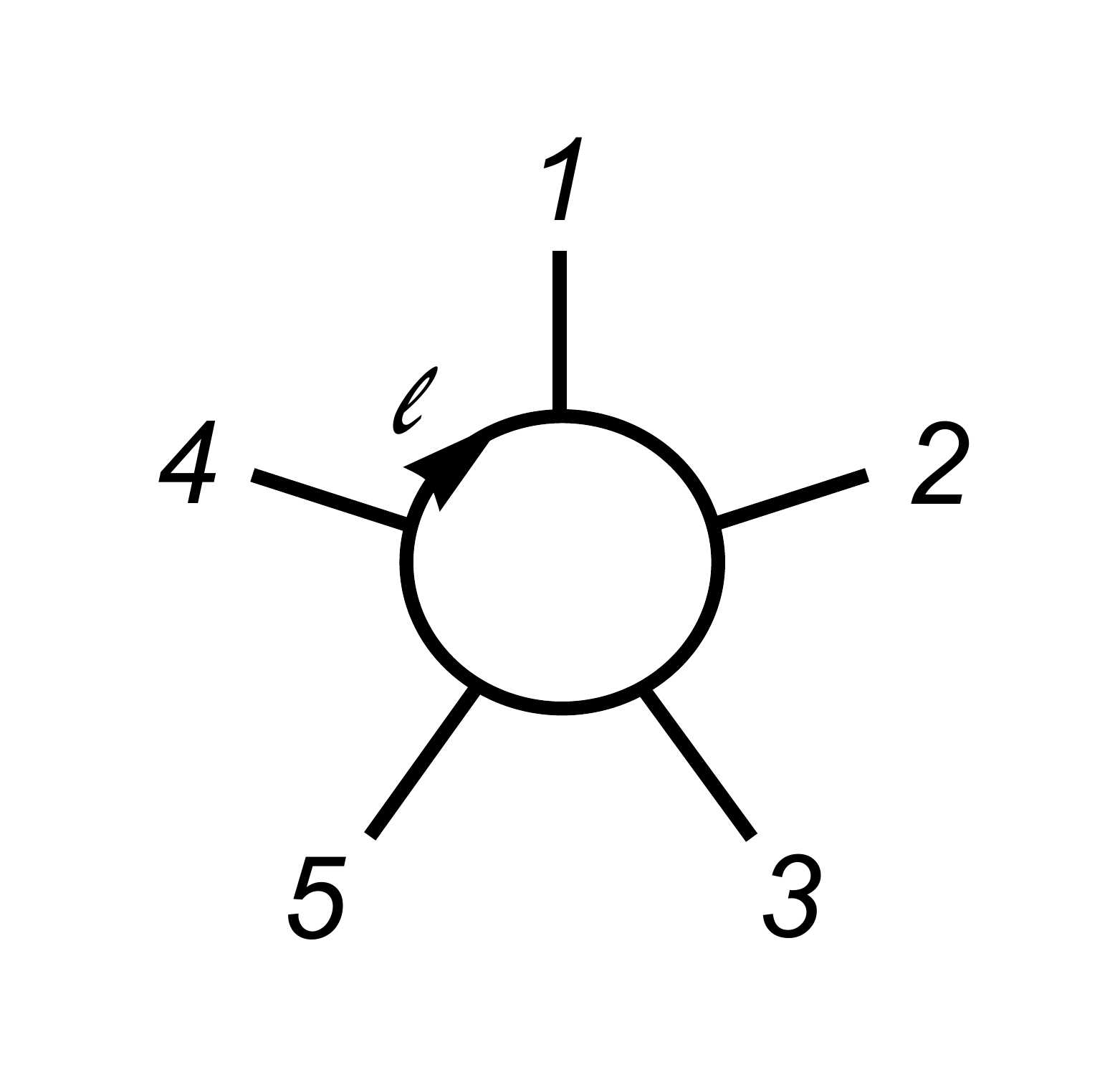}}
 +
  \hspace{-0.5cm}
\parbox[c]{9.0em}{\includegraphics[scale=0.25]{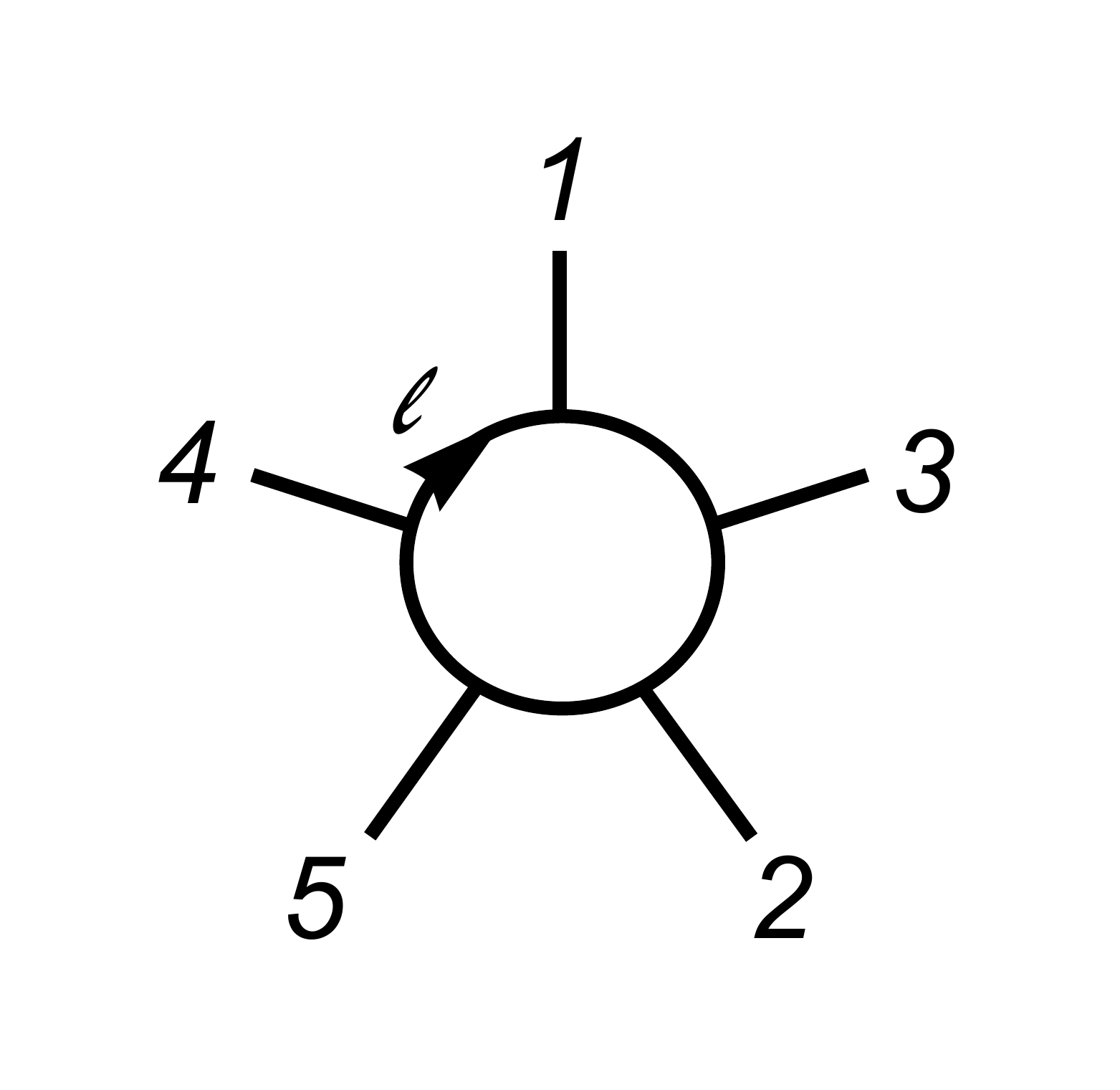}}
+
 \hspace{-0.5cm}
\parbox[c]{9.0em}{\includegraphics[scale=0.25]{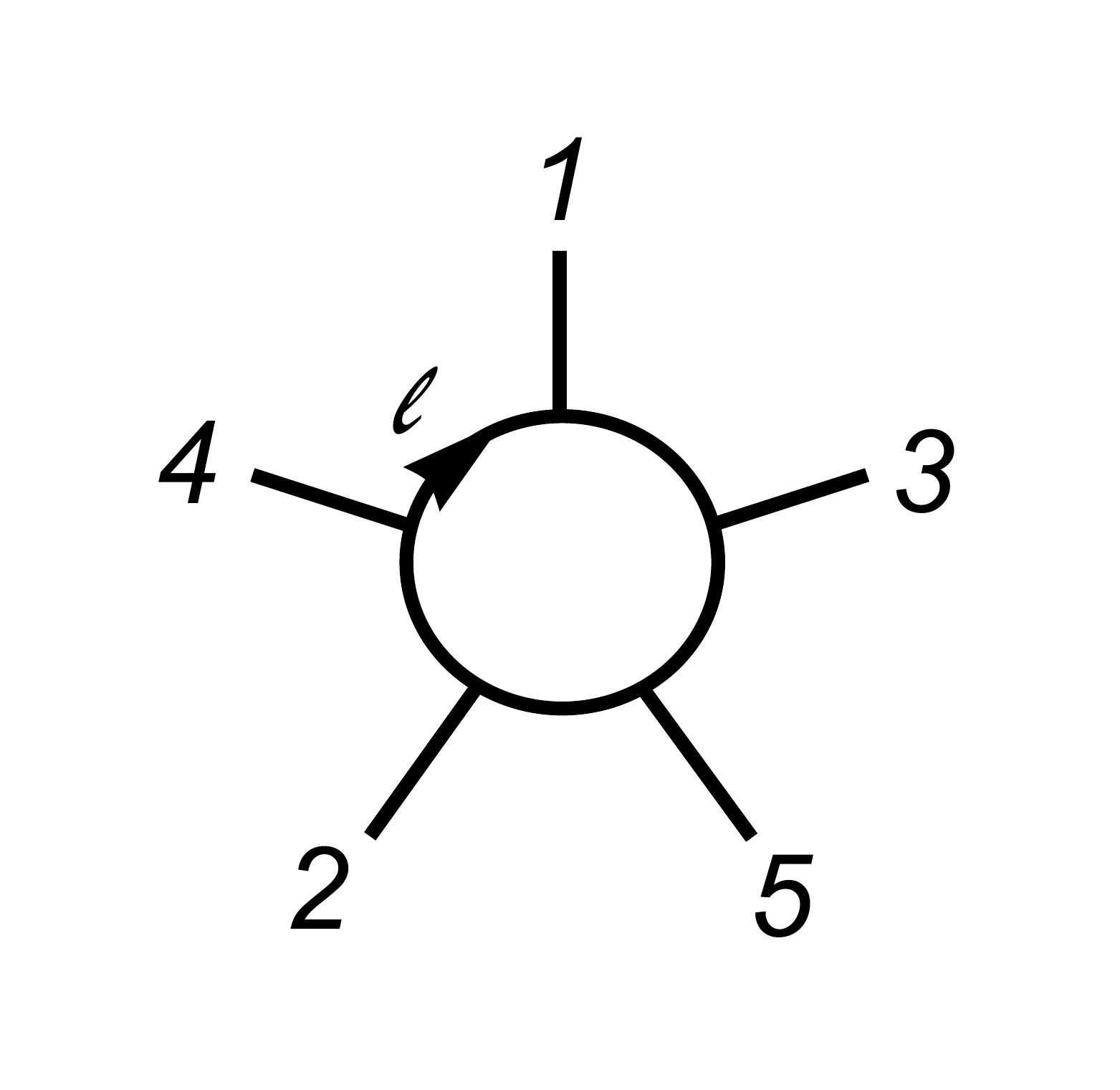}}
+{\rm (7\, terms)} 
\nonumber
\end{eqnarray}
\vskip-0.8cm
\begin{eqnarray}\label{5ptB-r1}
\hspace{0.2cm}
=  \overrightarrow{ \, \, \left[ 1,2\right] \sh \left[3,5,4\right] \, \,} \,   ,
\end{eqnarray}
\vskip-0.9cm
\begin{eqnarray}\label{}
\hspace{-1.5cm}
 \frac{1}{2^{6}} \int d\Omega \, s_{a_1 b_1}\int d\mu^{\rm t}_{5+4}   
 \hspace{-0.2cm}
 \parbox[c]{7.8em}{\includegraphics[scale=0.23]{5pts-NP-NP-2.pdf}}
 &=& 
 \hspace{-0.5cm}
\parbox[c]{9.0em}{\includegraphics[scale=0.25]{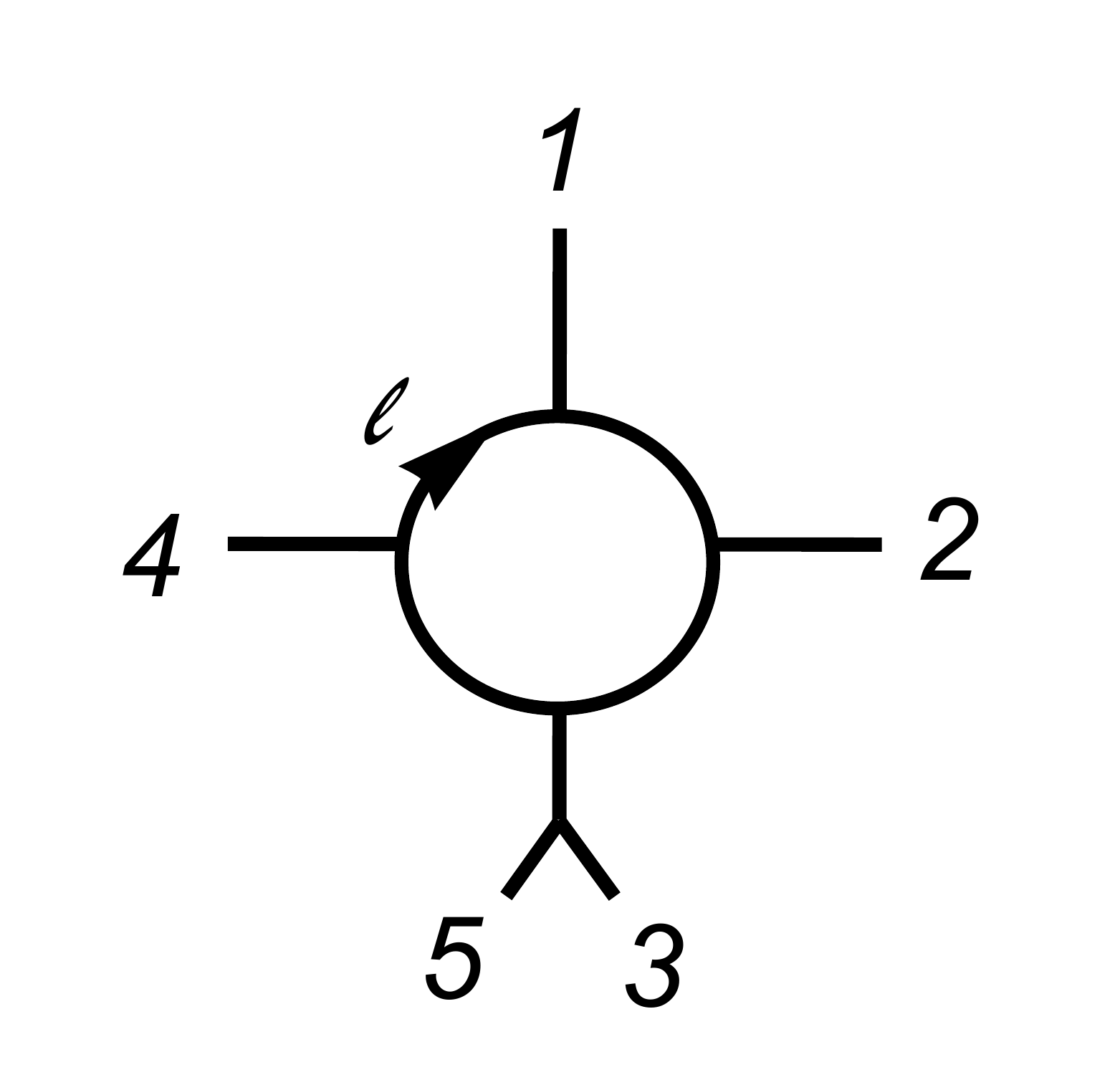}}
 +
  \hspace{-0.5cm}
\parbox[c]{9.0em}{\includegraphics[scale=0.25]{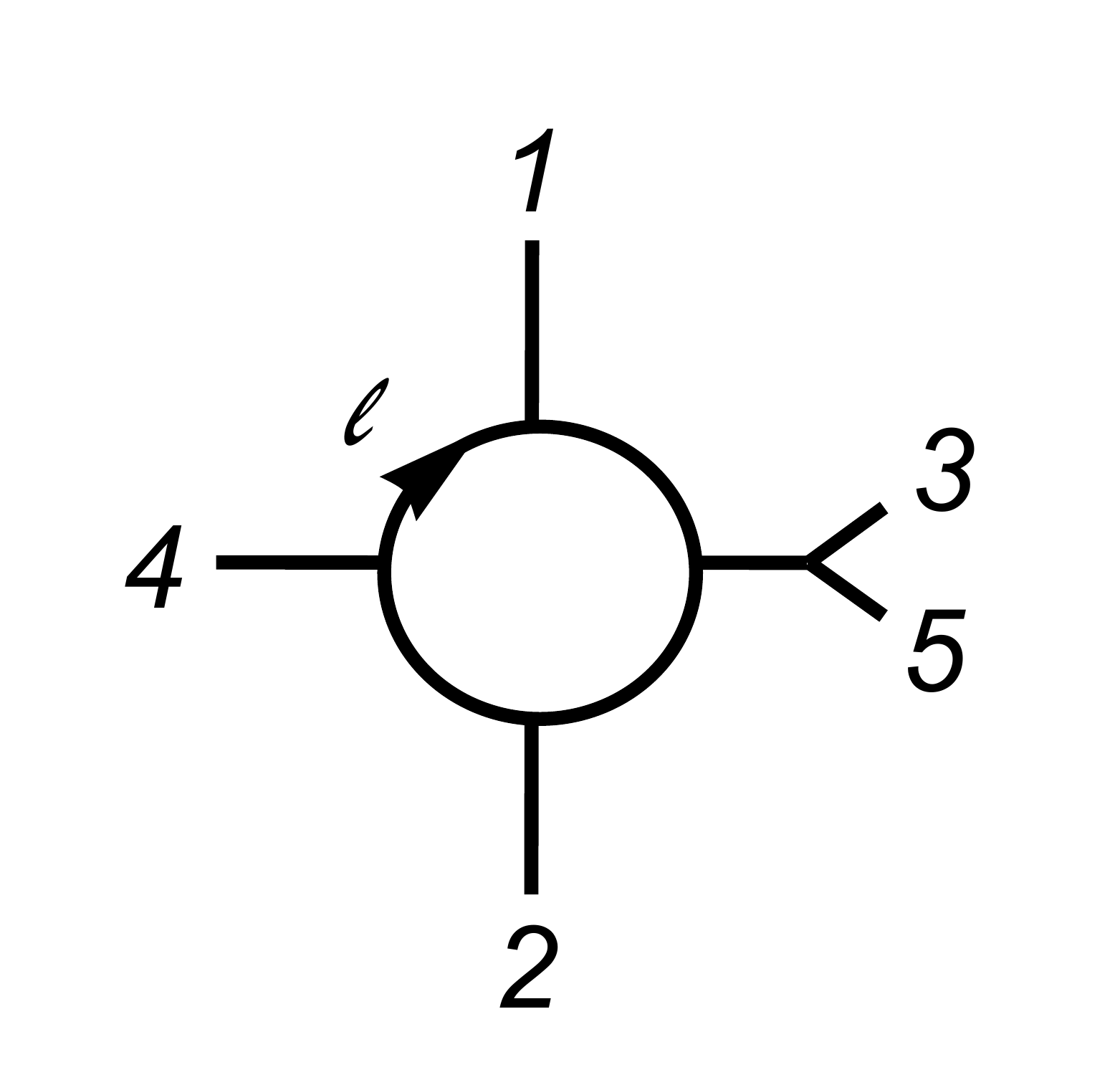}}
+
 \hspace{-0.5cm}
\parbox[c]{9.0em}{\includegraphics[scale=0.25]{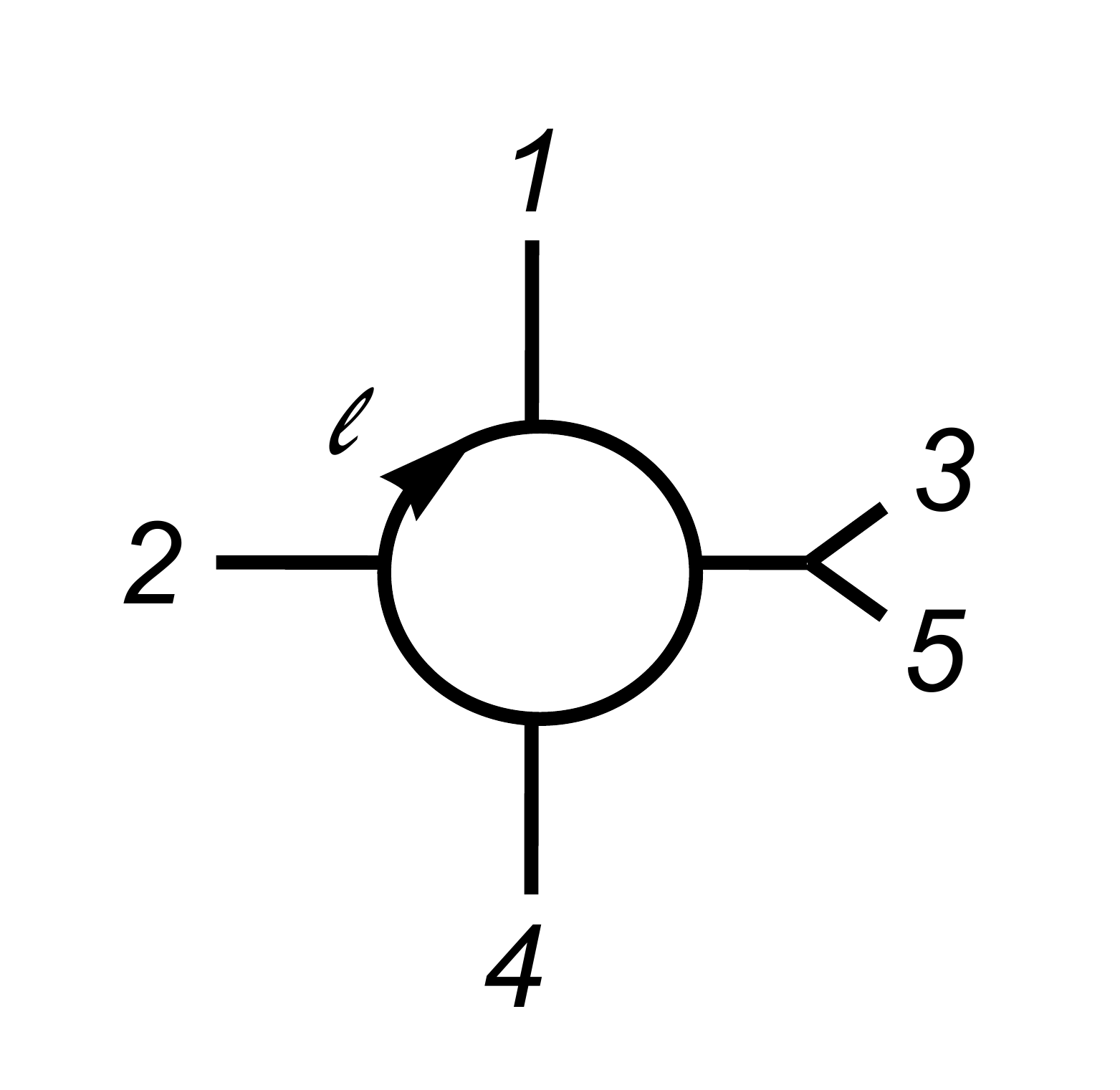}}
+ {\rm (3\, terms)}  
\nonumber
\end{eqnarray}
\vskip-0.6cm
\begin{eqnarray}\label{5ptB-r2}
\hspace{2.2cm}
= \overrightarrow{ \, \, \left[ 1,2,\{3,5\},4\right] \, \,} +\overrightarrow{ \, \, \left[ 1,\{3,5\},2,4\right] \, \,}  + \overrightarrow{ \, \, \left[ 1,\{3,5\},4,2\right] \, \,}  +\cdots = \overrightarrow{ \, \, \left[ 1,2\right] \sh \left[\{3,5\},4\right] \, \,} \, ,\qquad\quad
\end{eqnarray}
\vskip-0.9cm
\begin{eqnarray}\label{}
\hspace{-1.5cm}
 \frac{1}{2^{6}} \int d\Omega \, s_{a_1 b_1}\int d\mu^{\rm t}_{5+4}   
 \hspace{-0.2cm}
 \parbox[c]{7.8em}{\includegraphics[scale=0.23]{5pts-NP-NP-3.pdf}}
 &=& 
 \hspace{-0.5cm}
\parbox[c]{9.0em}{\includegraphics[scale=0.25]{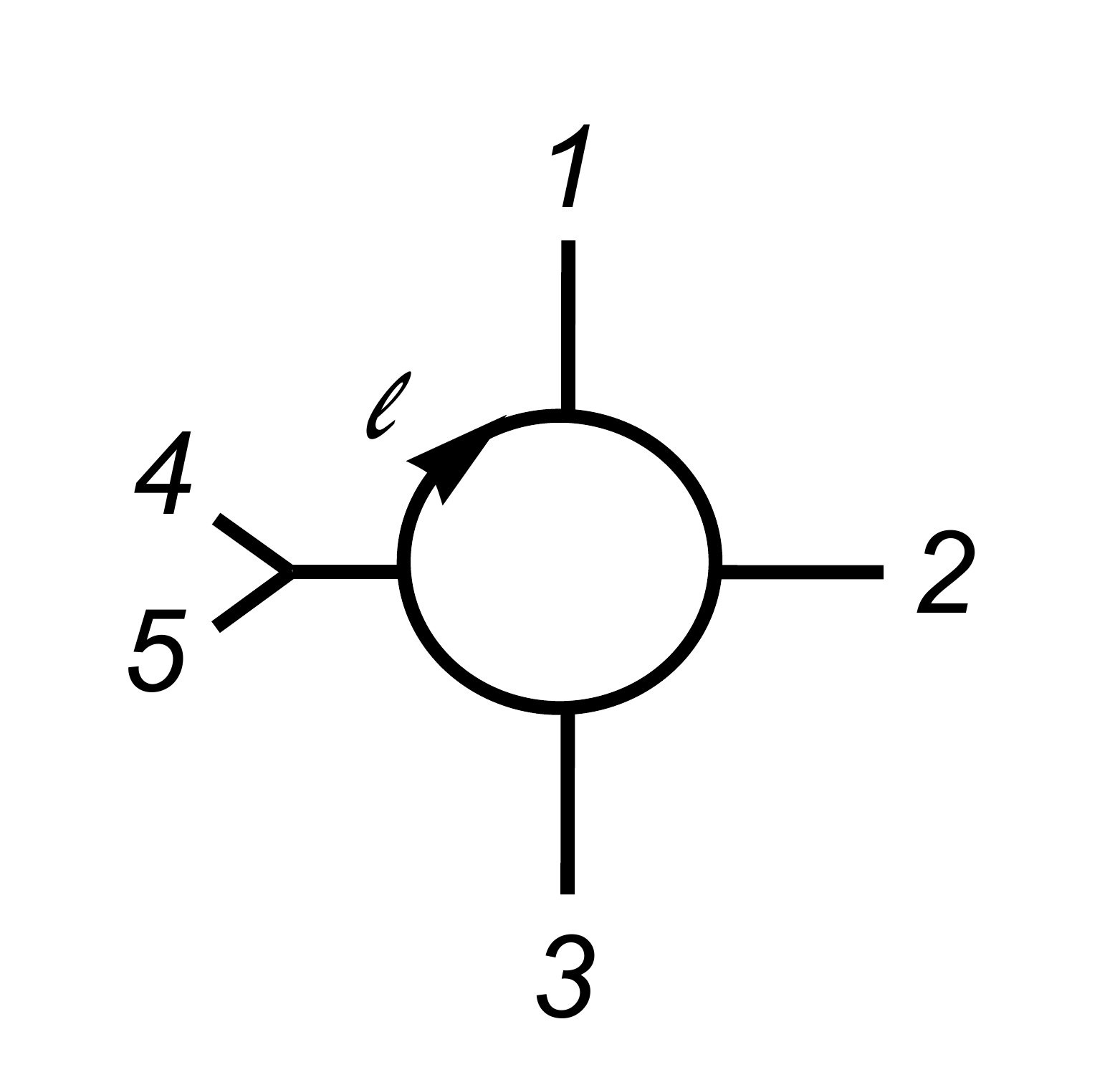}}
 +
  \hspace{-0.5cm}
\parbox[c]{9.0em}{\includegraphics[scale=0.25]{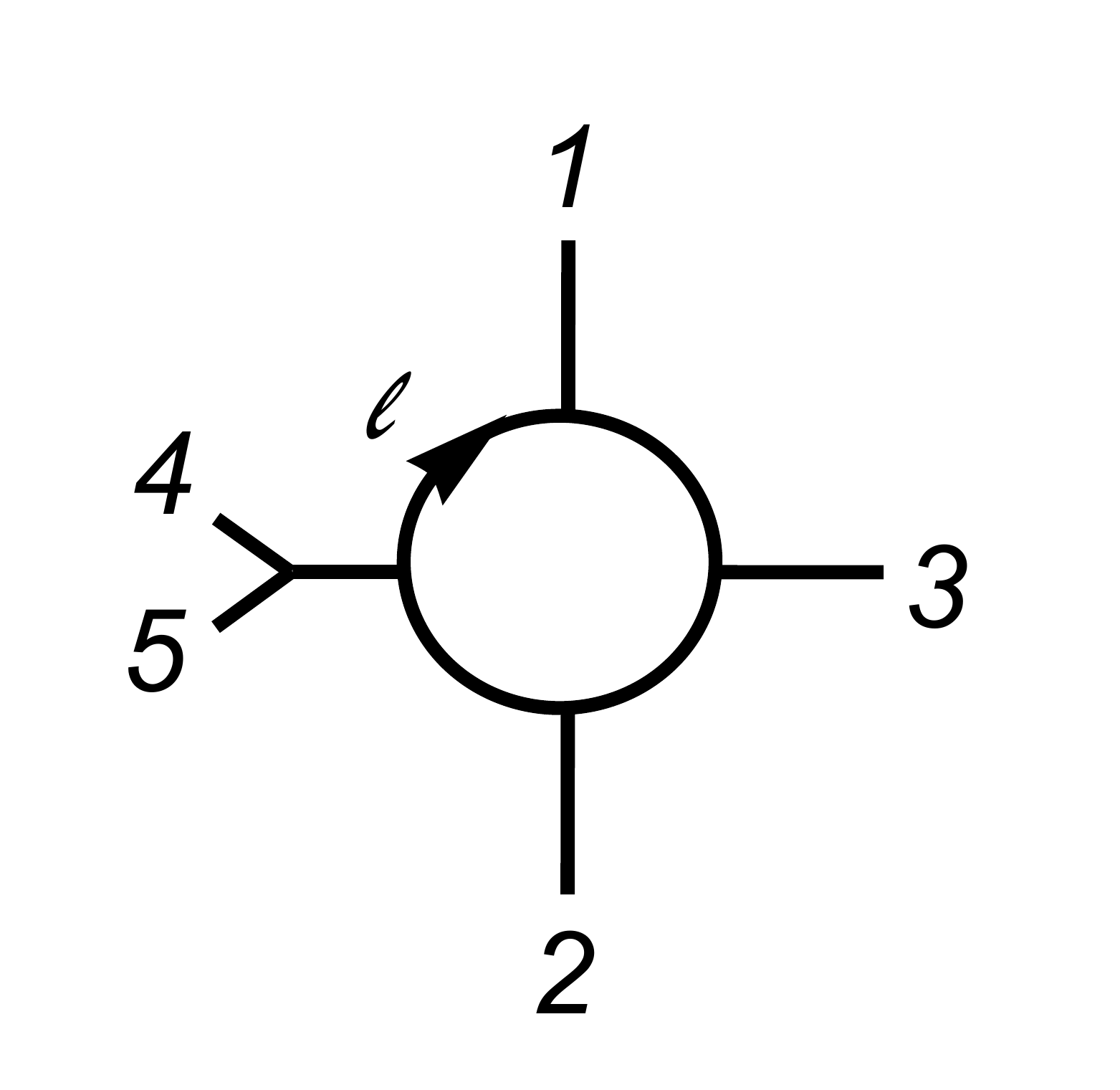}}
+
 \hspace{-0.5cm}
\parbox[c]{9.0em}{\includegraphics[scale=0.25]{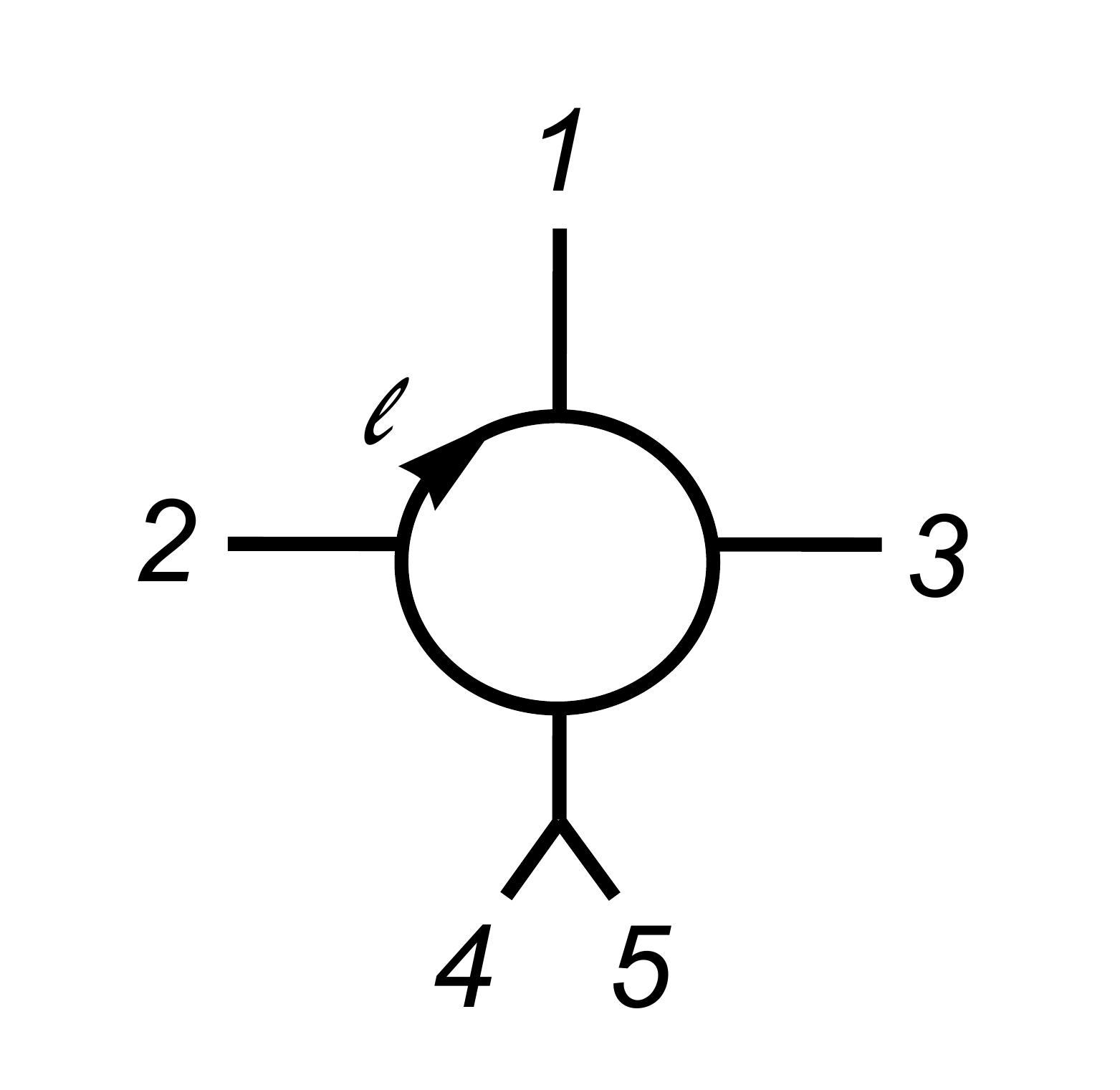}}
+{\rm (3\, terms)}
\nonumber
\end{eqnarray}
\vskip-0.6cm
\begin{eqnarray}\label{5ptB-r3}
\hspace{2.2cm}
= \overrightarrow{ \, \, \left[ 1,2,3,\{5,4\}\right] \, \,} +\overrightarrow{ \, \, \left[ 1,3,2\{5,4\}\right] \, \,}  + \overrightarrow{ \, \, \left[ 1,3,\{5,4\},2\right] \, \,}  +\cdots = \overrightarrow{ \, \, \left[ 1,2\right] \sh \left[3,\{5,4\}\right] \, \,} \, ,\qquad\quad
\end{eqnarray}
where we have denoted $\{3,5\}$, $\{5,4\}$ and $\{4,3 \}$ the tree level sector.  Therefore, the final answer is given by
\vskip-0.5cm
\begin{eqnarray}\label{5ptsNPNP-result}
\mathfrak{M}_5^{\rm (1-NP;NP)}[1,2 | 3,4,5 \, ;  \, 1,2|3,4,5 ] &=& 
\mathbf{S} [1,2|3,5,4]\, 
+ \,\mathbf{S} [1,2|5,4,3]\, 
+ \, \mathbf{S} [1,2|4,3,5] \, 
 \non
&&
\mathbf{S} [2,1|3,5,4]\, 
+ \,\mathbf{S} [2,1|5,4,3]\, 
+ \, \mathbf{S} [2,1|4,3,5] \,
\, , \qquad\non
\end{eqnarray}
\vskip-0.05cm\noindent
where we have defined $\mathbf{S} [a_1, a_2 | a_3,a_4,a_5 ] $ as
\vskip-0.5cm
\begin{eqnarray}\label{}
\mathbf{S} [a_1, a_2 | a_3,a_4,a_5 ] \equiv
\overrightarrow{ \, \, \left[ a_1,a_2\right] \sh \left[a_3,a_4,a_5\right] \, \,}  
+  \overrightarrow{ \, \, \left[ a_1,a_2\right] \sh \left[\{a_3,a_4\},a_5\right] \, \,} 
+  \overrightarrow{ \, \, \left[ a_1,a_2\right] \sh \left[a_3,\{a_4,a_5\}\right] \, \,} .
\qquad\quad 
\end{eqnarray}
\vskip-0.05cm\noindent

It is simple to check the total number of Feynman diagrams in \eqref{5ptsNPNP-result} is\footnote{Under the equivalence relation given by the loop momentum shifting in \eqref{fey-1loop-T}, there are 12 nonequivalent {\it pentagons} and 18 nonequivalent {\it boxes}. In addition, note that these 12 nonequivalent {\it pentagons} are in perfect agreement with the expansion given in \eqref{ex-expansion} for ${\rm PT}^{(1)}_{\, a_1:a_2}[ 1,2|3,4,5 ] $.
},  $ 60\,\,
{\it pentagons}\, + \,  72\,\, {\it boxes }\, = \, 132$, while in the CHY representation we only have 18  CHY-graphs (equation \eqref{5ptsNP-2ex}). 

Finally, looking at the results obtained in this section, it is interesting to note  that the amplitudes, $\mathfrak{M}_4^{\rm (1-NP;P)}[1,2 | 3,4 \, ;  \, 1,2,3,4 ] $ and $\mathfrak{M}_5^{\rm (1-NP;P)}[1,2 | 3,4,5 \, ;  \, 1,2,3,4,5 ] $, can be found from the intersection\footnote{This  equality is given up to an overall  sign.} 
\begin{align}
&\hspace{-0.5cm}
\mathfrak{M}_4^{\rm (1-NP;P)}[1,2 | 3,4 \, ;  \, 1,2,3,4 ] = \mathfrak{M}_4^{\rm (1-NP;NP)}[1,2 | 3,4 \, ;  \, 1,2 |3,4 ] \, \cap \,  \mathfrak{M}_4^{\rm (1-P;P)}[1,2 , 3,4 \, ;  \, 1,2,3,4 ] \nonumber , \\ 
&\hspace{-0.5cm}
\mathfrak{M}_5^{\rm (1-NP;P)}[1,2 | 3,4,5 \, ;  \, 1,2,3,4,5 ] = \mathfrak{M}_5^{\rm (1-NP;NP)}[1,2 | 3,4,5 \, ;  \, 1,2 |3,4,5 ] \, \cap \,  \mathfrak{M}_5^{\rm (1-P;P)}[1,2 , 3,4,5 \, ;  \, 1,2,3,4,5 ]  \nonumber ,
\end{align}
such as it was done in the planar case, \cite{He:2015yua,Gomez:2017cpe}.  Although we do not have  a formal proof, there are evidences that the previous intersection relation could be applied to higher number of points, therefore we conjecture the following general relation, up to an overall sign
\begin{align}
&\mathfrak{M}_n^{\rm (1-NP;NP)}[\a | \b \, ;  \, \g | \delta ] = \mathfrak{M}_n^{\rm (1-NP;NP)}[\a | \b \, ;  \, \a | \b ] \, \cap \,  \mathfrak{M}_n^{\rm (1-NP;NP)}[\g | \delta  \, ;  \,  \g | \delta ] \, \, . 
\end{align}
where the ordered lists, $\a=\{ \a_1,\ldots, \a_i \},  \, \b= \{\b_1,\ldots, \b_j \} , \, \g=\{ \g_1,\ldots, \g_k \}$ and $\delta=\{ \delta_1,\ldots, \delta_m \}$, satisfy the conditions, $\a\,\cap\, \b = \g \,\cap\, \delta =\emptyset$ and $\a\,\cup\, \b = \g \,\cup\, \delta =\{ 1,2, \ldots, n\}$.

In the following section we will generalize the previous results to an arbitrary number of points. Additionally, it is useful to remind that all computations were checked numerically.

\section{General non-planar CHY-graphs}\label{nonplanarG}

We have found a new type of CHY-graphs at one-loop, the {\it non-planar CHY-graphs} (or {\it butterfly graphs}). Those type of graphs could be identified as a generalization of the planar case obtained in \cite{Gomez:2017cpe}. In addition, let us recall that when the CHY-integrand associated to a planar CHY-graph is integrated, its result is a sum of Feynman diagrams at one-loop. This fact is summarized by the equality
\vskip-1.4cm
\begin{eqnarray}\label{planar-gen}
\frac{1}{2^{N+1}} \int d\Omega \times s_{a_1 b_1} \times
 \int d\mu^{\rm t}_{N+4}  
\hspace{-0.4cm}
 \parbox[c]{14em}{\includegraphics[scale=0.34]{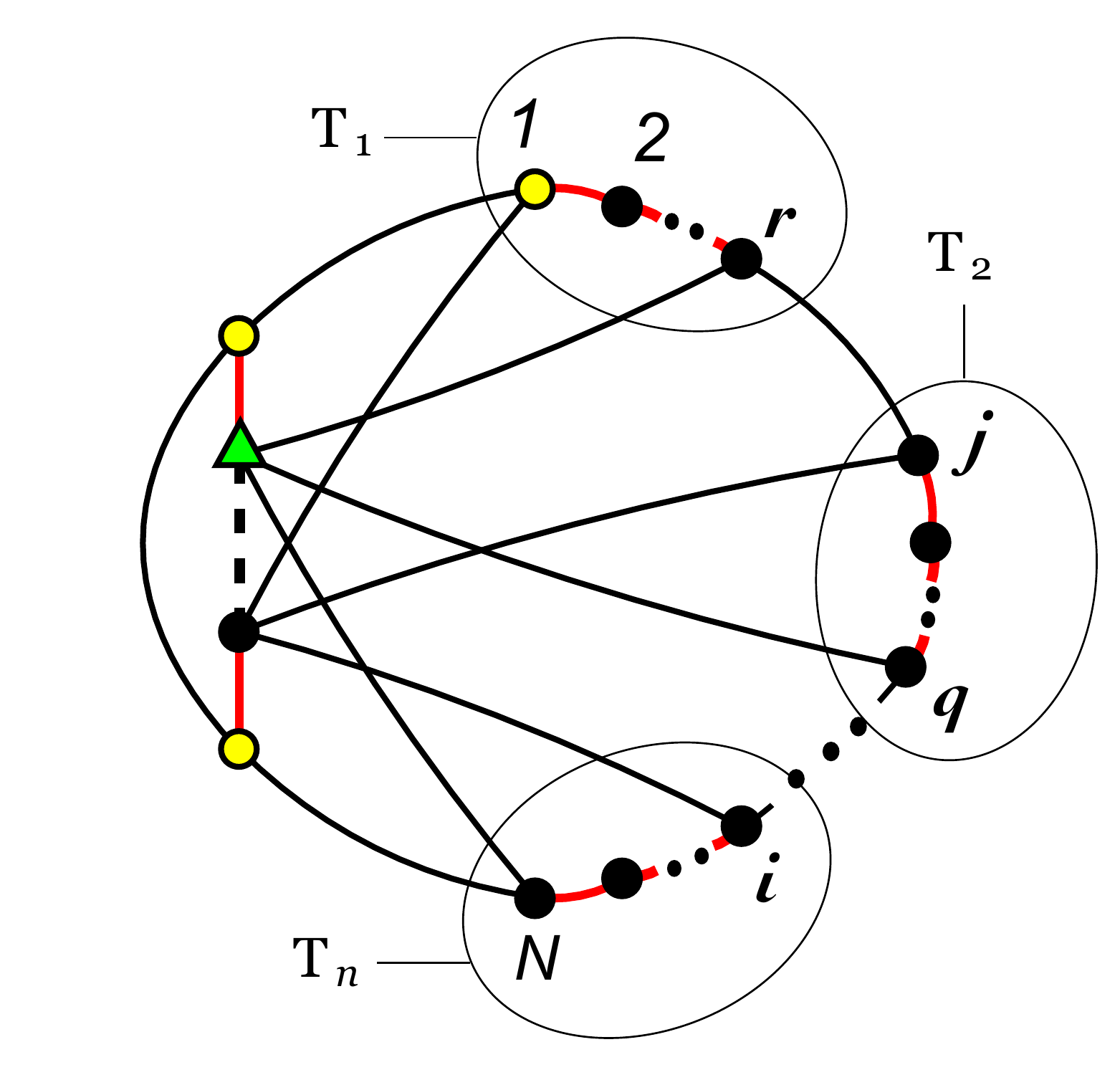}}
 =
 \hspace{-0.21cm}
  \parbox[c]{15.8em}{\includegraphics[scale=0.38]{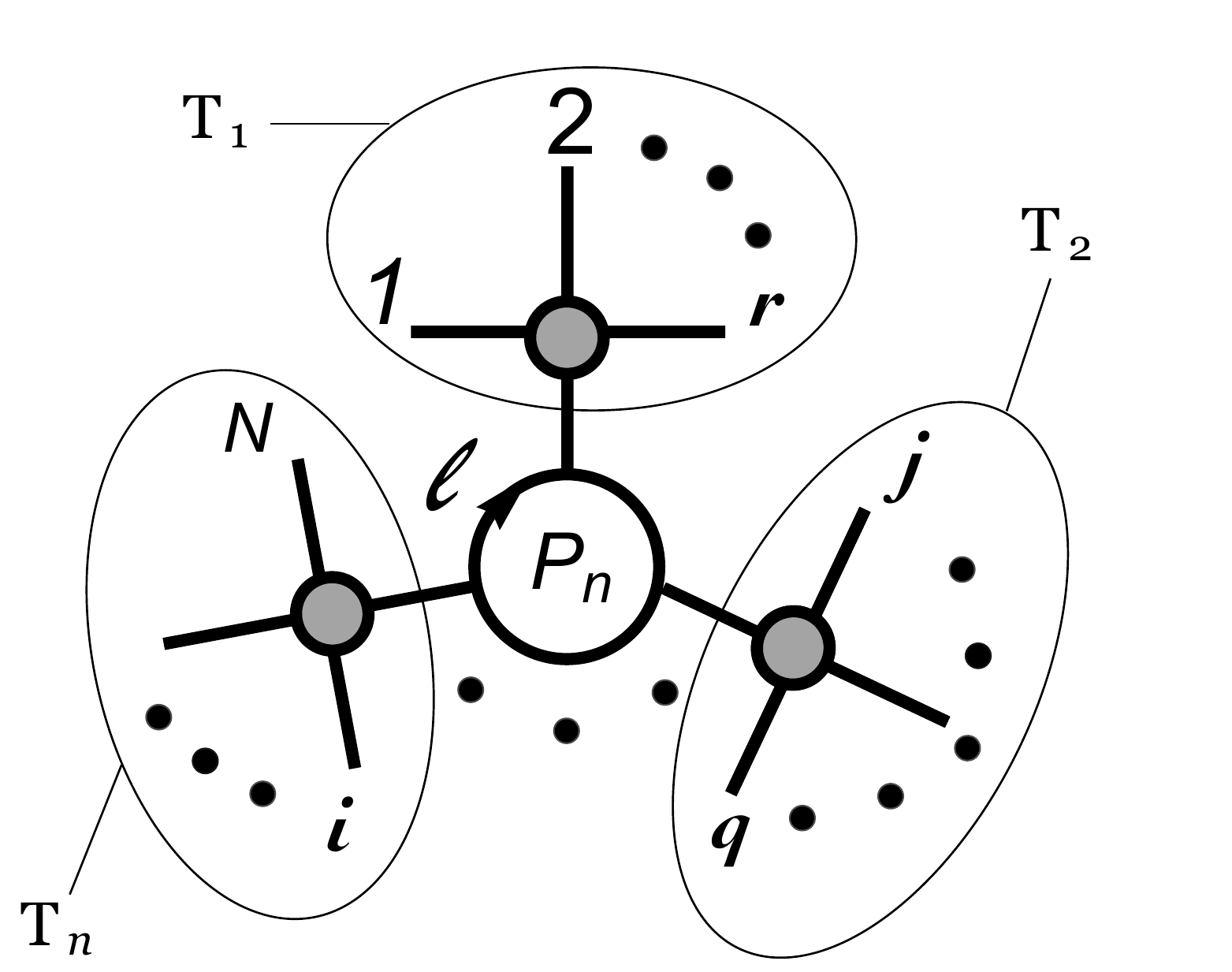}}\,,\non  
\end{eqnarray}
\vskip-0.35cm\noindent
where the grey circles  mean the sum over all possible trivalent planar diagrams\footnote{As it is very well known, at tree-level there is a map between the CHY-graphs and Feynman diagrams given by
\vskip-1.6cm
\begin{eqnarray}\label{}
 \int d\mu^{\rm t}_{n}  
\hspace{-0.4cm}
 \parbox[c]{9.0em}{\includegraphics[scale=0.2]{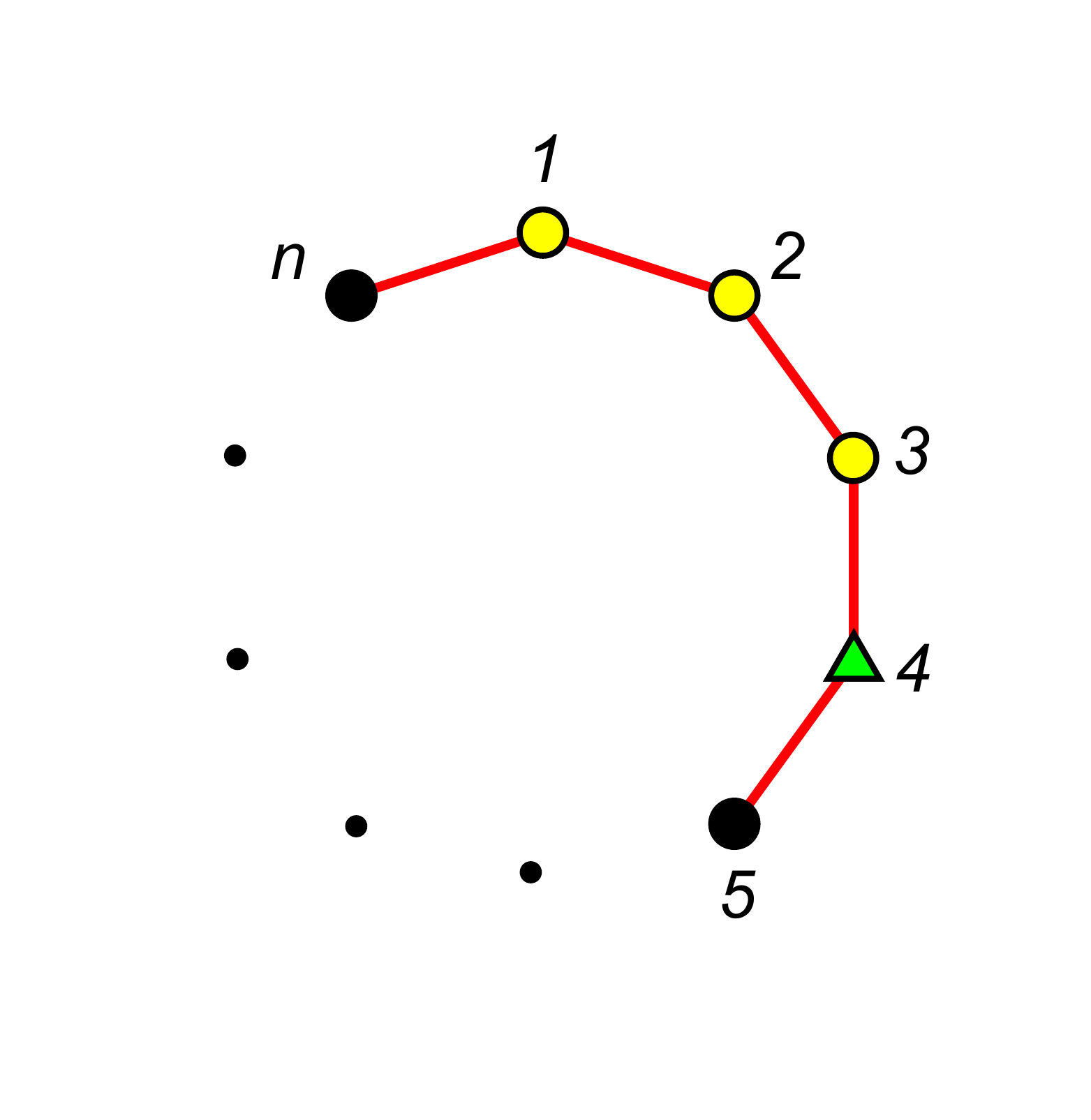}}
 =
 \hspace{-1.2cm}
  \parbox[c]{12.4em}{\includegraphics[scale=0.3]{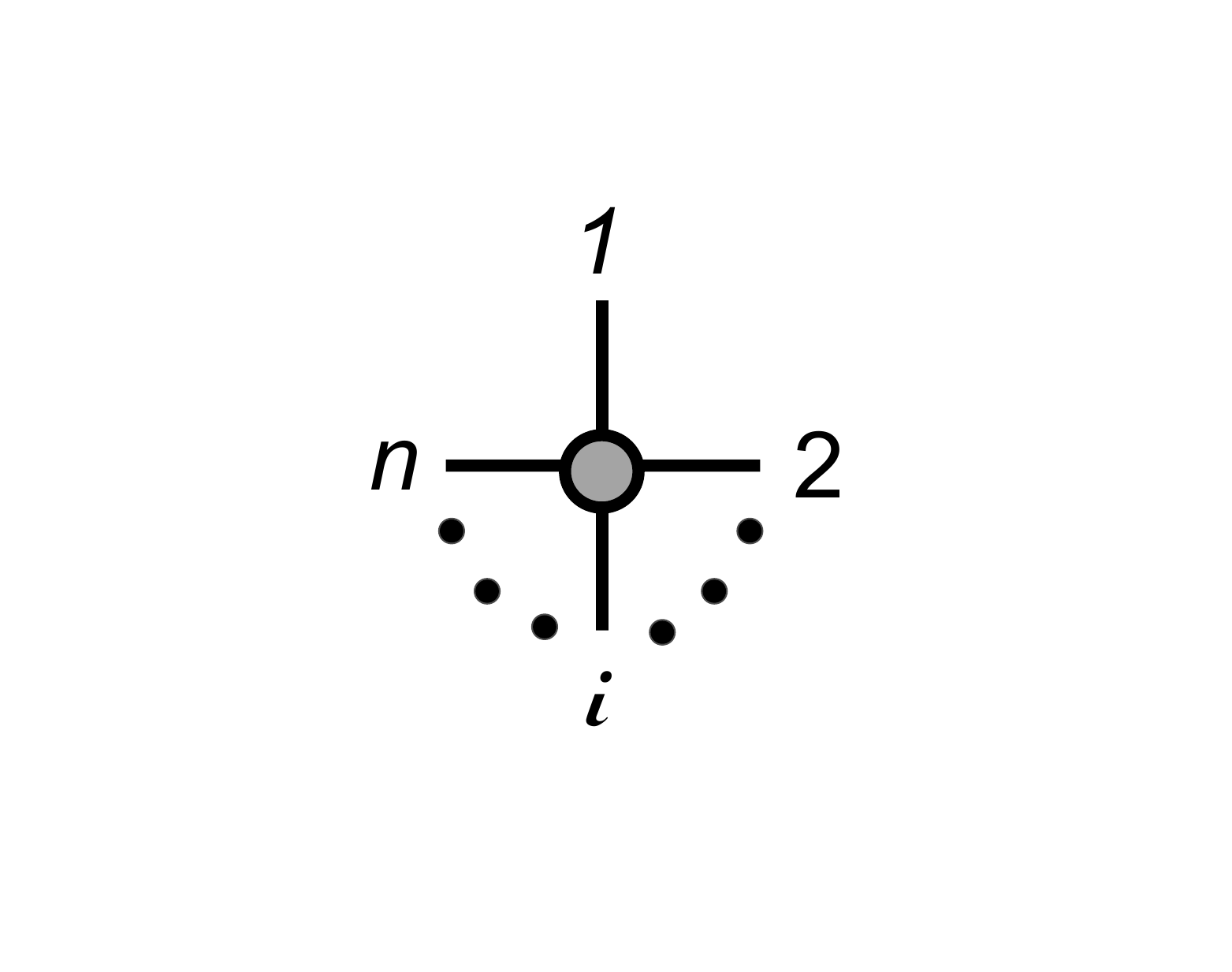}} \, \, \, ,
\end{eqnarray}
\vskip-1.0cm\noindent
where  the grey circle  means the sum over all possible trivalent planar diagrams with the ordering $(1,2,\ldots, n)$.
} ($T_i$), the symbol ``$P_n$'' in the loop circle denotes a regular polygon of $n$-edges
and  $``N"$ is total number of particles.

From the butterfly graphs obtained in \eqref{4pts-2ex}  and \eqref{5ptsNP-2ex}, we generalize 
the planar CHY-graph in \eqref{planar-gen} to the non-planar case. Additionally, by using the $\L-$algorithm, it is straightforward to compute this new kind of graph.  Thus,  a general butterfly graph and its result in terms of Feynman diagrams is given by the expression
\vskip-1.1cm
\begin{eqnarray}\label{MOSTGEN}
\frac{1}{2^{N+1}} \int d\Omega  s_{a_1 b_1} 
 \int d\mu^{\rm t}_{N+4}  
\hspace{0.1cm}
 \parbox[c]{14.2em}{\includegraphics[scale=0.34]{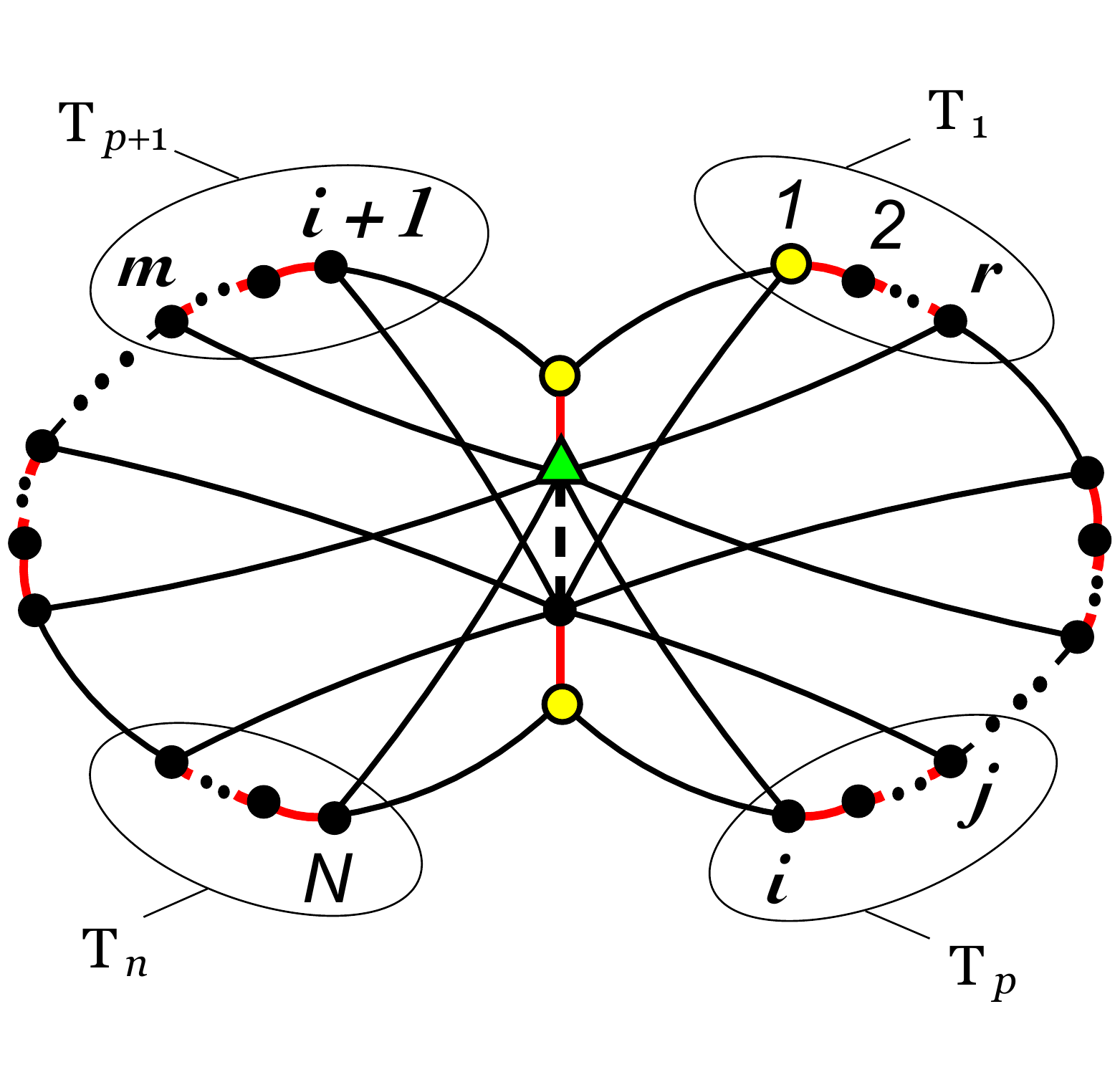}}
&=&
\hspace{-0.5cm}
 \parbox[c]{14em}{\includegraphics[scale=0.38]{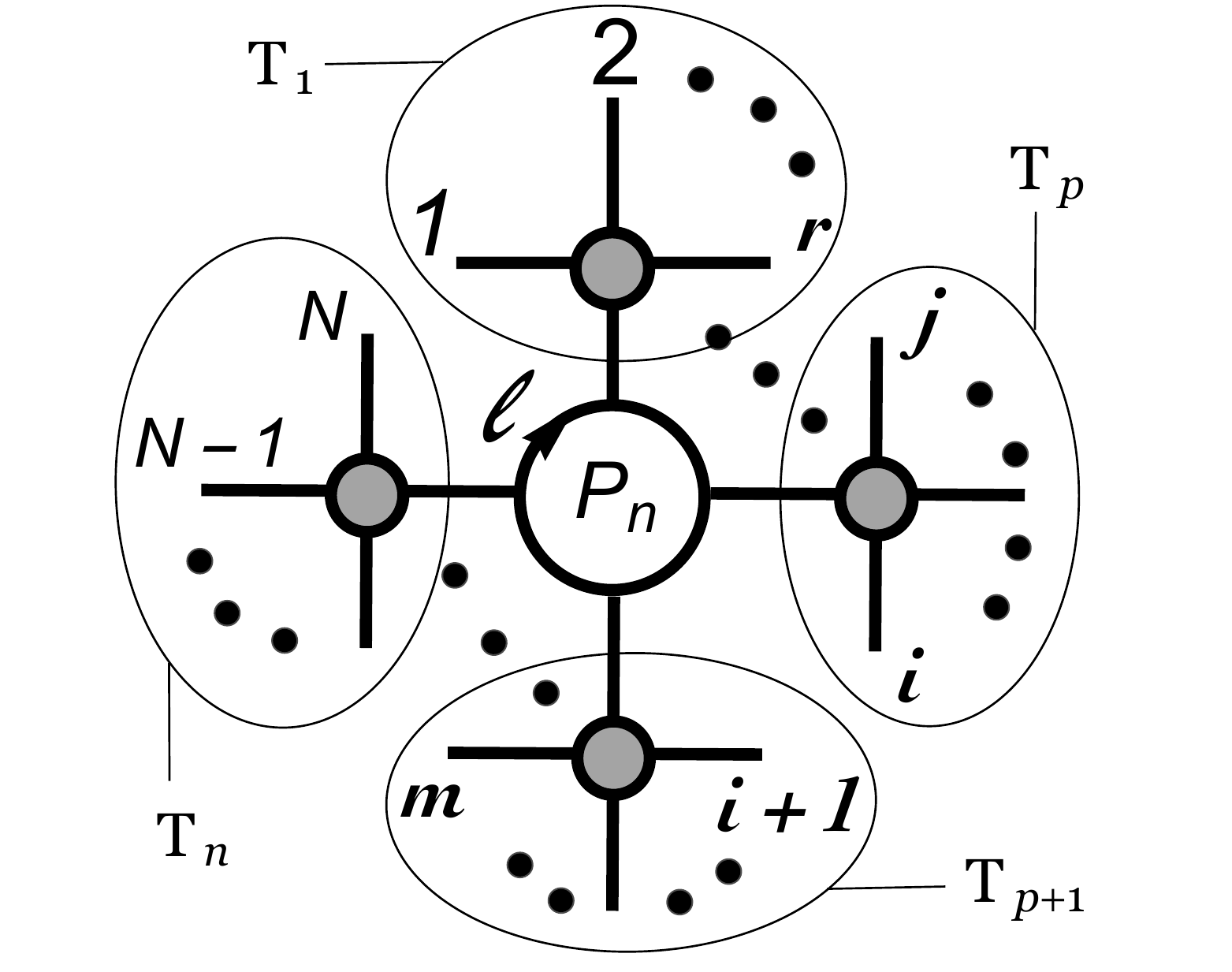}} \nonumber\\ 
&+&
\left\{ \, \left[  \left(   
\begin{matrix}
n \\
p
\end{matrix}
\right)  -1 \right] \, {\rm terms} \,\right\}  ,
\qquad\quad  \nonumber\\
&=&
\overrightarrow{ \,\,\left[ T_1,\ldots , T_p \right] \sh \left[ T_{p+1},\ldots , T_n \right] \,\,}  ,
\qquad\quad\non
\end{eqnarray}
\vskip-0.1cm\noindent
where, as in \eqref{planar-gen},  the grey circles  mean the sum over all possible trivalent planar diagrams ($T_i$), ``$P_n$''  denotes a regular polygon of $n$-edges and ``$N$" is the total number of particles. 

Finally, using the results found in \cite{Gomez:2017cpe}  and the  CHY-graphs representation obtained for $\mathfrak{M}_4^{\rm 1-NP:NP}[1,2 | 3,4 \, ; \, 1,2|3,4 ]$ and $\mathfrak{M}_5^{\rm 1-NP:NP}[1,2 | 3,4,5 \, ; \, 1,2|3,4,5 ]$  in \eqref{4pts-2ex}  and \eqref{5ptsNP-2ex} respectively, we formulate a general expression for $\mathfrak{M}_N^{\rm 1-NP;NP}[\a|\b \, ; \, \a | \b] $. To be precise, up to an overall sign, we propose the following expression
\begin{eqnarray}\label{M-Ngen}
\hspace{-0.3cm}
\mathfrak{M}_N^{\rm 1-NP:NP}[\a | \b \, ; \, \a | \b] = \frac{1}{2^{N+1}} \int d\Omega \,\, s_{a_1 b_1} 
 \int d\mu^{\rm tree}_{N+4}  \left[
 \sum_{a=2 \atop b=2}^{|\a | ,\,  |\b|} \sum_i{\rm NPchy}^{a;b}_{(\a | \b)}[[i]] \, + \, {\rm cyc}(\a) \times   {\rm cyc}(\b)\,\,
\right], \qquad
\end{eqnarray}
where the ordered lists $\a$ and $\b$ are given by, $\a = \{ 1,2,\ldots , p \}$ and $\b = \{ p+1,p+2,\ldots , N \}$, $|\a|$ and $|\b|$ are the lengths of the lists, i.e. $|\a| = p$ and $| \b | = N-p$,  and we have defined the set, ${\rm NPchy}^{a:b}_{(\a | \b)}$,  as
\vskip-1.0cm
\begin{eqnarray}
{\rm NPchy}^{a:b}_{(\a | \b)}:=\left\{
{\rm All\,\, possible\,\, non{\rm -}planar\,\, CHY\,graphs\,\, with\,\, the \,\, form}  \right.
\hspace{-0.0cm}
\parbox[c]{11em}{\includegraphics[scale=0.27]{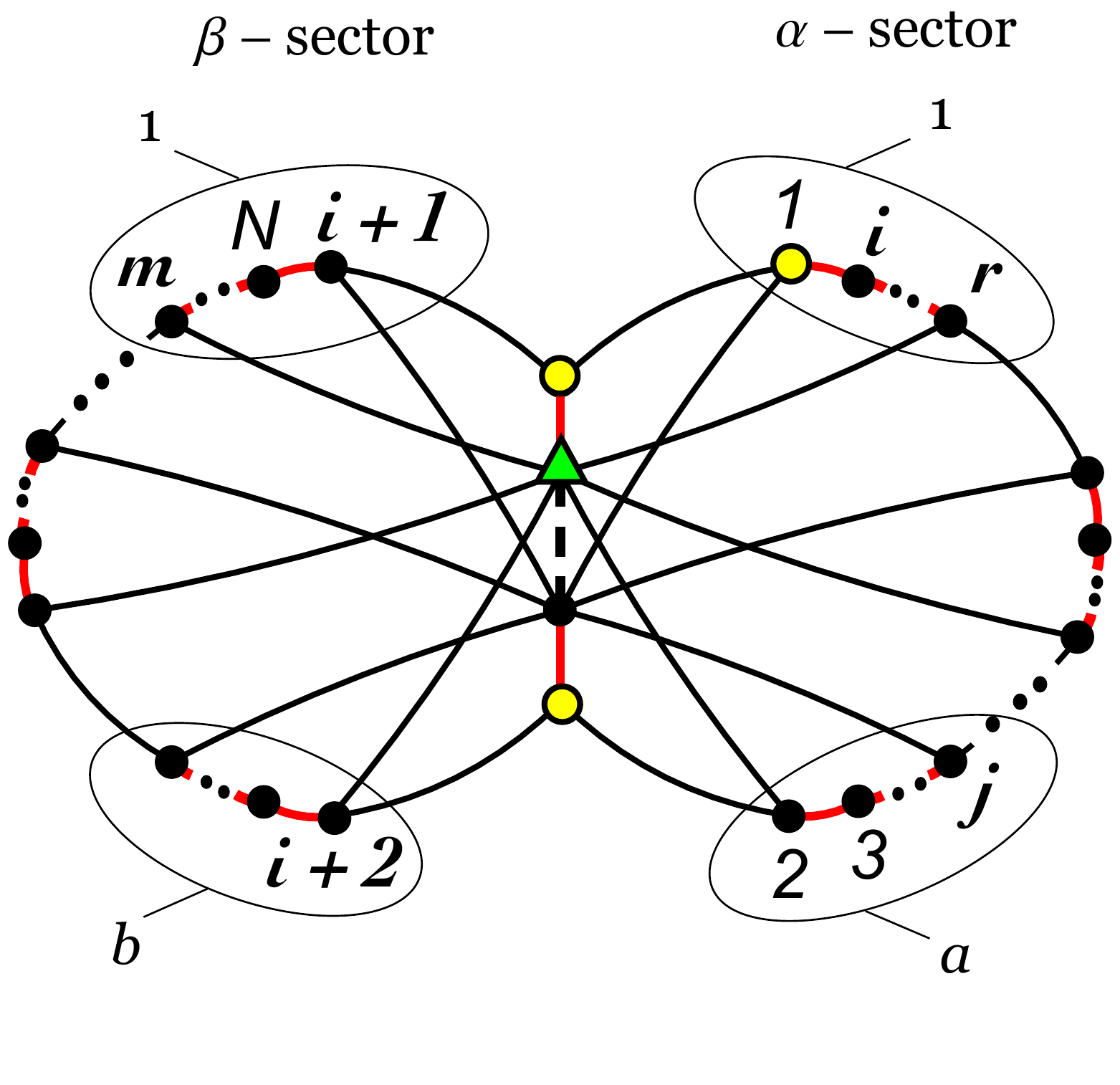}}
\left.
\right\}\, , \qquad\label{NPchydefi}
\end{eqnarray}
\vskip-0.7cm\noindent
being ${\rm NPchy}^{a:b}_{(\a | \b)}[[i]]$ the $i-$th element in ${\rm NPchy}^{a:b}_{(\a | \b)}$. For instance
\vskip-0.51cm 
\begin{eqnarray}
\hspace{-3.8cm}
{\rm NPchy}^{3:3}_{(1,2,3 | 4,5,6)} 
=\left\{
 \parbox[c]{9.3em}{\includegraphics[scale=0.23]{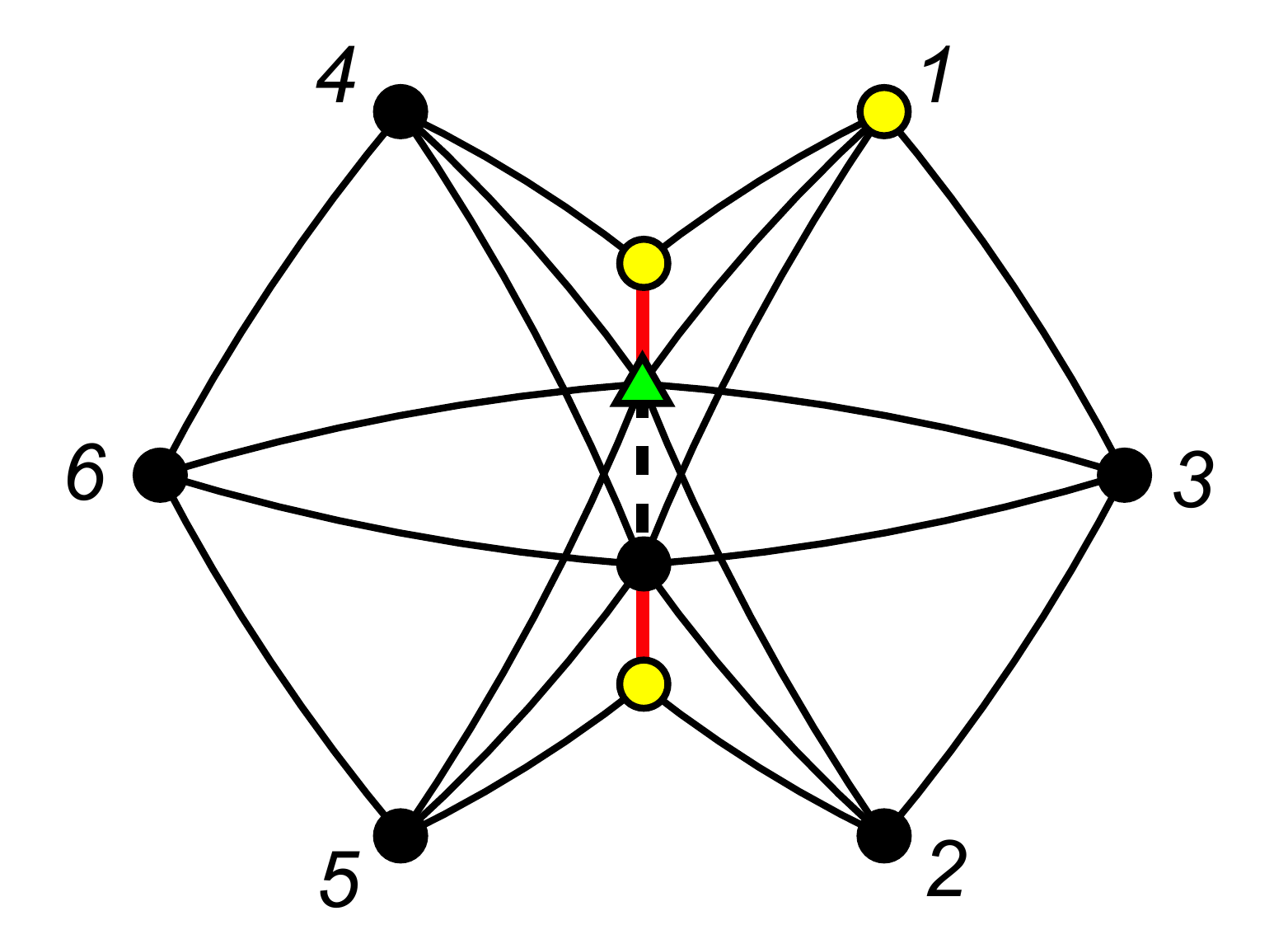}}
\right\}\, \, ,
\end{eqnarray}
\vskip-0.4cm 
\begin{eqnarray}
{\rm NPchy}^{3:2}_{(1,2,3 | 4,5,6)} 
=\left\{
 \hspace{0.1cm}
 \parbox[c]{9.5em}{\includegraphics[scale=0.21]{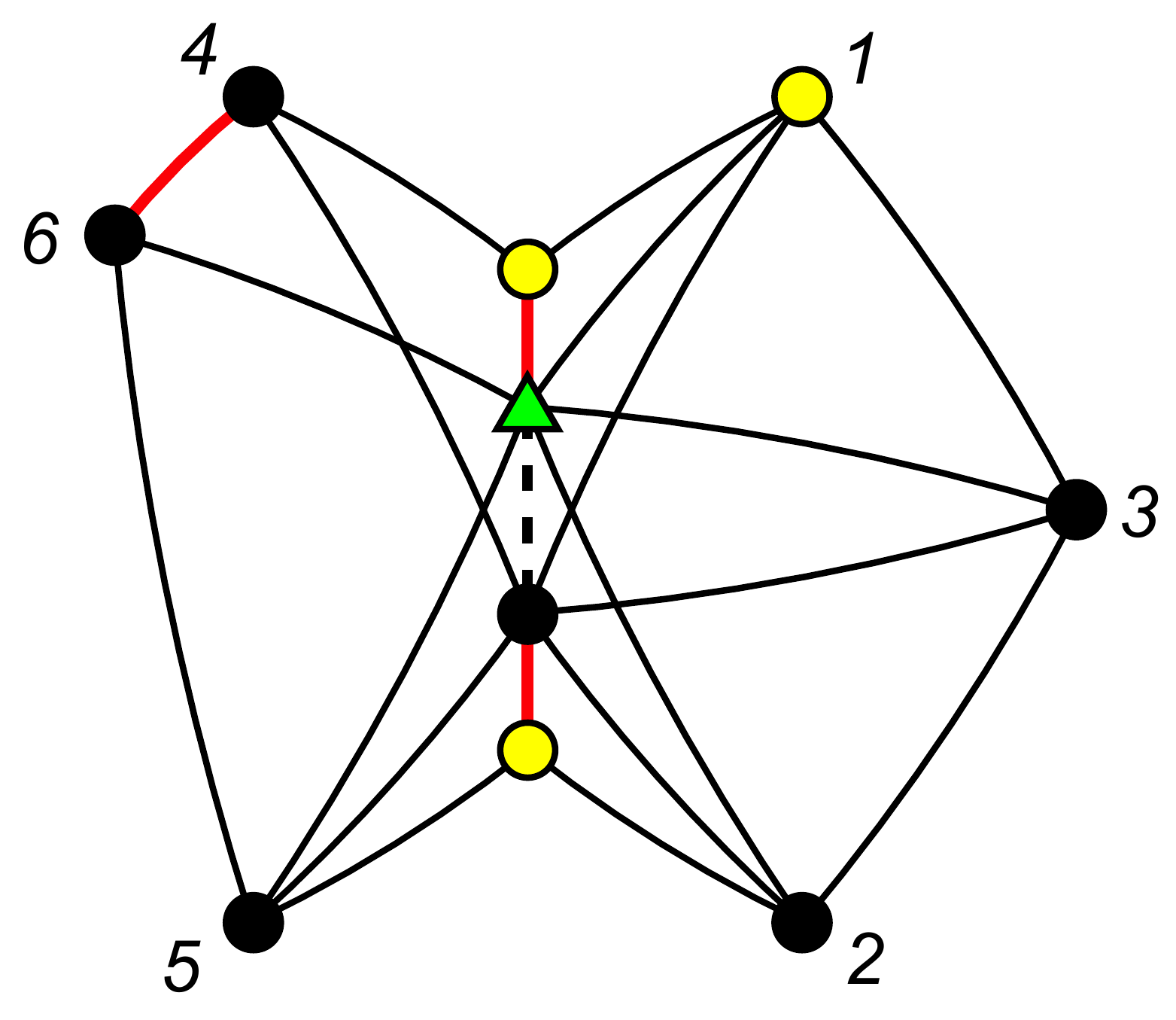}},
 \hspace{0.1cm}
 \parbox[c]{9em}{\includegraphics[scale=0.21]{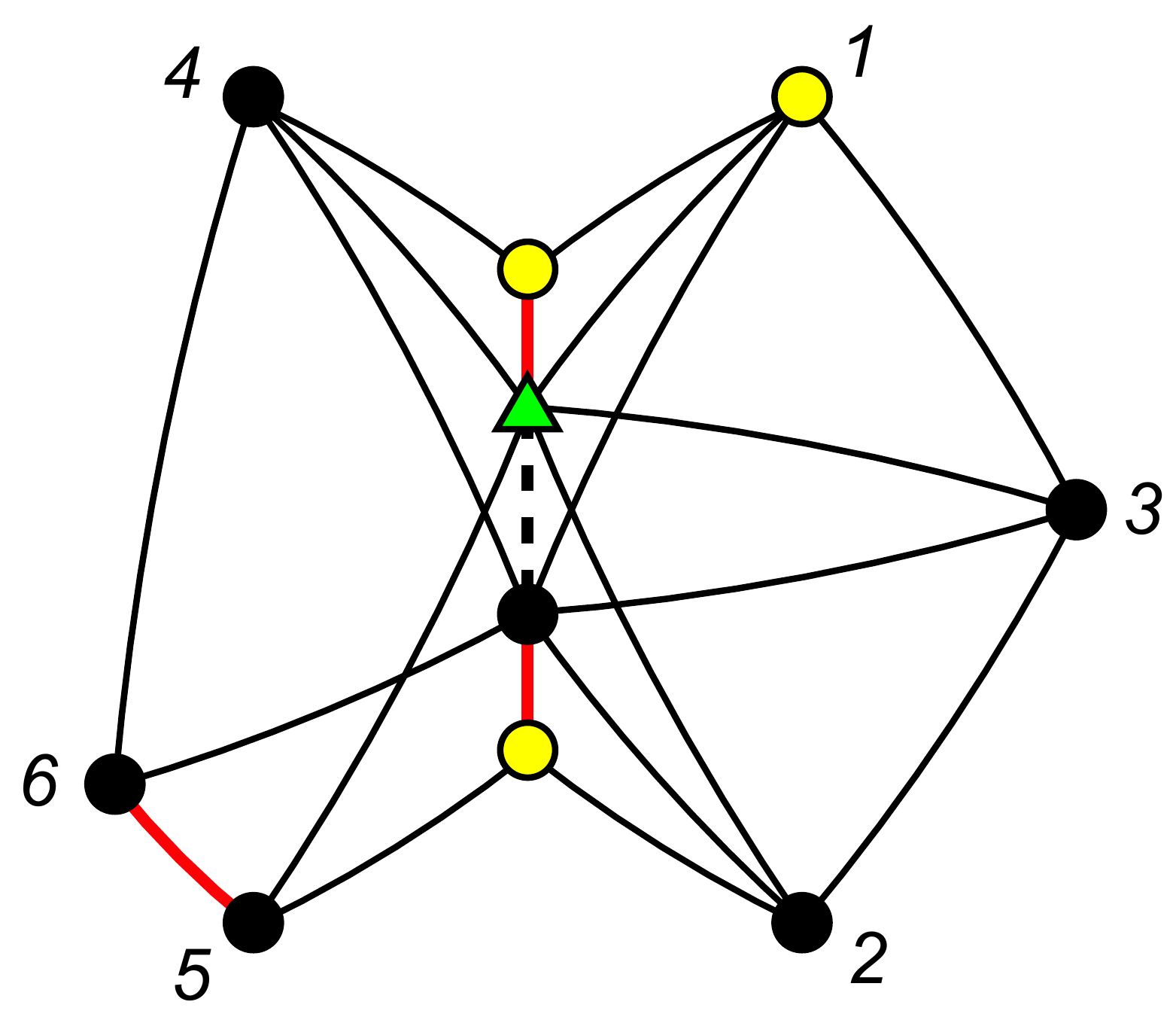}}
\right\}\,.
\end{eqnarray}
\vskip-0.1cm\noindent
This is clear that \eqref{M-Ngen} is in agreement with \eqref{4pts-2ex}  and \eqref{5ptsNP-2ex}.

We can now summarise the results obtained in this work in the following and final section.

\section{Discussions}\label{discussions}

In this paper we continued the program to obtain one-loop quadratic Feynman propagators directly from CHY. Studying the one-loop Parke-Taylor factors found in \cite{Gomez:2017cpe}, we found the well known KK relations for amplitudes in momentum space, now in the context of functions of $\sigma-variables$ in the CHY formalism. These relations allowed a generalization that leaded us also to obtain a generalization to the {\it multi-trace one-loop Parke-Taylor factors}, which as a special case have the double-trace one, or the {\it non-planar one-loop Parke-Taylor factors}. 

The KK relations have been traditionally used to obtain the subleading order contributions at one-loop in terms of leading ones. What we have done here, is exploit the KK relations and as a consequence of them obtain a remarkable result, the {\it non-planar CHY-graphs}. There is not an equivalent of these graphs in the traditional formalism and among the advantages they offer we have:
\begin{itemize}
\item The subleading order can now be written directly in terms of these graphs, since they encode all the information for this order in a fewer number of graphs in comparison to Feynman diagrams.
\item Easier computation and generalization for a higher number of points.
\item Using intersections between graphs the amplitudes for mixed orderings can be obtained in a straightforward way.
\end{itemize} 

We applied the non-planar CHY-graphs in the study of the bi-adjoint scalar theory at one-loop. We have found all there is to know, as far as we understand, for this theory at that loop order, at the level of Feynman integrands with on-shell external particles. 

There are several directions to move on from the developments we have done in this paper. We are ready to apply all the technology we have developed in the study of the Yang-Mills theory and gravity at one-loop with quadratic Feynman propagators, the KK relations we have found here play an important role in a version of the BCJ duality we are currently working on. The study of the BCJ duality have already been done for the case of linear propagators \cite{He:2017spx,Geyer:2017ela}, so it will be very interesting to compare. 

Another direction is going to a higher number of loops for the Parke-Taylor factors, and study relations among them, that is another work in progress. There is not much study done in this part at the moment. 

Since we have found a way to encode all the one-loop information in a fewer number of CHY-graphs(integrands), it would be desirable to be able, or understand the viability, of performing the loop integration before the contour integration in the $\sigma-variable$, this could lead to a new and more compact way of calculating loop corrections.

\subsection*{Acknowledgements}

We would like to thank E. Bjerrum-Bohr for reading the manuscript and discussions. H.G. would like to thank to J. Bourjaily, and
P. Damgaard for discussions. H.G. is very grateful to the Niels Bohr Institute - University of Copenhagen for hospitality.
The work of N.A. was supported by IBS (Institute for Basic Science) under grant No. IBS-R012-D1. 
The work of  H.G.  is supported by USC grant DGI-COCEIN-No 935-621115-N22.

\appendix

\section{Four-point non-planar CHY-graph computation}
\label{4-p-app-chy}
In this appendix we will work out an example using the $\L-$algorithm in order to compute non-planar CHY-graphs. Since that the  $\L-$algorithm is a graphical method we can omit the integral $\int\, d\mu_{4+4}^{\rm t}$.

The simplest non-trivial example is the four-point case computation obtained in \eqref{4pts-2ex}. Applying the $\L-$rules presented in \cite{Gomez:2016bmv}, we identify that there are three non-zero cuts given by
\vskip-0.5cm
\begin{eqnarray}\label{ex-appen-cuts1}
 \parbox[c]{8.4em}{\includegraphics[scale=0.23]{4pts_12-34.pdf}}
 = 
\hspace{-0.4cm}
    \parbox[c]{7.6em}{\includegraphics[scale=0.23]{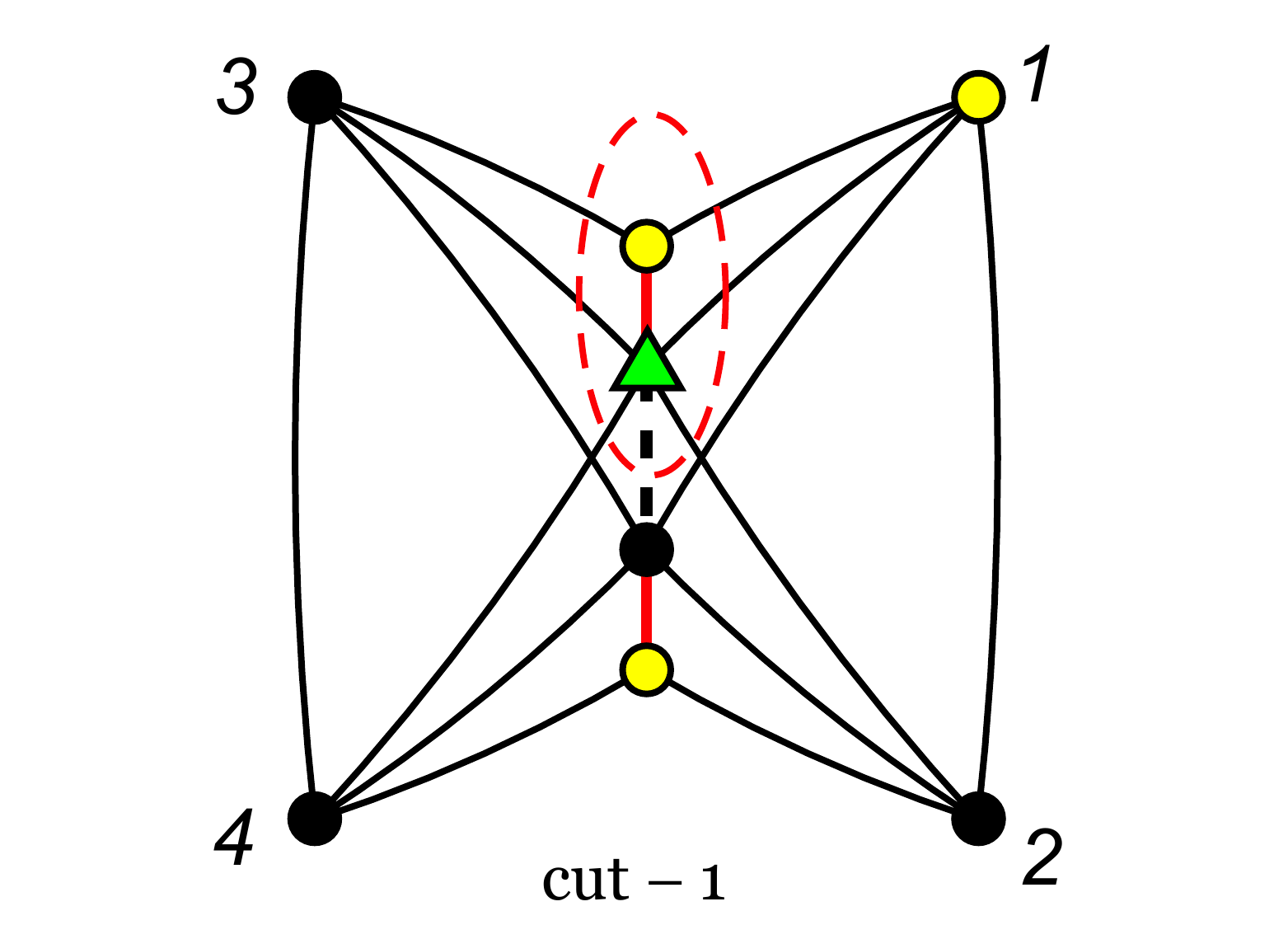}} 
    +  
\hspace{-0.3cm}
 \parbox[c]{7.5em}{\includegraphics[scale=0.23]{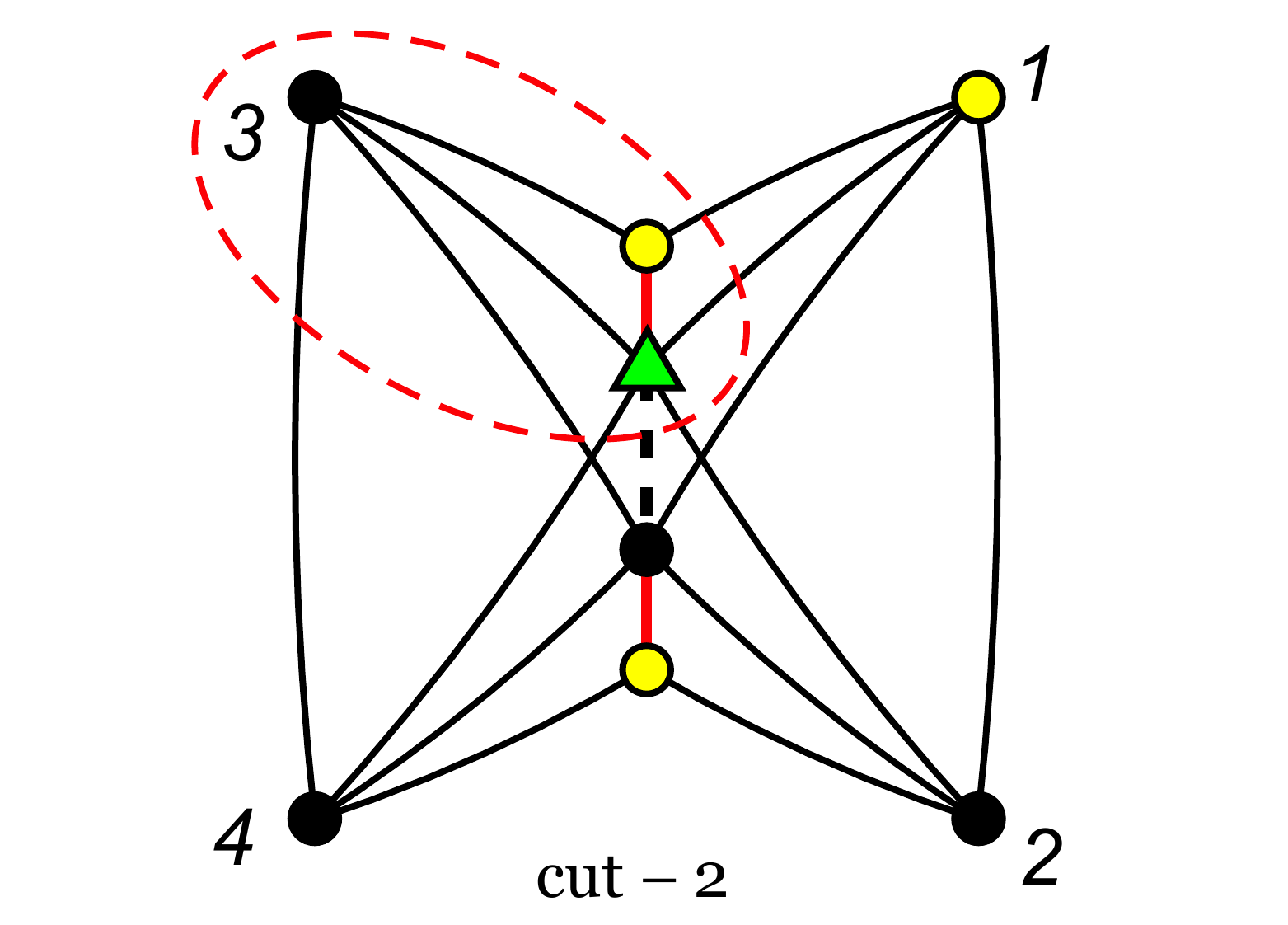}} +
\hspace{-0.3cm}
  \parbox[c]{7.5em}{\includegraphics[scale=0.23]{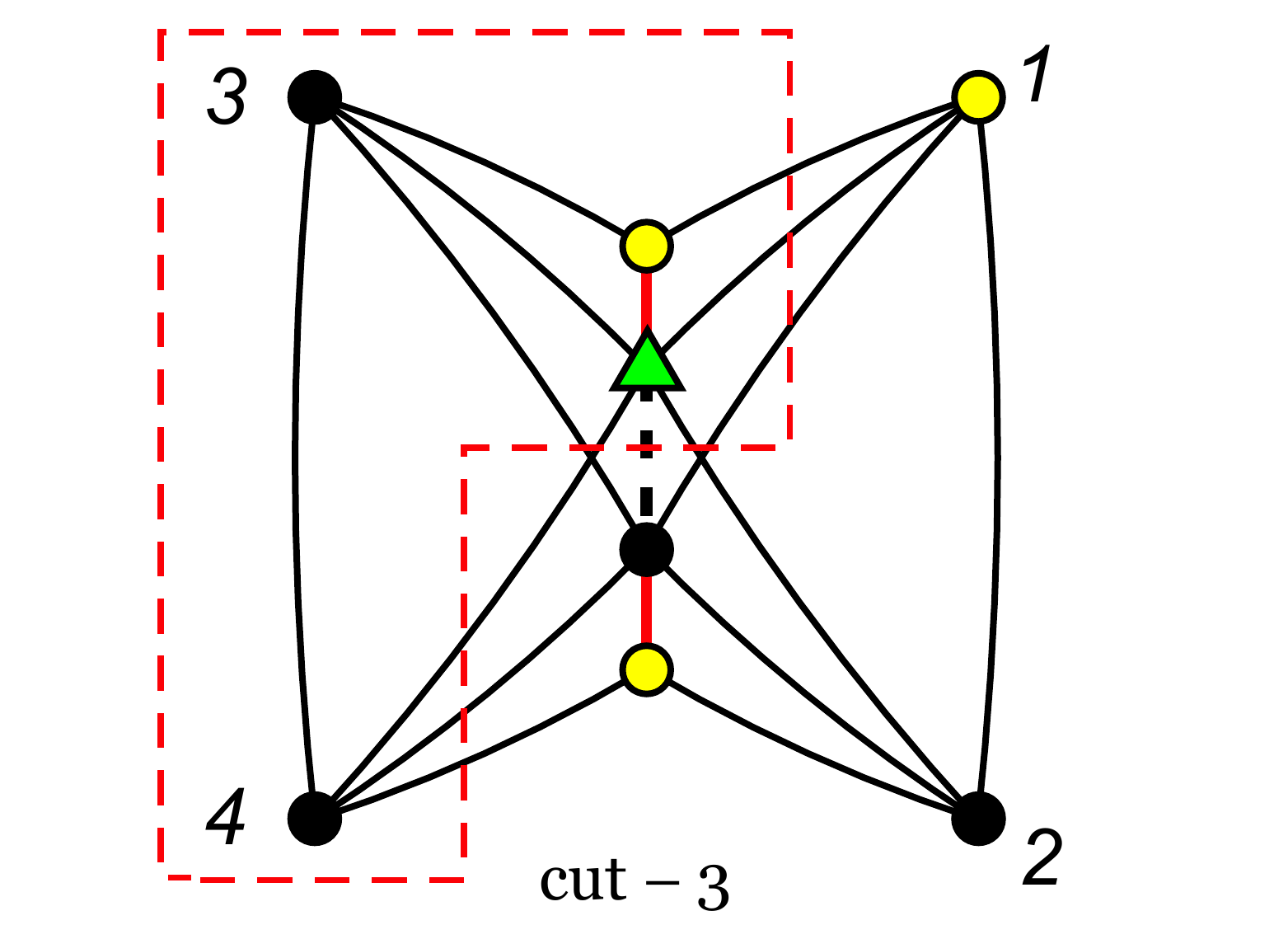}} \,\,
\,\, .
\end{eqnarray}
\vskip-0.05cm\noindent
Note that in this paper we have used this gauge fixing for all graphs, which was very useful in the planar case \cite{Gomez:2017cpe}. However, for the non-planar case this gauge is not as efficient. Although it is not evident at first sight, the cuts in \eqref{ex-appen-cuts1} generate non-trivial CHY-subgraphs with spurious poles. Thus, we are going to choose a new gauge fixing that it is able to produce CHY-subgraphs with only physical poles. 

Let us fix the punctures $\{ \s_1, \s_2, \s_3\}$ by  ${\rm PSL}(2,\mathbb{C})$, and the puncture $\{ \s_4\}$ by scaling symmetry. Under this gauge fixing, the non-planar graph becomes 
\vskip-0.5cm
\begin{eqnarray}\label{}
 \parbox[c]{8.4em}{\includegraphics[scale=0.23]{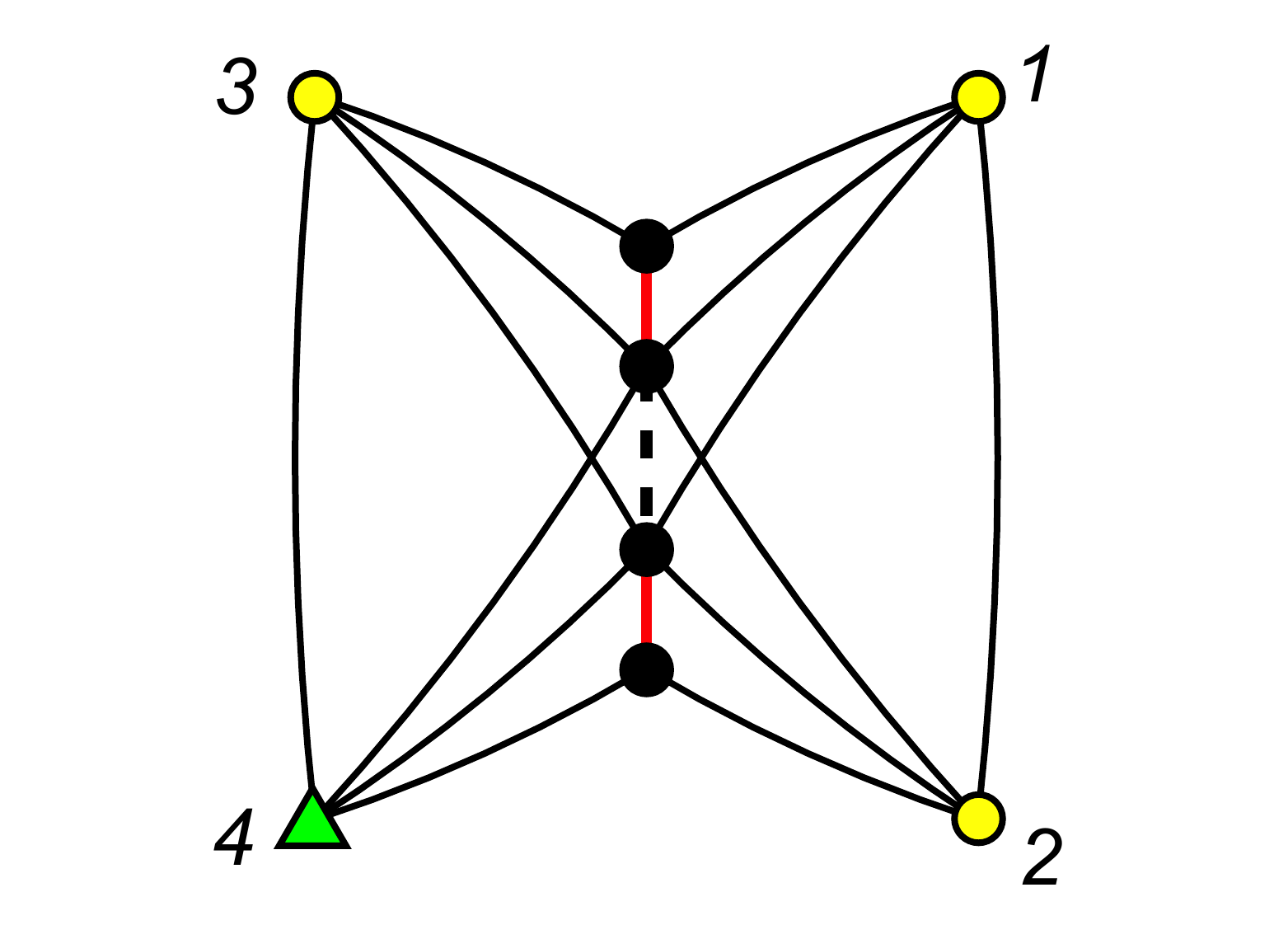}}
 = 
\hspace{-0.4cm}
    \parbox[c]{8.6em}{\includegraphics[scale=0.23]{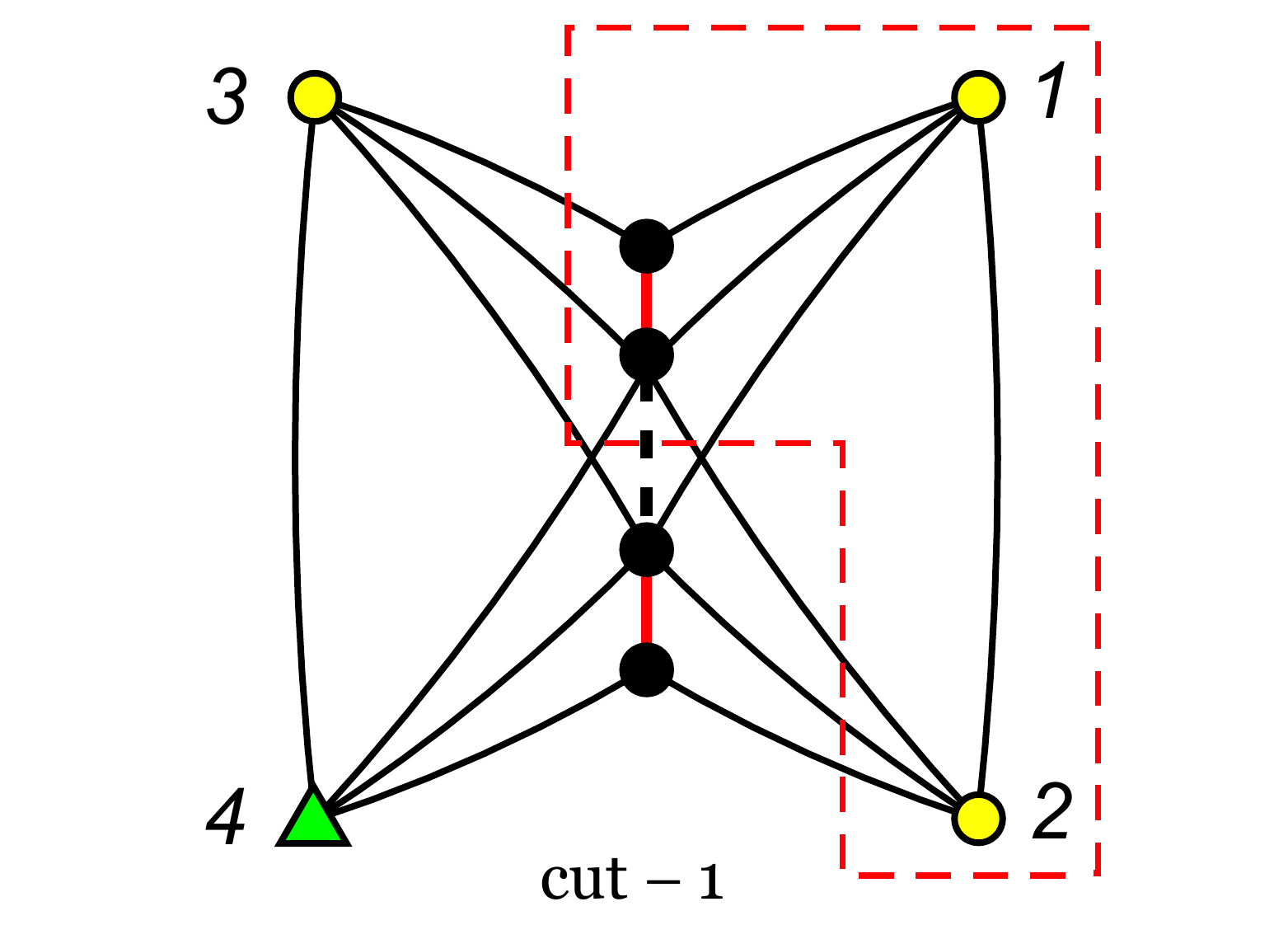}} 
    +  
\hspace{-0.3cm}
 \parbox[c]{7.5em}{\includegraphics[scale=0.23]{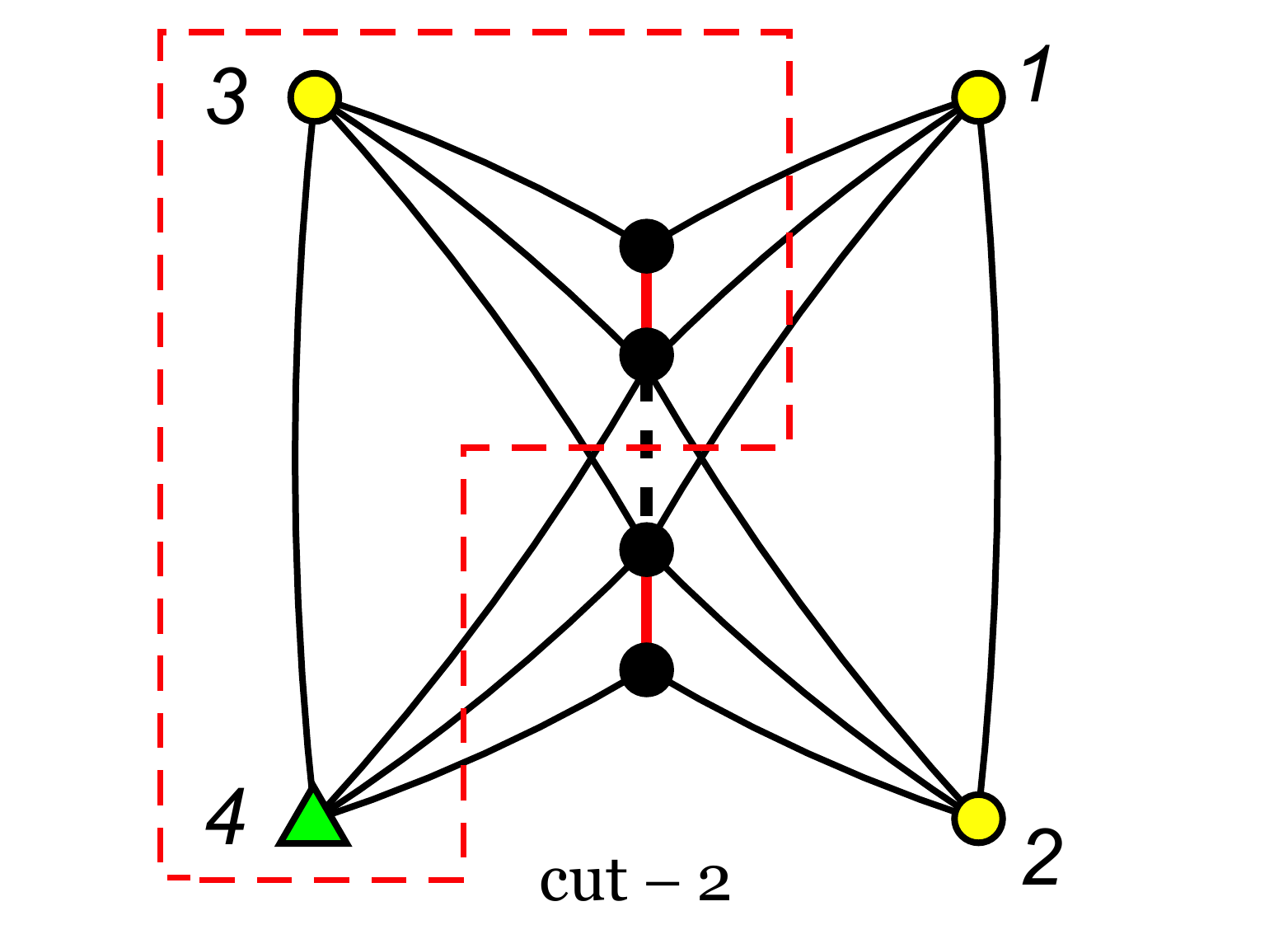}} +
\hspace{-0.3cm}
  \parbox[c]{7.5em}{\includegraphics[scale=0.23]{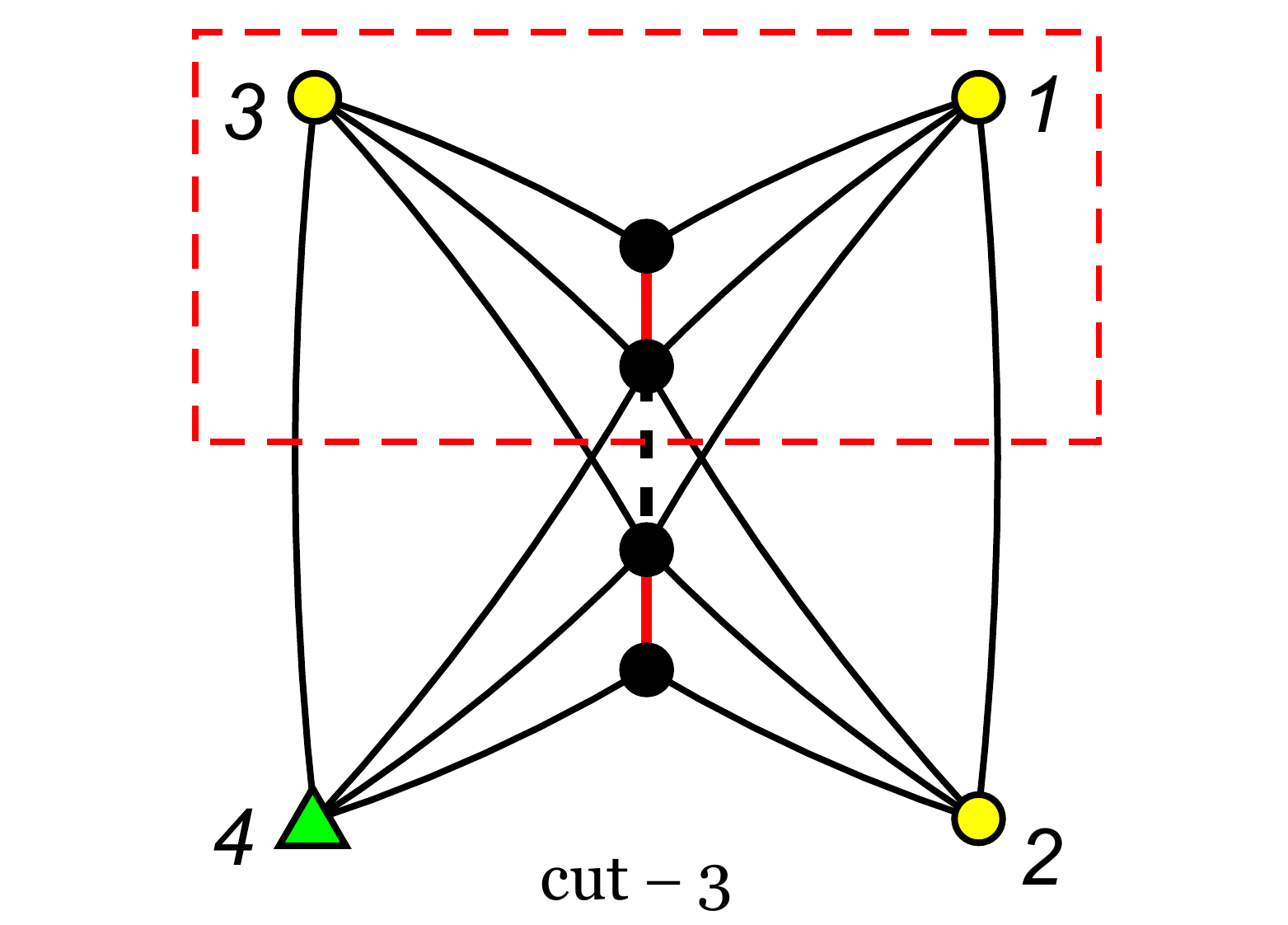}} \,\quad .
\end{eqnarray}
\vskip-0.05cm\noindent
The above cuts are straightforward to be computed and  their result is   
\vskip-0.5cm
\begin{eqnarray}\label{cut1-r1}
    \parbox[c]{8.6em}{\includegraphics[scale=0.23]{cut-1g2.pdf}} 
    =  
\frac{2}{(k_{a_1}+k_{b_1}+k_{1}+k_{2})^2}    \times
    \left(
\hspace{-0.1cm}
 \parbox[c]{6.5em}{\includegraphics[scale=0.19]{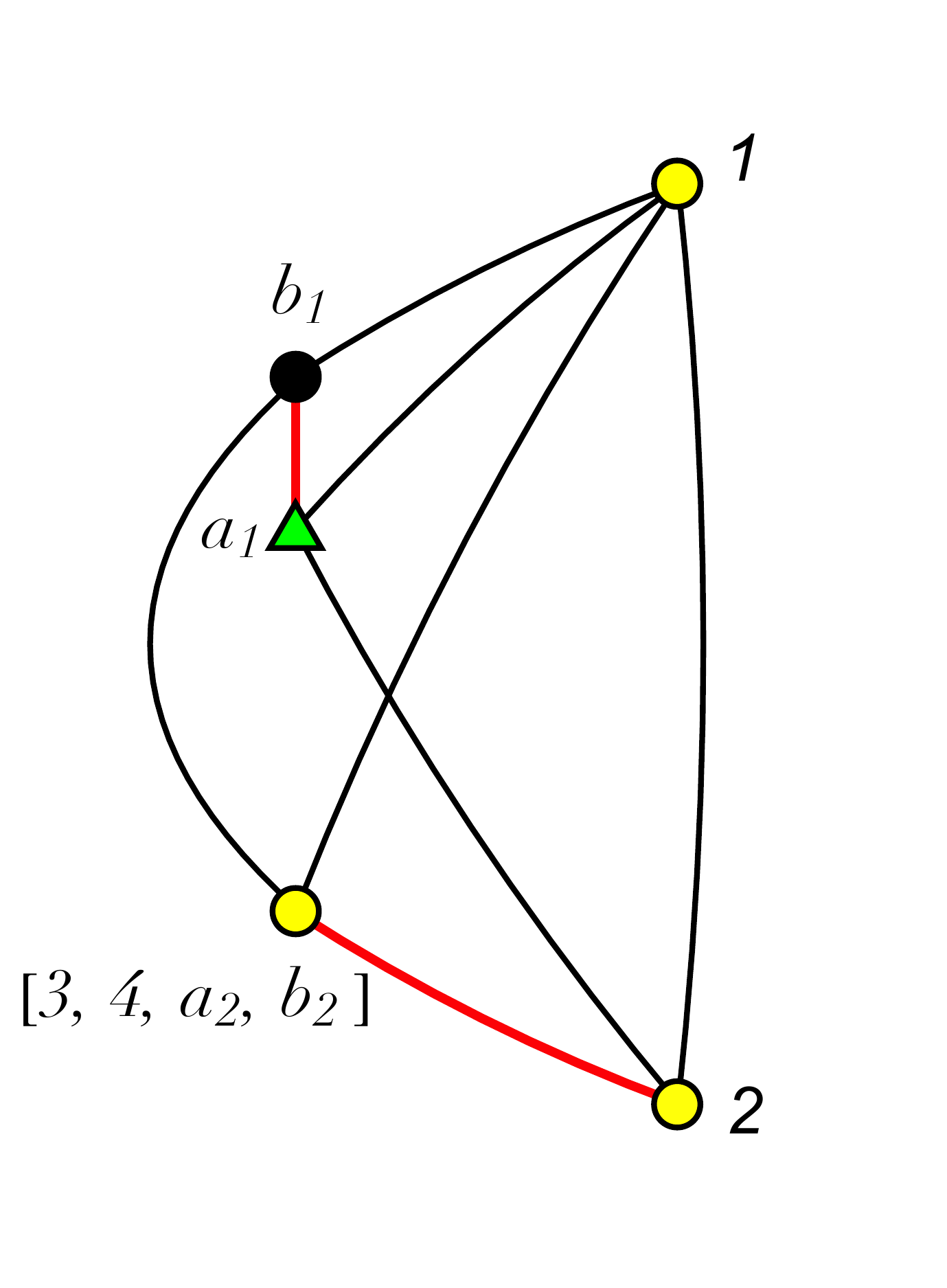}} 
\right) 
 \times
 \left(
\hspace{-0.1cm}
  \parbox[c]{6.5em}{\includegraphics[scale=0.19]{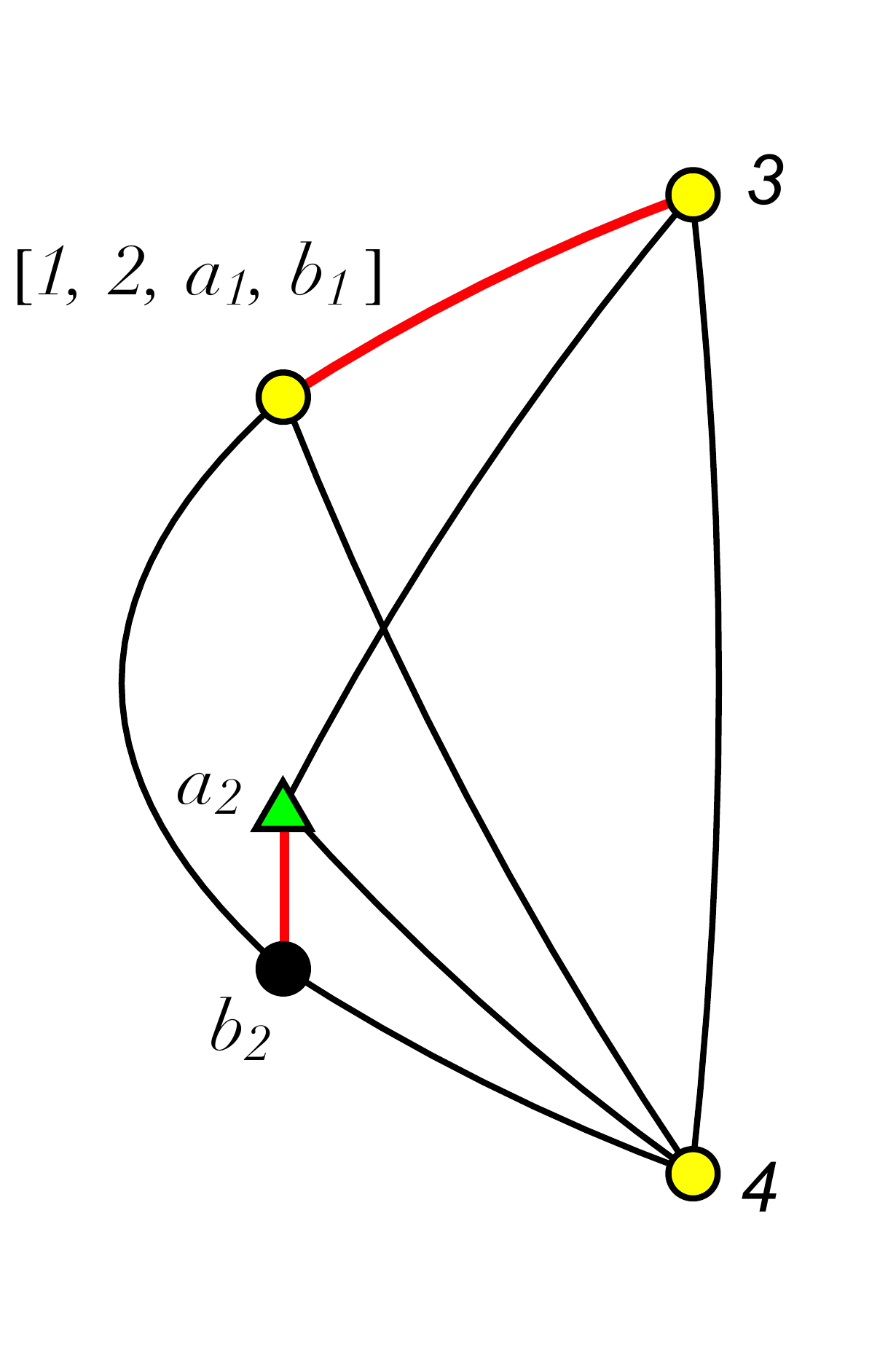}} 
\right)  \,  , \qquad
\end{eqnarray}
\vskip-0.2cm
\begin{eqnarray}\label{cut2-r1}
    \parbox[c]{8.6em}{\includegraphics[scale=0.23]{cut-2g2.pdf}} 
    =  
\frac{2}{(k_{a_1}+k_{b_1}+k_{3}+k_{4})^2}    \times
    \left(
\hspace{-0.1cm}
 \parbox[c]{6.5em}{\includegraphics[scale=0.19]{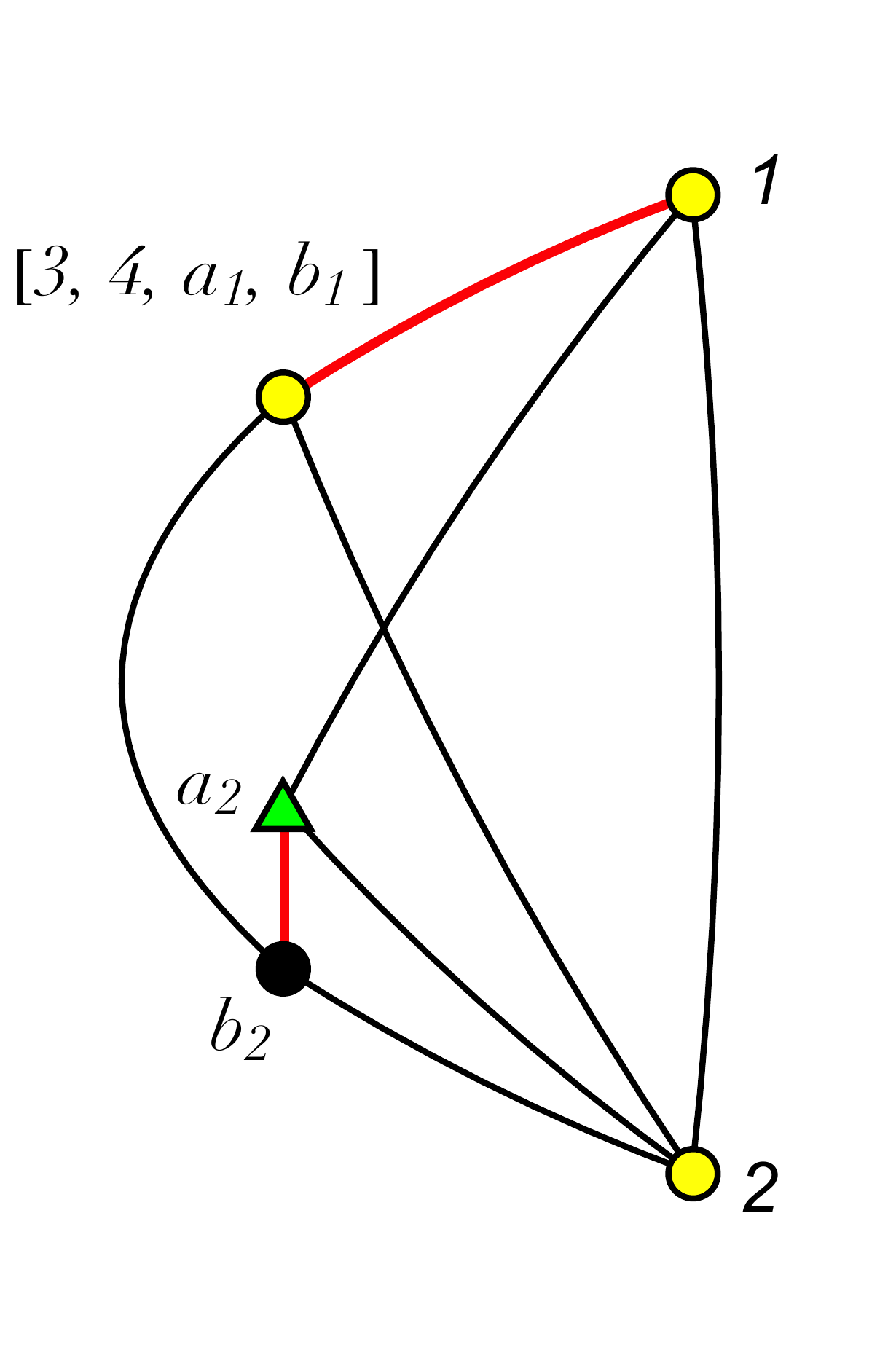}} 
\right) 
 \times
 \left(
\hspace{-0.1cm}
  \parbox[c]{6.5em}{\includegraphics[scale=0.19]{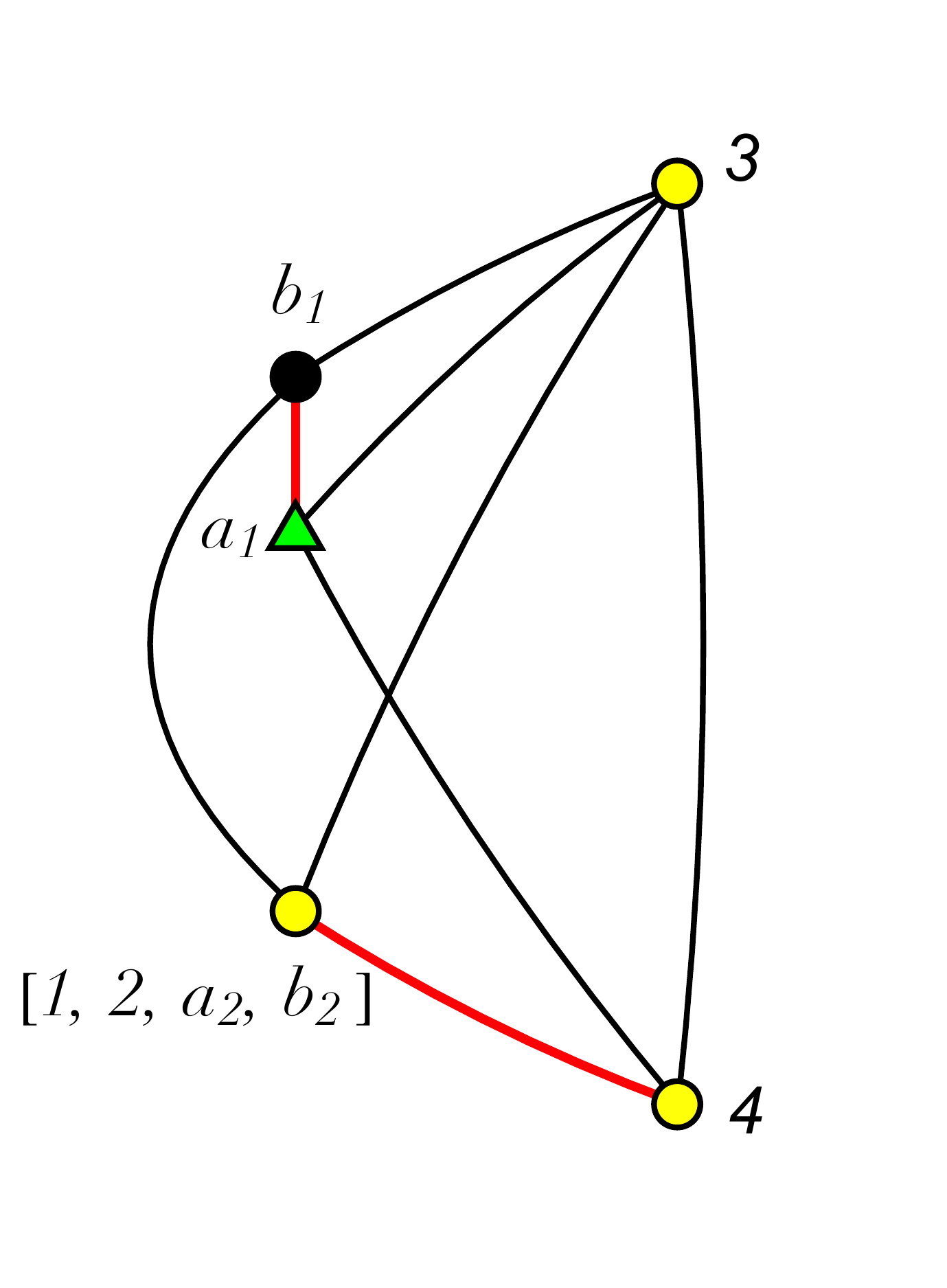}} 
\right)  \,  , \qquad
\end{eqnarray}
\vskip-0.6cm
\begin{eqnarray}\label{cut3-r1}
    \parbox[c]{8.6em}{\includegraphics[scale=0.23]{cut-3g2.pdf}} 
    =  
\frac{2}{(k_{a_1}+k_{b_1}+k_{1}+k_{3})^2}    \times
    \left(
\hspace{-0.1cm}
 \parbox[c]{7.5em}{\includegraphics[scale=0.19]{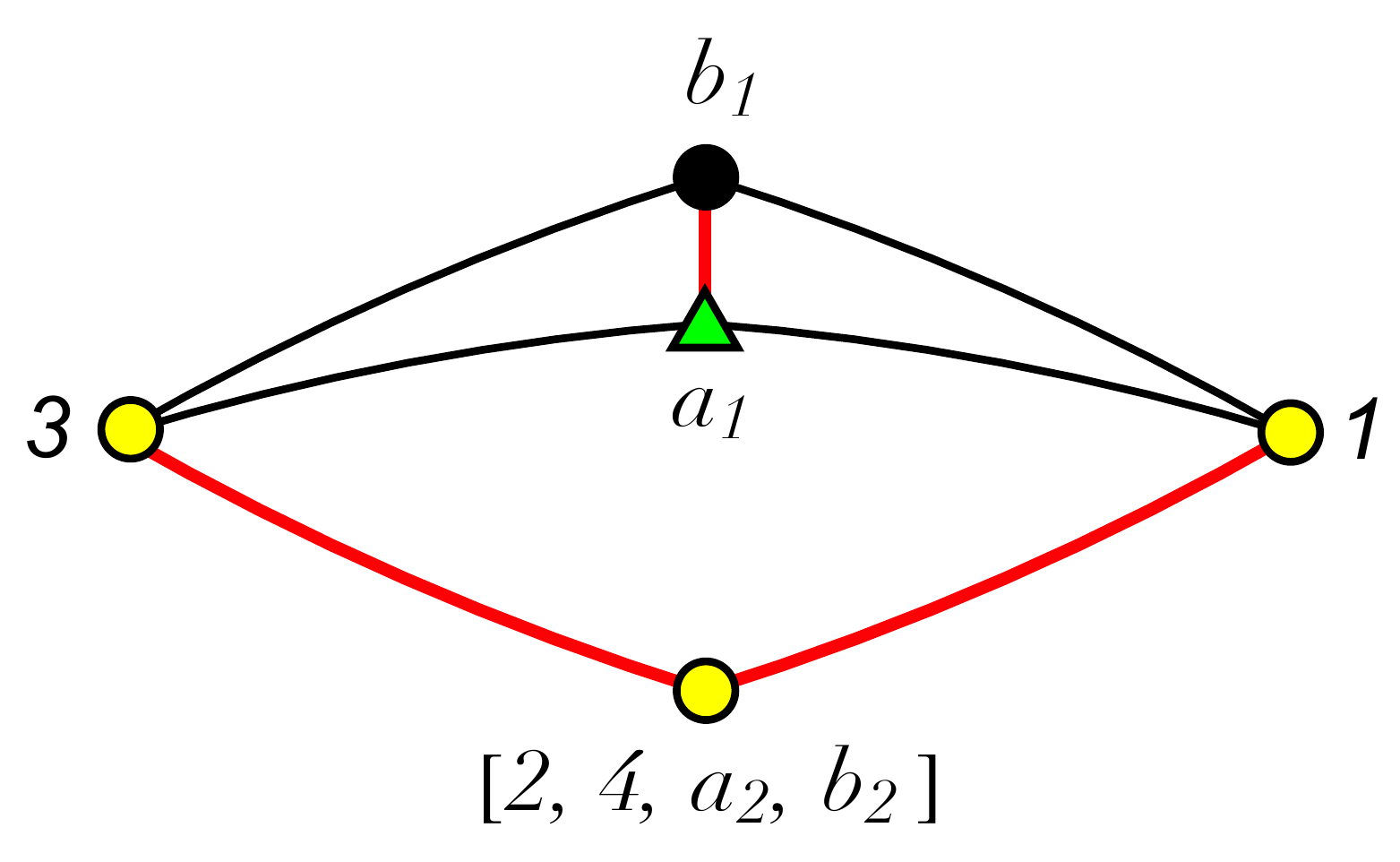}} 
\right) 
 \times
 \left(
\hspace{-0.1cm}
  \parbox[c]{7.5em}{\includegraphics[scale=0.19]{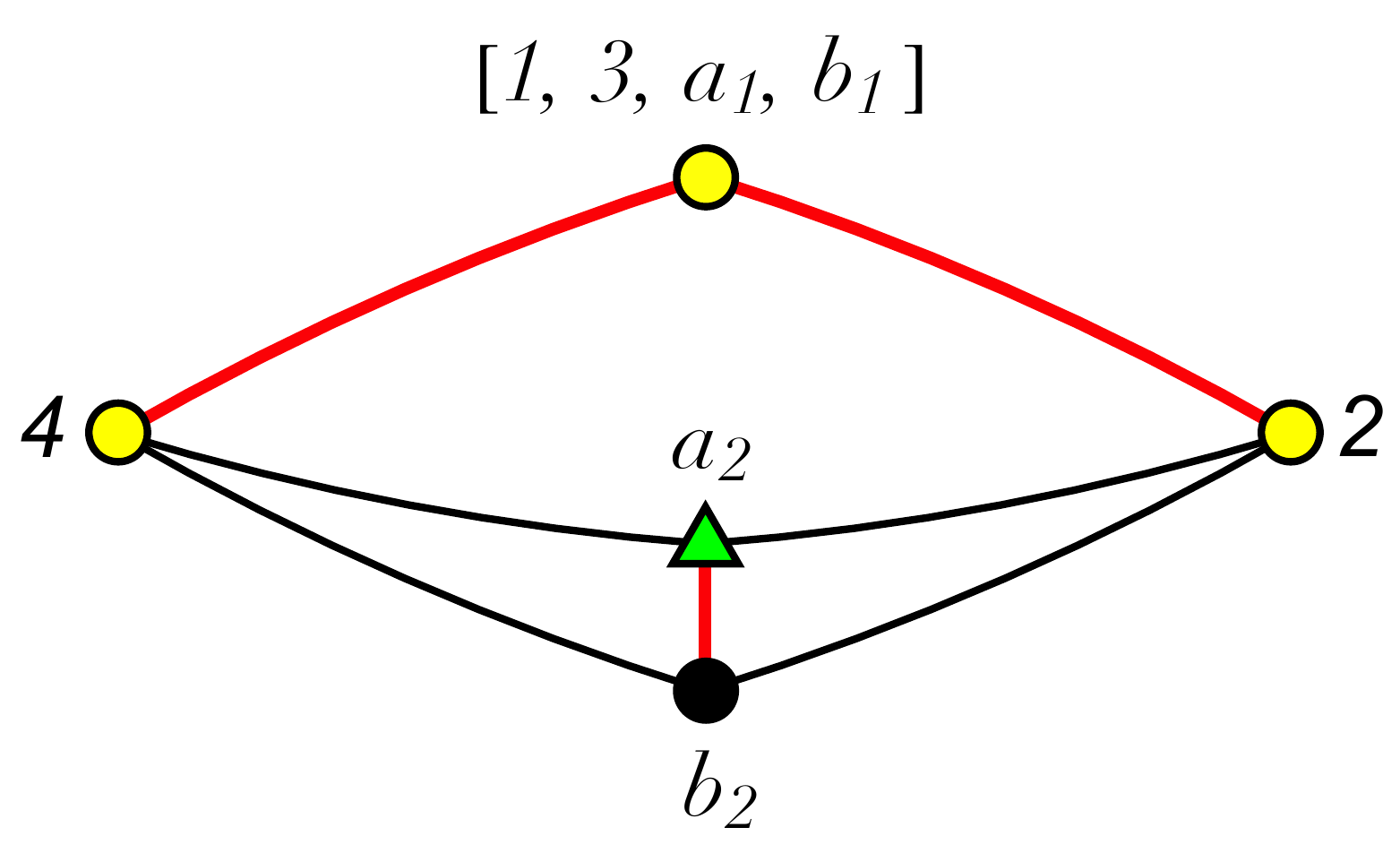}} 
\right)  \,  . \qquad
\end{eqnarray}
\vskip-0.00cm\noindent
Note that the CHY-subgraphs obtained in \eqref{cut1-r1} and \eqref{cut2-r1} have the same structure as the one given in \eqref{planar-gen}, which was computed in datail in  \cite{Gomez:2017cpe}. So, the final answer for cut-1 and cut-2 is
\begin{align}
{\rm cut-1} = \frac{2^5}{s_{a_2b_2}\,s_{a_1b_1} \, s_{1a_1b_1} \, s_{12a_1b_1}\, s_{123a_1b_1} } \, ,
\nonumber\\
{\rm cut-2} = \frac{2^5}{s_{a_2b_2}\,s_{a_1b_1} \, s_{3a_1b_1} \, s_{34a_1b_1}\, s_{341a_1b_1} } \, .
\nonumber
\end{align}
On the other hand, the subgraphs in \eqref{cut3-r1} have the same form as ones studied in \cite{Gomez:2017lhy}. The computation of these subgraphs is straightforward and the total result of the cut-3 is
\begin{equation}
{\rm cut-3} = \frac{2^5}{s_{a_2b_2}\,s_{a_1b_1} \, s_{13a_1b_1}}\left(  \frac{1}{ s_{1a_1b_1}} + \frac{1}{ s_{3a_1b_1}} \right)\times  \left(  \frac{1}{ s_{132a_1b_1}} + \frac{1}{ s_{134a_1b_1}} \right)  \, . \nonumber
\end{equation}
Adding the cuts, ${\rm cut-1 + cut-2 + cut-3} $, we obtain the final result for this four-point non-planar CHY-graph, namely
\vskip-0.5cm
\begin{eqnarray}\label{}
&&\hspace{-1.0cm}
\parbox[c]{8.0em}{\includegraphics[scale=0.23]{4pts_12-34g2.pdf}}
 = 
\frac{2^5}{s_{a_2b_2}\, s_{a_1b_1} } \left( \frac{1}{ s_{1a_1b_1} \,  s_{12a_1b_1}\, s_{123a_1b_1}    }
+\frac{1}{ s_{1a_1b_1} \,  s_{13a_1b_1}\, s_{132a_1b_1}    }
+\frac{1}{  s_{1a_1b_1} \,  s_{13a_1b_1}\, s_{134a_1b_1}    }
\right. \nonumber \\
&&
\hspace{3.0cm}
\left.
+ \frac{1}{  s_{3a_1b_1} \,  s_{31a_1b_1}\, s_{312a_1b_1}    }
+\frac{1}{ s_{3a_1b_1} \,  s_{31a_1b_1}\, s_{314a_1b_1}    }
+\frac{1}{  s_{3a_1b_1} \,  s_{34a_1b_1}\, s_{341a_1b_1}    }
\right). \nonumber
\end{eqnarray}
\vskip-0.05cm\noindent 
Therefore, by carrying out the integral, $\frac{1}{2^5}\int d\Omega \, s_{a_1b_1}$, it is trivial to check the previous expression becomes \eqref{4ptB-r}.

Finally, note that this gauge fixing can be applied to higher number of points.

\section{Four-point amplitude from the Feynman rules}
\label{4-p-app-fey}

In this appendix we calculate the four-point amplitude from the standard Feynman rules to compare with our final results based on the master formulas we obtained in previous sections. We have the following Lagrangian for the bi-adjoint scalar field theory
\begin{equation}
\mathcal{L}=\half\partial^\mu \Phi^{aa'}\partial_\mu \Phi^{aa'}+\frac{1}{3!}f^{abc}\tilde{f}^{a'b'c'}\Phi^{aa'}\Phi^{bb'}\Phi^{cc'}\,,
\end{equation}
in which we have the following Lie algebra 
\begin{equation}
[T^{a},T^{b}]=if^{abc}T^c~~~~,~~~~[\tilde{T}^{a},\tilde{T}^{b}]=i\tilde{f}^{abc}\tilde{T}^c\,,
\end{equation}
with two sets of generators $\{T^a\}$ and $\{\tilde T^a\}$ and their corresponding structure constants, $f^{abc}$ and $\tilde{f}^{abc}$ \cite{Mafra:2016ltu}. In this theory we only have the following three-vertex 
\begin{figure}[h]
\centering
\includegraphics[width=6cm]{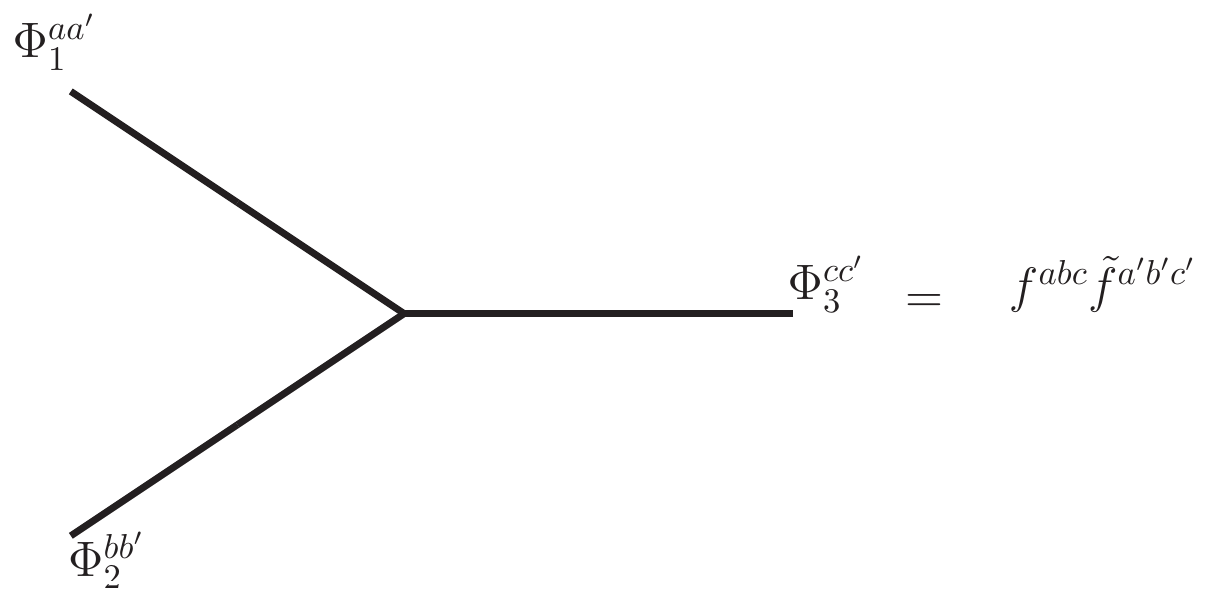}
\caption{Standard Feynman three-vertex in the bi-adjoint theory.}
\label{fig1}
\end{figure}
to be used for constructing tree- and loop-level diagrams, see \cite{Cachazo:2013iea} for some examples in tree-level amplitudes. In this paper we are interested in one-loop amplitudes in bi-adjoint scalar field theory, especially its non-planar contributions. The first one-loop diagram with non-planar counterpart in this theory is the four-point amplitude which will be constructed using the above three-vertex to compare with the CHY results in section \ref{four-point-chy}. After sewing four copies of the three-vertex in Fig. \ref{fig1} 
\begin{figure}[h]
\centering
\includegraphics[width=4cm]{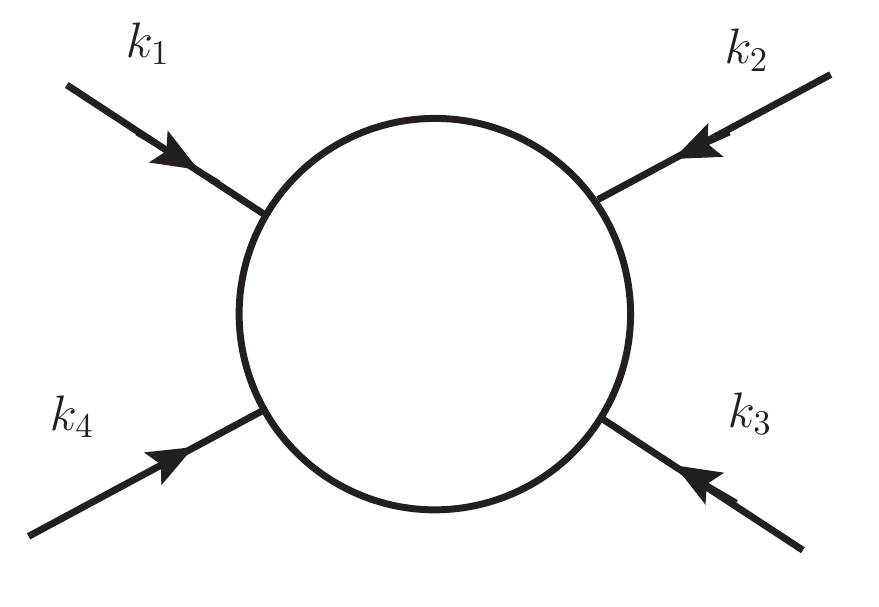}
\caption{Irreducible contribution to the one-loop four-point amplitude.}
\label{fig2}
\end{figure}
we get the irreducible contribution to the four-point amplitude depicted in Fig. \ref{fig2} which leads to the following representation
\begin{eqnarray}
\mathcal{A}_{\rm Irr}[k_1,k_2,k_3,k_4]&=&\delta^D\Big(\sum_{i=1}^4k_i\Big)\sum_{b,c,d,e=1}^{N^2}f^{ca_1b}\,f^{ba_2d}\,f^{da_3e}\,f^{ea_4c}\sum_{b',c',d',e'=1}^{\tilde{N}^2}\tilde{f}^{c'a'_1b'}\,\tilde{f}^{b'a'_2d'}\,\tilde{f}^{d'a'_3e'}\,\tilde{f}^{e'a'_4c'}\nonumber\\
&&\times\int \frac{d^Dl}{(2\pi)^D}\frac{1}{l^2[l+k_2]^2[l+k_2+k_3]^2[l-k_1]^2}\,.\non
\label{irr-4-point}
\end{eqnarray}
After using identities in (\ref{sum-identities}) for the natural ordering of the external scalars in the loop (canonical ordering: $1234$) we get 
\begin{eqnarray}
\sum_{b,c,d,e=1}^{N^2}f^{ca_1b}\,f^{ba_2d}\,f^{da_3e}\,f^{ea_4c}&=&\sum_{b,c,d,e}\tr(T^{c}[T^{a_1},T^b])\,\tr(T^{b}[T^{a_2},T^d])\,\tr(T^d[T^{a_3},T^e])\,\tr(T^e[T^{a_4},T^c])\non
&=&2N\tr(T^{a_1}T^{a_2}T^{a_3}T^{a_4})+2\tr(T^{a_1}T^{a_2})\tr(T^{a_3}T^{a_4})\non
&&+2\tr(T^{a_1}T^{a_3})\tr(T^{a_2}T^{a_4})+2\tr(T^{a_1}T^{a_4})\tr(T^{a_2}T^{a_3})\,,\non
\end{eqnarray}
and
\begin{eqnarray}
\sum_{b',c',d',e'=1}^{\tilde{N}^2}\tilde{f}^{c'a'_1b'}\,\tilde{f}^{b'a'_2d'}\,\tilde{f}^{d'a'_3e'}\,\tilde{f}^{e'a'_4c'}
&=&2\tilde{N}\tr(\tilde{T}^{a'_1}\tilde{T}^{a'_2}\tilde{T}^{a'_3}\tilde{T}^{a'_4})+2\tr(\tilde{T}^{a'_1}\tilde{T}^{a'_2})\tr(\tilde{T}^{a'_3}\tilde{T}^{a'_4})\non
&&+2\tr(\tilde{T}^{a'_1}\tilde{T}^{a'_3})\tr(\tilde{T}^{a'_2}\tilde{T}^{a'_4})+2\tr(\tilde{T}^{a'_1}\tilde{T}^{a'_4})\tr(\tilde{T}^{a'_2}\tilde{T}^{a'_3})\,,\non
\end{eqnarray}
where $N^2$ ($\tilde{N}^2$) is the dimension of $U(N)$ ($U(\tilde{N})$). Now if we exclude the fully planar part of (\ref{irr-4-point}) the rest of the amplitude gives
\begin{eqnarray}
\mathcal{A}'_{\rm Irr}[k_1,k_2,k_3,k_4]&=&\delta^D\Big(\sum_{i=1}^4k_i\Big) {\bf T'}^{1234}\int \frac{d^Dl}{(2\pi)^D}\frac{1}{l^2[l+k_2]^2[l+k_2+k_3]^2[l-k_1]^2}\,.\non
\label{irr-4-point-np}
\end{eqnarray}
We define 
\begin{eqnarray}
{\bf T'}^{1234}&=&{\bf T'}^{\rm 1234-P;NP}+{\bf T'}^{\rm 1234-NP;P}+{\bf T'}^{\rm 1234-NP;NP}\,,
\end{eqnarray}
in which  
\begin{eqnarray}
{\bf T}^{\rm 1234-P;NP}&=&4N\tr(\T^{a_1}\T^{a_2}\T^{a_3}\T^{a_4})\Big\{\tr(\Ti^{a_1}\Ti^{a_2})\tr(\Ti^{a_3}\Ti^{a_4})+\tr(\Ti^{a_1}\Ti^{a_3})\tr(\Ti^{a_2}\Ti^{a_4})\non
&&\hspace{6cm}+\tr(\Ti^{a_1}\Ti^{a_4})\tr(\Ti^{a_2}\Ti^{a_3})\Big\}\,,\non
\end{eqnarray}
represents the mixed planar-nonplanar (${\rm P;NP}$) part,
\begin{eqnarray}
{\bf T}^{\rm 1234-NP;P}&=&4\tilde{N}\tr(\Ti^{a_1}\Ti^{a_2}\Ti^{a_3}\Ti^{a_4})\Big\{\tr(\T^{a_1}\T^{a_2})\tr(\T^{a_3}\T^{a_4})+\tr(\T^{a_1}\T^{a_3})\tr(\T^{a_2}\T^{a_4})\non
&&\hspace{6cm}+\tr(\T^{a_1}\T^{a_4})\tr(\T^{a_2}\T^{a_3})\Big\}\,,\non
\end{eqnarray}
for the mixed nonplanar-planar (${\rm NP;P}$) part and
\begin{eqnarray}
{\bf T}^{\rm 1234-NP;NP}&=&4\Big\{\tr(\T^{a_1}\T^{a_2})\tr(\T^{a_3}\T^{a_4})+\tr(\T^{a_1}\T^{a_3})\tr(\T^{a_2}\T^{a_4})+\tr(\T^{a_1}\T^{a_4})\tr(\T^{a_2}\T^{a_3})\Big\}\non
&&\times\Big\{\tr(\Ti^{a_1}\Ti^{a_2})\tr(\Ti^{a_3}\Ti^{a_4})+\tr(\Ti^{a_1}\Ti^{a_3})\tr(\Ti^{a_2}\Ti^{a_4})+\tr(\Ti^{a_1}\Ti^{a_4})\tr(\Ti^{a_2}\Ti^{a_3})\Big\}\,,\non 
\label{canonical}
\end{eqnarray}
for the nonplanar-nonplanar (${\rm NP;NP}$) contribution of the color structure. In total we have six inequivalent diagrams for the one-loop four-point amplitude with the following orderings 
\begin{equation}
 S_4/\mathbb{Z}_4=\Big\{\{1234\},\{1243\},\{1324\},\{1342\},\{1423\},\{1432\}\Big\}\,,
 \label{orderings}
\end{equation}
which only the first ordering has been shown in (\ref{irr-4-point}). The full four-point amplitude (irreducible contribution) is the sum of all the orderings in (\ref{orderings}), which can be written as ($\mathcal{A}'_{\rm Irr}[i,j,k,l]\equiv\mathcal{A}'_{\rm Irr}[k_i,k_j,k_k,k_l]$)
\begin{eqnarray}
\mathcal{A}'_{\rm Irr}&=&\mathcal{A}'_{\rm Irr}[1,2,3,4]+\mathcal{A}'_{\rm Irr}[1,2,4,3]+\mathcal{A}'_{\rm Irr}[1,3,2,4]\non
&&+\mathcal{A}'_{\rm Irr}[1,3,4,3]+\mathcal{A}'_{\rm Irr}[1,4,2,3]+\mathcal{A}'_{\rm Irr}[1,4,3,2]\,.\non
\label{full-ampf}
\end{eqnarray}
Note that the ${\bf T}^{\rm 1234-NP;NP}$ is invariant under exchanging any two external scalars, which indicates its appearance for all six orderings.\\

In the following we present the CHY results from (\ref{full-amp}) for this amplitude to compare with the results from the Feynman rules in (\ref{full-ampf}). ${\bf m}_4^{1-{\rm loop}}$ in (\ref{full-amp}) has two parts, one for the color decomposition and one for the momentum integral. If we exclude the planar contribution (which has been discussed in \cite{Gomez:2017cpe}) in ${\bf m}_4^{1-{\rm loop}}$ we have:
\begin{eqnarray}
I^{({\rm 1-P;NP})}&=& 4( N)  \sum_{p=1 }^{3}
 \sum_{\pi\in S_{4}/\mathbb{Z}_{4}   }
  \sum_{\g\in S_{p}/\mathbb{Z}_{p} \atop  \d\in S_{4-p}/\mathbb{Z}_{4-p}  }
{\rm Tr}(T^{i_{\pi_1}} \cdots T^{i_{\pi_4}}  ) \times 
m_{4}^{(\rm 1-P;NP)}[\pi \, ; \, \g | \d ]
\non
&&\hspace{4cm}\times
{\rm Tr}(\tilde T^{i_{\g_1}} \cdots  \tilde T^{i_{\g_p}}  ) \times {\rm Tr}(\tilde T^{i_{\d_{p+1}}} \cdots  \tilde T^{i_{\d_4}}  ) \nonumber \\
&=&4N\tr(T^{a_1}T^{a_2}T^{a_3}T^{a_4}  )
\times\Big\{
{\rm Tr}(\tilde {T}^{a_1}\tilde{T}^{a_2} ) \times {\rm Tr}(\tilde{T}^{a_3}\tilde{T}^{a_4}  )m_{4}^{(\rm 1-P;NP)}[1,2,3,4 \, ; \,1,2| 3,4 ]\non
&&\hspace{4cm}+{\rm Tr}(\tilde {T}^{a_1}\tilde{T}^{a_3} ) \times {\rm Tr}(\tilde{T}^{a_2}\tilde{T}^{a_4}  )m_{4}^{(\rm 1-P;NP)}[1,2,3,4 \, ; \,1,3| 2,4 ]\non
&&\hspace{4cm}+{\rm Tr}(\tilde {T}^{a_1}\tilde{T}^{a_4} ) \times {\rm Tr}(\tilde{T}^{a_2}\tilde{T}^{a_3}  )m_{4}^{(\rm 1-P;NP)}[1,2,3,4 \, ; \,1,4| 2,3 ]\Big\} \non
&&+{\rm five~more~permutations}\,,\non
\end{eqnarray}
which coincides with the color structure of (\ref{irr-4-point-np}) after replacing ${\bf T'}^{1234}$ with ${\bf T'}^{\rm 1234-P;NP}$ and considering other permutations in (\ref{full-ampf}). Now the loop integral appearing in (\ref{4p-NP-P2}) (after some momentum shifts) corresponds to what we have in (\ref{irr-4-point-np}). We also used the fact that 
\begin{eqnarray}
&&\mathfrak{M}_{4}^{(\rm 1-P;NP)}[1,2,3,4 \, ; \,1,2| 3,4 ]\equiv\mathfrak{M}_{4}^{(\rm 1-P;NP)}[1,2,3,4 \, ; \,1,3| 2,4 ]\equiv \mathfrak{M}_{4}^{(\rm 1-P;NP)}[1,2,3,4 \, ; \,1,4| 2,3 ]\,.\non
\end{eqnarray}
The same discussion holds for the ${\rm NP;P}$ part. 

Now, the most nontrivial part is the ${\rm NP;NP}$ contribution. For this piece of the amplitude from (\ref{full-amp}) we get 
\begin{eqnarray}
I^{({\rm 1-NP;NP})}&=& 4\sum_{p=1 \atop q=1 }^{3}
  \sum_{\pi\in S_{p}/\mathbb{Z}_{p}, \, \g\in S_{q}/\mathbb{Z}_{q} \atop  \rho\in S_{4-p}/\mathbb{Z}_{4-p}  , \, \d\in S_{4-q}/\mathbb{Z}_{4-q}}
{\rm Tr}(T^{i_{\pi_1}} \cdots T^{i_{\pi_p}}  ) \times 
{\rm Tr}(T^{i_{\rho_{p+1}}} \cdots T^{i_{\rho_4}}  ) \times
m_{4}^{(\rm 1-NP;NP)}[\pi | \rho \, ; \, \g | \d ]\non
&&\hspace{5cm}\times {\rm Tr}(\tilde T^{i_{\g_1}} \cdots  \tilde T^{i_{\g_q}}  ) \times {\rm Tr}(\tilde T^{i_{\d_{q+1}}} \cdots  \tilde T^{i_{\d_4}} )\non
&=&4\tr(T^{a_1}T^{a_2})\tr(T^{a_3}T^{a_4})\Big\{m_{4}^{(\rm 1-NP;NP)}[1,2 | 3,4 \, ; \,1,2 | 3,4 ] \tr(\tilde{T}^{a_1}\tilde{T}^{a_2})\tr(\tilde{T}^{a_3}\tilde{T}^{a_4})\non
&& \hspace{4cm}+m_{4}^{(\rm 1-NP;NP)}[1,2 | 3,4 \, ; \,13 | 24 ] \tr(\tilde{T}^{a_1}\tilde{T}^{a_3})\tr(\tilde{T}^{a_2}\tilde{T}^{a_4})\non
&&\hspace{4cm}+m_{4}^{(\rm 1-NP;NP)}[1,2 | 3,4 \, ; \,14 | 23 ] \tr(\tilde{T}^{a_1}\tilde{T}^{a_4})\tr(\tilde{T}^{a_2}\tilde{T}^{a_3})\Big\}\non
&&+4\tr(T^{a_1}T^{a_3})\tr(T^{a_2}T^{a_4})\Big\{m_{4}^{(\rm 1-NP;NP)}[1,3 | 2,4 \, ; \,1,2 | 3,4 ] \tr(\tilde{T}^{a_1}\tilde{T}^{a_2})\tr(\tilde{T}^{a_3}\tilde{T}^{a_4})\non
&& \hspace{4cm}+m_{4}^{(\rm 1-NP;NP)}[1,3 | 2,4 \, ; \,13 | 24 ] \tr(\tilde{T}^{a_1}\tilde{T}^{a_3})\tr(\tilde{T}^{a_2}\tilde{T}^{a_4})\non
&&\hspace{4cm}+m_{4}^{(\rm 1-NP;NP)}[1,3 | 2,4 \, ; \,14 | 23 ] \tr(\tilde{T}^{a_1}\tilde{T}^{a_4})\tr(\tilde{T}^{a_2}\tilde{T}^{a_3})\Big\}\non
&&+4\tr(T^{a_1}T^{a_4})\tr(T^{a_2}T^{a_3})\Big\{m_{4}^{(\rm 1-NP;NP)}[1,4 | 2,3 \, ; \,1,2 | 3,4 ] \tr(\tilde{T}^{a_1}\tilde{T}^{a_2})\tr(\tilde{T}^{a_3}\tilde{T}^{a_4})\non
&& \hspace{4cm}+m_{4}^{(\rm 1-NP;NP)}[1,4 | 2,3 \, ; \,13 | 24 ] \tr(\tilde{T}^{a_1}\tilde{T}^{a_3})\tr(\tilde{T}^{a_2}\tilde{T}^{a_4})\non
&&\hspace{4cm}+m_{4}^{(\rm 1-NP;NP)}[1,4 | 2,3 \, ; \,14 | 23 ] \tr(\tilde{T}^{a_1}\tilde{T}^{a_4})\tr(\tilde{T}^{a_2}\tilde{T}^{a_3})\Big\}\,.\non
\label{4-p-NP-NP2}
\end{eqnarray}
We have already calculated the momentum integrals appearing here in (\ref{4pt-NP-NP}), it  contains all the six inequivalent momentum integrals we have for the four-point amplitude, which means all $m_{4}^{(\rm 1-NP;NP)}[\cdots] $ in (\ref{4-p-NP-NP2}) are equivalent. After factorizing the momentum integral, (\ref{4-p-NP-NP2}) is exactly what we have for the ${\rm NP;NP}$ contribution from the standard computation in ${\bf T}^{\rm 1234-NP;NP}$ after considering all permutations and momentum integrals in (\ref{full-ampf}). 

To summarize, in this appendix we have shown the exact equivalence for the non-planar contribution to the four-point amplitude between our formula in the CHY side (\ref{full-amp}) and the one from the standard method based on the Feynman rules for the bi-adjoint scalar field theory in (\ref{full-ampf}). The same discussion applies to higher order amplitudes at one-loop level.

\bibliographystyle{JHEP}
\bibliography{mybib}
\end{document}